 \documentclass[journal, onecolumn,draftcls,12pt]{IEEEtran} 
\usepackage{amsmath,amsfonts}
\usepackage{algorithmic} 
\usepackage{algorithm, color}
\usepackage{array}

\usepackage[font=small]{caption}
\usepackage{textcomp}
\usepackage{stfloats}
\usepackage{url}
\usepackage{subfigure}  

\usepackage{verbatim}
\usepackage{epsfig,graphicx}
\usepackage{cite} 
\usepackage[T1]{fontenc}
\usepackage{amssymb}

\newtheorem{Theo}{\bf Theorem}
\newtheorem{Lem}{\bf Lemma} 

\newtheorem{remark}{Remark}
\newtheorem{Proposition}{\bf Proposition} 
\newtheorem{Def}{\bf Definition} 
\usepackage{makecell}
\usepackage{threeparttable} 
\usepackage{multirow} 
\usepackage{booktabs}
\usepackage{extarrows}

\makeatletter

\newcommand{\Rmnum}[1]{\expandafter\@slowromancap\romannumeral #1@}
\makeatother
\usepackage{autobreak} 
\allowdisplaybreaks 

\IEEEoverridecommandlockouts

\title{Demand Private Coded Caching: Small Cache Size}
\author{Qinyi Lu, Nan Liu, Wei Kang, and  Chunguo Li
\thanks{(Corresponding author: Nan Liu) 
	This article was presented in part at the ITW conference \cite{luitw2024}. 
	
	Q. Lu, W. Kang and C. Li are with the School of Information Science and Engineering, Southeast University, Nanjing 211189, China. 
	(e-mail: \{qylu, wkang, chunguoli\}@seu.edu.cn).
	N. Liu is with the National Mobile Communications Research Laboratory, Southeast University, Nanjing 211189, China. (e-mail: nanliu@seu.edu.cn). }}%
\begin{document} 
\maketitle  

\begin{abstract}  
	We investigate the demand private coded caching problem, which is an $(N,K)$ coded caching problem with $N$ files, $K$ users, each equipped with a cache of size $M$, and an additional privacy constraint on user demands, i.e., each user can not gain any information about the demands of other users. We focus on  scenarios where the size of users' caches is small, aiming to further characterize the fundamental limits of this problem. We first present a new virtual-user-based achievable scheme for arbitrary number of users and files, and two MDS-code-based achievable schemes for  the case $N \le K$. With a newly derived  converse  bound for the case $N \le K$, these  proposed schemes lead  to  the optimal  memory-rate tradeoff of the demand private coded caching problem for $M \in \big[0, \frac{N}{(K+1)(N-1)} \big] $ where   $N \le K \le 2N-2$, and the optimal  memory-rate tradeoff for  $M \in \big[0, \frac{1}{K+1} \big] $ where $ K > 2N-2$.  Moreover, for the case of 2 files and arbitrary number of users,  by deriving another new converse bound, the optimal  memory-rate tradeoff is characterized  for $M\in \big[0,\frac{2}{K}\big] \cup \big[\frac{2(K-1)}{K+1},2\big]$.  Finally, we provide the optimal  memory-rate tradeoff of the demand private coded caching problem for 2 files and 3 users.  
\end{abstract}    

 
\section{Introduction}
Maddah Ali and Niesen proposed  the coded caching problem from the perspective of information theory and studied its fundamental  limits in \cite{MaddahAli2014}.
There are $K$ users and one server, who has access to $N$ files of equal size in the system. Each user has a cache which can store $M$ files and is connected to the server through an error-free shared link.   
In the placement phase, the cache of each user is filled by a function of $N$ files. 
In the delivery phase, each user requests one file from the server, and the server broadcasts a signal to all $K$ users based on the received requests. 
The goal is to minimize the worst-case  rate of the broadcast signal, given a cache size $M$, while ensuring that each user can correctly decode the required file.   
The caching and delivery scheme  proposed  in  \cite{MaddahAli2014} is proved to be order optimal and 
the optimal memory-rate tradeoff for the coded caching problem is still open for general $N$ and $K$. 
Several works had made efforts to characterize the optimal  memory-rate tradeoff by proposing improved achievable schemes \cite{Wan2016, Chen2016, Mohammadi2016, Mohammadi2017, Qian2018, Tian2018,  Gomez2018, Kumar2019, Shao2019ITW, Shao2019Tcom, Wan2020 } or proving tighter bounds \cite{Sengupta2015, Ajaykrishnan2015, Tian2016, Ghasemi2017, Wang2018, Qian2019,Kumar2023}.
In particular, the scheme proposed by Yu \emph{et al.}, referred to as \emph{the YMA scheme},  
is proved to be optimal under the condition of uncoded placement \cite{Qian2018}.   
 
In addition, the coded caching problems under extended models, such as \emph{nonuniform demands} \cite{Niesen2017,Zhang2018,Deng2022,Deng2022Uncoded},  \emph{decentralized}  users \cite{MaddahAli2015,Wei2017,Reisizadeh2018},   \emph{Device-to-Device} systems \cite{Ji2016,Yapar2019,Wang2023D2D}, and \emph{asynchronous} systems  \cite{Niesen2015,Ghasemi2020,Yang2022}, have also been explored.
In this paper, we focus on the privacy of users based on the coded caching model presented in \cite{MaddahAli2014}. Therefore, we do not discuss these extended models in detail. 

As analyzed in \cite{Aravind2020, Wan2021}, schemes designed for traditional coded caching problem, referred to as  \emph{non-private schemes},  
leak users' demands to other users within the system. 
This is undesirable for participating users as it violates their privacy. Several works have investigated the coded caching problem with the constraint of demand privacy \cite{Chinmay2022,Aravind2020,  Aravind2022,Wan2021,Yan2021,Namboodiri2021,Gholami2023}. 
More specifically,  Gurjarpadhye  \emph{et al.}  proposed three schemes and the best performance of these three schemes together is shown to be order optimal for the demand private coded caching problem\cite{Chinmay2022}. 
In particular, first, a coded caching scheme designed without considering demand privacy is proposed   
in \cite[Lemma 1]{Chinmay2022}, which we call \emph{the GRK non-private scheme},   
where there are $N$ files, $NK$ users  and the demands are restricted to a set of special  demands of size $N^K$.  Then, a coded caching scheme  considering demand privacy is proposed in \cite[Theorem 2]{Chinmay2022} by using the GRK non-private    scheme and the  idea of \emph{virtual users},    i.e.,  user's demand privacy is ensured by creating $NK-K$ virtual users and having each file requested by exactly $K$ users out of a total of $NK$ users. 
In addition, the optimal memory-rate tradeoff with $N \ge K = 2$ for the demand private coded caching problem  is found\cite{Chinmay2022}.  
  
Other related works on coded caching with demand privacy include  \cite{Aravind2020, Aravind2022}, where Aravind \emph{et al.} mainly focused on  reducing subpacketization  for the problem studied in \cite{Chinmay2022},   and as a result, they proposed new schemes using placement delivery arrays (PDAs). 
In \cite{Wan2021}, Wan and Caire  considered the scenario where users request multiple files and proposed two schemes, one based on virtual users and the other based on a novel MDS-based cache placement 
which significantly reduces subpacketization.  
By using the MDS-based scheme, the optimal  memory-rate tradeoff  on $ M \ge \min \big\{ \frac{2K-1}{2K}, \frac{2^{K-1}}{2^{K-1}+1} \big\} N$ is  characterized  in \cite[Theorem 6]{Wan2021}.      
In \cite{Yan2021}, Yan and Tuninetti considered demand privacy against colluding users in both scenarios: Single File Retrieval (SFR) and Linear Function  Retrieval (LFR), and provided a new converse bound for the SFR scenario.  
The optimal memory-rate tradeoff when $ M \le \frac{1}{K(N-1)+1}$ for the SFR scenario is characterized in \cite{Namboodiri2021}. 
In \cite{Gholami2023}, Gholami \emph{et al.} constructed a new private coded caching scheme based on any Private Information Retrieval (PIR)  scheme and investigated the coded caching problem with simultaneously private demands and caches.  
Several demand private problems under different coded caching models such as Multi-access \cite{Wan2021Multi, Namboodiri2022, Chinnapadamala2022}, Device-to-Device \cite{Wan2020D2D, Wan2020D2DConverse, Wan2022D2D}, with a wiretapper\cite{Yan2021Security, Yan2022Secure}, and  with offline users \cite{Yan2022Secure,Ma2023, ma2024coded} have also been investigated.

In this paper, we focus on the demand private coded caching problem studied in \cite{Chinmay2022}.    
Table \ref{result} summarizes the optimality results achieved in the previous works.  
Just like the coded caching problem without privacy constraint, the optimal   memory-rate tradeoff for general demand private coded caching problem is still open.      
 \begin{table*}  
	\begin{center}  
		\caption{Summary of Existing Optimal Results for Demand Private Coded Caching Problem}   
		\label{result}
		 \vspace{-1mm}
		\begin{threeparttable} 
			 \setlength{\tabcolsep}{4pt}
			\begin{tabular}{|c|c|c|c|}   
				\hline  
				Reference  & Cache Size  &  Optimal  Memory-rate Tradeoff  & Condition \\
				\hline
				\multirow{2}{*}{\cite[Theorem 6]{Chinmay2022} } & \multirow{2}{*}{$   [0,N] $ }  & $  	 R_{2,2}^{*p}(M)     = \max \big \{ 2-2M, \frac{5-3M}{3},  \frac{2-M}{2} \big \} $   &   $N = K=2$      \\ 
				\cline{3-4} 
				& & $  	 R_{N,2}^{*p}(M)     = \max \big \{ 2- \frac{3}{N}M, \frac{2N+1-3M}{N+1},  \frac{N-M}{N} \big \} $  & $N > K=2$  \\
				\hline   
				\cite[Theorem 4]{Namboodiri2021} & $   \big[0,\frac{1}{K(N-1)+1}\big] $ & $  	R_{N,K}^{*p}(M) = N(1-M) $  &   $N \le K$        \\
				\hline   
				\cite[Theorem 6]{Wan2021} & $    \big[ \min \big\{ \frac{2K-1}{2K}, \frac{2^{K-1}}{2^{K-1}+1} \big\}N, N\big] $    & $  	R_{N,K}^{*p}(M) = 1- \frac{M}{N} $  &  -      \\
				\hline
				\multirow{4}{*}{This paper}    
				& $[0, \frac{1}{K+1}]$  &  $R_{N,K}^{*p}(M) = N(1-M)$ & $N \le K$     \\
				\cline{2-4} 
				& $[\frac{1}{K+1}, \frac{N}{(K+1)(N-1)}]$  &  $  	 R_{N,K}^{*p}(M) = N-\frac{1}{K+1} - (N-1)M   $  &   $ N \le K \le 2N-2 $  \\
				\cline{2-4}
				& $[0,\frac{2}{K}] \cup [\frac{2(K-1)}{K+1},2] $  &  See \eqref{cor1} in  Theorem  \ref{Theop2}
				&   $K \ge N = 2$   \\
				\cline{2-4} 
				& $[0,N]$  & See \eqref{cor2} in  Theorem  \ref{Theop2}  &   $K = 3, N = 2$  \\
				\hline 
			\end{tabular}
		\end{threeparttable}
		\vspace{-1cm}
	\end{center}
\end{table*}
In this paper, we focus on scenarios where the size of users' caches is small, aiming to further elucidate the optimal memory-rate tradeoff for the demand private coded caching problem.  

\subsection{Main Contributions} 
The contributions and novelties of our paper are summarized as follows.  

$\bullet$  
We propose three novel achievable schemes as follows.   
\begin{enumerate} 
	\item  
	 Based on the idea of virtual users and with a newly proposed non-private scheme	for restricted demand subset, we present a new achievable demand private scheme for arbitrary number of users and files.    In terms of the proposed non-private scheme, while the idea of exchanging cache contents and the corresponding delivery signals  \cite[Lemma 1]{Aravind2020}  aids in the design of the proposed non-private scheme for restricted demand subset, our novelty lies in the correspondence rule between the cache content and the delivery signal, as well as the construction of delivery signals that cannot be directly obtained through exchanging. 
	\item     We  present two novel demand private schemes, achieving the memory-rate pairs $ \left( \frac{1}{K+1}, \frac{KN}{K+1} \right)$ and $\left( \frac{N}{(K+1)(N-1)},\frac{KN-1}{K+1}\right)$   for $N \le K$, respectively. The proposed two schemes are inspired by the cache placement design of a non-private scheme stated in \cite[Theorem 1]{Gomez2018}, and furthermore, 
	the techniques of MDS-based cache placement \cite{Wan2021} and  random permutation for delivery signals \cite{Chinmay2022}  are applied to ensure the consistency of the delivery signal structure across different demands, thereby guaranteeing demand privacy.  
	\end{enumerate}  
Numerical results show that the best performance of these   proposed schemes together outperforms existing achievable schemes when the cache size $M$ is small and $N \leq K$. 
  
$\bullet$ For the case where the number of users is greater than the number of files, we derive a new information-theoretic converse bound, which is obtained by  suitably combining entropy terms containing different cache contents and delivery signals, and using techniques such as induction and recursion. 
The newly derived converse improves upon the previously known  bounds, and for the case of $N \le K \le 2N-2$, it  characterizes the optimal  memory-rate tradeoff on  $M \in \big[0, \frac{N}{(K+1)(N-1)} \big] $ in conjunction with the  above proposed schemes.

$\bullet$  For the case of two files and arbitrary number of users, we further derive a new converse bound through 
swapping the cache contents and delivery signals in the proof of the previously obtained converse bound. As a result, the optimal  memory-rate tradeoff for the case of $N=2$ on $M\in \big[0,\frac{2}{K}\big] \cup \big[\frac{2(K-1)}{K+1},2\big]$ is obtained.  
As for the case of 2 files and 3 users, we show that the  proposed virtual-user-based scheme, together with the achievability result in \cite[Theorem 2]{Chinmay2022}, and the proposed converse meet, thus characterizing the optimal  memory-rate tradeoff for any $M \in [0,N]$.

Table \ref{result} summarizes the optimal  memory-rate tradeoff results obtained in this paper.

\subsection{Paper Organization}
The remainder of this paper is organized as follows. In Section \ref{System}, we provide the system model of the demand private problem.  
We present our results and briefly describe the sketch of the proof in Section \ref{Main}.  
The detailed proofs of achievable schemes and converse bounds can be found in Sections \ref{sec4} and \ref{seccon}, respectively,  while Section \ref{sec7} provides the conclusions. 

\subsection{Notations}  

We use $[a:b]$ to denote the set $\{a,a+1,\dots,b\}$,   $[a]=[0:a-1]$ to denote the set  $\{0,\dots,a-1\}$,
$[a, b]$ to denote the closed interval between two real numbers $a$ and $b$,  
and $\text{Unif}( [a:b])$ to denote the discrete uniform distribution  over set $[a:b]$.   
$ (X_i)_{i \in \mathcal{I}}$ denotes a vector composed of $X_i$ for each $i \in \mathcal{I}$, arranged in lexicographical order; 
$\left(X_{i,j}\right)_{i \in \mathcal{I}, j \in \mathcal{J}_i} = \left( \left( X_{i,j}\right)_{j \in \mathcal{J}_i} \right)_{i \in \mathcal{I}}$;   
$X_{\mathcal{I}}  =  \{X_i\}_{i\in \mathcal{I} }$ denotes a set of random variables;    
$\mathbf{e}_i \in \{0,1\}^{KN-N+1}$, $\mathbf{e}_i' \in [N]^{K}$   denote the unit vector where the $i$-th ($i$ starts from 0) element  is 1 and the other elements are 0;    
$\boldsymbol{1}_K = (1,\dots,1)$  denotes the $K$-length vector with all elements equal to $1$, and
$(a\boldsymbol{1}_{K-k},b\boldsymbol{1}_k)$ denotes the $K$-length vector composed of $K-k$ elements of $a$ followed by $k$ elements of $b$.    
Let $\oplus$ denote bit-wise XOR,  
and  $\oplus_q$ and $\ominus_q$ denote the addition and subtraction modulo $q$, respectively.

\section{System Model} \label{System}
 We study an $(N,K)$ demand private coded caching problem as follows. The system has a
 server connected to $K$  cache-aided users through an error-free shared link.
 The server has access to a library of $N$ independent files of $F$ bits each. Let $W_n$, $n \in [0:N-1]  = [N]$, represent these files and these files are uniformly distributed on $\{0,1\}^F$.  Each user $k \in  [0:K-1] = [K] $  has a  cache whose size is limited to  $MF$ bits. 
 The coded caching system operates in two phases as follows.
 \subsubsection{Placement Phase} 
 From the probability space $ \mathcal{P}$, the server generates a random variable $P$, 
 which is not known to any user.
 Then the server places at most $MF$ bits in user $k$'s cache through a cache function $\phi_k$. More specifically, for $\forall k \in [K]$, the cache function of the $k$-th user is  known to  all $K$ users in the system and is a map  as follows 
 \begin{align*}
 \phi_k: \mathcal{P} \times [0:2^F-1]^N \longrightarrow [0:2^{MF}-1] .
 \end{align*} 
 Let $Z_k$ denote the cache content for user $k$, so $Z_k$ is given by 
 \begin{align}   \label{sysZ}
 Z_k = \phi_k \left( W_{[N]},P \right) , \quad \forall k\in[K].
 \end{align}

 \subsubsection{Delivery Phase}
 In the delivery phase, each user demands one of the $N$ files. 
 Let $D_k$ represent the demand of user $k$, and the demand for all users in the system can be denoted by the vector $\boldsymbol{D} = (D_0,D_1, \dots, D_{K-1})$. 
 $D_{[K]}$ are all i.i.d. random variables and uniformly   distributed on $[N]$, and let $\mathcal{D} = [N]^K$ denote the set of all possible demand vectors in the system.  
 The files $W_{[N]}$, the randomness  $P$ and the demands $D_{[K]}$ are independent, i.e.,  
 \begin{align}  \label{sysindependent}
 H\left( D_{[K]}, P, W_{[N]} \right) =   \sum_{k=0}^{K-1} H(D_k) + H(P) + NF. 
 \end{align} 
 Note that the demands $D_{[K]}$ are also independent of the cache contents $Z_{[K]}$. As $Z_k$ is  a deterministic function of $W_{[N]}$ and $P$ (see \eqref{sysZ}), the independence between $D_{[K]}$ and $Z_{[K]}$ is implied by \eqref{sysindependent}.
 
 Each user sends its own demand to the server through a  private link. Upon receiving the demands of all users, the server broadcasts a signal $X_{\boldsymbol{D}}$  of size $RF$ bits and the signal $X_{\boldsymbol{D}}$ is created through an encoding function $\psi$. 
 The encoding function $\psi$ is a mapping as follows 
 \begin{align*}
 \psi : \mathcal{P} \times \mathcal{D} \times [0:2^F-1]^N \longrightarrow [0:2^{RF}-1],
 \end{align*} 
 where $R$ is referred to as the rate of the shared link.
 Hence, the delivery signal under demand $\boldsymbol{D}$, i.e., $X_{\boldsymbol{D}}$, is given by
 \begin{align*} 
 X_{\boldsymbol{D}} = \psi \left(W_{[N]},\boldsymbol{D},P  \right).  
 \end{align*} 
 The $(N,K,M,R)$-private coded caching scheme consists of $\phi_{[K]}$ and $ \psi$, and it must  satisfy the following correctness and privacy constraints.
 
 \emph{Correctness:}
 Each user $k\in[K]$ must decode its  demand file $W_{D_k}$ by using $(X_{\boldsymbol{D}},Z_k)$ with no error, i.e., 
 \begin{align}  \label{decoding}
 H\left( W_{D_k} | Z_k, X_{\boldsymbol{D}},D_k\right) = 0, \quad \forall k\in [K].
 \end{align} 
 
 \emph{Privacy:}
 The demand privacy requires that any user can not infer any information about the demand  $\boldsymbol{D}$ beyond its own demand  $D_k$ from all the information it possesses, i.e.,  
 \begin{align}  \label{privacy2}
 I \left(D_{[K]\setminus \{k\}};X_{\boldsymbol{D}},Z_{k} ,D_{k} \right) = 0, 
 \quad \forall k \in [K].
 \end{align}
 
 The pair $(M,R)$ is said to be achievable for the $(N,K)$ coded caching problem with demand privacy if there exists an $(N,K,M,R)$-private coded caching scheme for large enough $F$.
 The optimal memory-rate tradeoff for the $(N,K)$ coded caching problem with demand privacy is defined as 
 \begin{align}  \label{stsMR} 
 R_{N,K}^{*p}(M) = \inf \{R: \text{$(M,R)$  is achievable} \}. 
 \end{align}
 where the infimum is taken over all achievable $(N,K,M,R)$-private coded caching schemes.

 \begin{remark} We mainly consider the privacy constraint given by (\ref{privacy2}) in this paper. However, a stronger privacy constraint with colluding users is considered in \cite{Yan2021}, given by 
 	\begin{align}  \label{privacy1}
 		I \left(D_{[K]\setminus \mathcal{S}};X_{\boldsymbol{D}},Z_{\mathcal{S}},D_{ \mathcal{S}} \big| W_{[N]} \right) = 0, \text{ }   
 		\forall \mathcal{S} \subseteq  [K], \mathcal{S} \neq \emptyset,  
 	\end{align} 
 	where $\mathcal{S}$ denotes the set of colluding users, and the colluding set may be any subset of $[K]$. This constraint is stronger than (\ref{privacy2}) in two ways: 1) the users in set $\mathcal{S}$ are allowed to collude to attempt to learn about the demands of users not in $\mathcal{S}$; 2) the privacy of the users can not be leaked even when all users know all $N$ files.   
 \end{remark}  
 
\section{Main Results}  \label{Main}
The following are the main results of this paper.  
In Subsections \ref{main_A} and \ref{main_B}, we present three new achievable schemes, whose numerical evaluations are provided in Subsection \ref{main_C}.     
 Subsection \ref{main_D}  presents a new converse bound,  and by comparing the obtained converse and achievability results, the optimal memory-rate tradeoff is identified for the case where the cache size $M$ is small and $N \leq K$. 
Finally, Subsection \ref{main_E} focuses on the scenario in which \( N = 2 \). 
With an additional converse bound, the optimal memory-rate tradeoff is identified for the special case of \( N = 2 \) and \( K = 3 \), as well as for a special region of cache size, in the case of \( 2 \) files and an arbitrary number of users.

\subsection{A New Virtual User Scheme} \label{main_A}
The well-known idea of  virtual users, as provided in \cite{Chinmay2022}, presents a method for deriving a scheme for  the $(N,K)$ demand private coded caching problem from an existing scheme for  the $(N,NK)$ $\mathcal{D_{RS}}$-non-private coded caching problem, which is an $(N,NK)$ coded caching problem that considers demands in a specific set $\mathcal{D_{RS}}$  called the restricted demand subset.   
Recalling the definition of 
$\mathcal{D_{RS}}$  in \cite{Chinmay2022}, we have the following definition.  

	\begin{Def}[Restricted Demand Subset $\mathcal{D_{RS}}$]\label{Def1}
	The demands of the $NK$ users are denoted by  an $NK$-length vector $\overline {\boldsymbol{d}}$, where the $i$-th component of $\overline {\boldsymbol{d}}$  denotes the demand of user $i$. For any $\overline {\boldsymbol{d}} \in \mathcal{D_{RS}} $, $ \overline {\boldsymbol{d}}$ can be divided into $K$ subvector with $N$ length each, i.e., $\overline {\boldsymbol{d}} = \big(   \overline {d}^{(0)},  \overline {d}^{(1)}, \dots ,  \overline {d}^{(K-1)} \big) $,  
	where $\overline{d}^{(k)} = \big(\overline{d}_0^{(k)},\overline{d}_1^{(k)},\dots, \overline{d}_{N-1}^{(k)} \big)$ is an $N$-length vector and is a cyclic shift of the vector $(0,1,\dots,N-1)$ for any $ k \in [K]$.   
	\end{Def} 
	
	For $k\in [K]$, we further define $d_k = \overline{d}_0^{(k)} $ to denote the number of left cyclic shifts of $(0,1,\dots,N-1)$ needed to get $\overline{d}^{(k)}$.   
	We note that any $ \overline {\boldsymbol{d}} \in \mathcal{D_{RS}} $  can be uniquely represented by the vector $ \boldsymbol{d} = (d_0,d_1,\dots,d_{K-1})$, and $ \boldsymbol{d} \in [N]^K = \mathcal{D}$.   
	Taking \( N = 2 \) and \( NK = 4 \) as an example, all the demand vectors in \( \mathcal{D_{RS}} \) and their corresponding \( \boldsymbol{d} \)  are  presented  in Table \ref{Restricted}.    
\begin{table}
	\begin{center}  
		\caption{Restricted Demand Subset $\mathcal{D_{RS}}$  for $N=2$ and $NK=4$}     
		\label{Restricted}
		\begin{tabular}{|c|c|c|c|}
			\hline
			$ \overline {\boldsymbol{d}} $   & $\overline{d}^{(0)}  $ & $\overline{d}^{(1)}  $   & $  {\boldsymbol{d}}$ \\ 
			\hline
			$(0,1,0,1)$   & $(0,1)$  & $(0,1)$  &     	$(0,0)$   \\ 
			\hline
			$(0,1,1,0)$   & $(0,1)$  & $(1,0)$  &     	$(0,1)$   \\ 
			\hline
			$(1,0,0,1)$   & $(1,0)$  & $(0,1)$  &     	$(1,0)$   \\ 
			\hline
			$(1,0,1,0)$   & $(1,0)$  & $(1,0)$  &     	$(1,1)$   \\ 
			\hline 
		\end{tabular}
	\end{center}
	  \vspace{-1cm}
	\end{table} 
	
 We first propose a new achievable scheme for the $(N,NK)$ $\mathcal{D_{RS}}$-non-private coded caching problem in Lemma 1, then using the idea of virtual users, we transform the new achievable scheme for the $(N,NK)$  $\mathcal{D_{RS}}$-non-private coded caching into a new achievable scheme for the demand private coded caching problem in Theorem \ref{achD}.          
\begin{Lem} \label{lemach1} 
	For an $(N,NK)$ $\mathcal{D_{RS}}$-non-private coded caching problem,  the following memory-rate pairs  
	\begin{align*}   
		\left(M,R \right) = &   \left(  \frac{\tbinom{NK-K+1}{r+1}-\tbinom{NK-K-N+1}{r+1}}{\tbinom{NK-K+1}{r}}
		,\frac{Nr}{NK-K+1} \right) , 
	\end{align*}
	where $r \in [0:NK-K+1]$ are achievable.
\end{Lem}

\emph{\quad Sketch of Proof:}
	The proposed scheme is designed based on the YMA scheme for $N$ files and $NK-K+1$ users.  
	More specifically, we establish a one-to-one mapping between the $NK - K + 1$ demands from $\mathcal{D_{RS}}$ and users in the YMA scheme.  
	Following this mapping, we further define a corresponding demand in the YMA scheme for each user in the $(N, NK)$ $\mathcal{D_{RS}}$-non-private coded caching problem, such that   the cache content of each user is given by the delivery signal under the corresponding demand.   
	The delivery signals under the $NK-K+1$ demands from $\mathcal{D_{RS}}$ are given by  the cache contents of the corresponding users in the YMA scheme, while the delivery signals for the remaining demands are constructed  additionally to ensure correctness. 
	The detailed proof can be found in Section \ref{secach}.
	\hfill \mbox{ {\small$\blacksquare$}}    

\begin{remark}  
	Let $M^{\text{GRK}}$ and $R^{\text{GRK}}$ denote the cache size and rate of \emph{the GRK non-private scheme}  
	proposed in \cite[Lemma 1]{Chinmay2022}, the $(M,R)$ pair of the  proposed non-private scheme in Lemma \ref{lemach1} satisfies: $(M,R) = \big(R^{\text{GRK}}, M^{\text{GRK}}\big)$,   i.e., the cache size and delivery signal rate are swapped.    
\end{remark} 
 
 Similar to the proof of \cite[Theorem 1]{Chinmay2022},  the scheme in  Lemma \ref{lemach1} can be converted to a scheme for the $(N,K)$ demand private coded caching problem with the same memory-rate pairs.    
 This is stated in the following theorem, which presents  the achievability result of the   proposed virtual user scheme.     	
\begin{Theo}  \label{achD} 
	For an $(N,K)$ demand private coded caching problem, the lower convex envelope of the following memory-rate pairs
	\begin{align}  
		\left(M,R \right) = &   \left(  \frac{\tbinom{NK-K+1}{r+1}-\tbinom{NK-K-N+1}{r+1}}{\tbinom{NK-K+1}{r}}
		,\frac{Nr}{NK-K+1} \right),  \nonumber 
	\end{align}
	where $r \in [0:NK-K+1] $ is achievable.
\end{Theo}  
\begin{IEEEproof} 
	\cite[Theorem 1]{Chinmay2022} shows that if there exists a scheme for the $(N,NK)$ $\mathcal{D_{RS}}$-non-private
	coded caching problem achieving the memory-rate pair $(M,R)$, then there exists a   scheme for the $(N,K)$ demand private coded caching problem achieving the same memory-rate pair $(M,R)$.  
	Thus, a demand private scheme achieving the memory-rate pairs  given in Theorem \ref{achD} can be obtained by using the scheme in Lemma \ref{lemach1} as a black box, while the lower convex envelope of the above memory-rate pairs can be achieved by memory sharing \cite{MaddahAli2014}.  
\end{IEEEproof} 

\begin{remark} 
	 Since our scheme in Theorem \ref{achD} is obtained
	 using a scheme for the $(N,NK)$ $\mathcal{D_{RS}}$-non-private coded caching problem, our scheme also satisfies the stronger privacy constraint  \eqref{privacy1}, according to  \cite[Remark 2]{Chinmay2022}. 
\end{remark}

\subsection{Schemes Based on MDS Code and Random Permutation}  \label{main_B} 
Given that a virtual user scheme needs to satisfy the demands of \(NK\) users, it leads to inefficient delivery signals  and a high level of sub-packetization  compared to a scheme designed for \(K\)  users, thus necessitating a new demand private scheme  from a different perspective.    
From \cite[Theorem 4]{Wan2021}, we note that using MDS code and random permutation in the placement phase helps generate more equivalent segments of files than the number of partitions, and from \cite[Scheme D]{Chinmay2022}, we note that using random permutation in the delivery phase helps increase the consistency of delivery signals under different demands. 
Based on the above observation, we similarly introduce MDS code  and random permutation into the demand private scheme, thereby proposing the two schemes in Theorem \ref{achAB}.	 

 \begin{Theo} \label{achAB}  
For an $(N,K)$ demand private coded caching problem where $N \le K $, the lower convex envelope of memory-rate pairs 
\begin{align*}
 \left(\frac{N}{q(K+1)},  N - \frac{N}{K+1}  \frac{N+1}{q+1} \right), \quad q =  N, N-1,   
\end{align*} 
 and $(0,N)$ is achievable.  
 \end{Theo} 	
\begin{remark}     
 	  The memory-rate pairs above are also achievable for the case of \( N > K \). However, in the case of $N>K$, the memory-rate pairs shown in Theorem \ref{achAB} perform worse than the trivial scheme in \cite[Theorem 3]{Chinmay2022}. Hence, we only focus on the case \( N \leq K \) in Theorem \ref{achAB}.   
 \end{remark}  
 \emph{\quad Sketch of Proof:}  
 We first propose two new achievable schemes, which achieve memory-rate pairs  $\left(\frac{1}{K+1},\frac{KN}{K+1}\right) $ for $K \ge N$, and $\left(\frac{N}{(K+1)(N-1)},\frac{KN-1}{K+1}\right) $ for  $K \ge N  \ge 3$, respectively.     
 	In addition, the memory-rate pair $\left( \frac{2}{K+1},\frac{2K-1}{K+1}\right)$ for   $K\ge N = 2$,  and the  pair $(0,N)$ for arbitrary $N$ and $K$  have been proved to be achievable in \cite[Theorem 2]{Chinmay2022}.   
  	Hence, based on these results,   the proof of Theorem \ref{achAB} is complete, as the lower convex envelope of the above memory-rate pairs is achieved through memory sharing \cite{MaddahAli2014}. 
 	The main idea of the two  proposed  schemes, which are based on MDS code and random permutation, is outlined below.   
 	The design of cache contents and delivery signals under the demand vector $ \boldsymbol{D} $, where  all $N$ files are requested, is inspired by the non-private scheme achieving $\left(\frac{N}{Kq},  N - \frac{N}{K}  \frac{N+1}{q+1} \right)$ with $q=N,N-1$ stated in \cite[Theorem 1]{Gomez2018}.    
 	By aligning with the structure of the delivery signals under the demand vector $\boldsymbol{D}$,  where all $N$ files are requested, we carefully  design the delivery signals under the remaining demand vectors.   
  	Then, by using the MDS-based cache placement where the indices of MDS-coded segments are encrypted and by shuffling the order of certain segments in the delivery signal, we ensure that the delivery signals are equivalent under different demands from the perspective of a certain user. 
 	The details of the idea will be elucidated in the following example. 
  \subsubsection*{Example 1}  We describe a new scheme for $N = K = 2$ which achieves the memory-rate pair $(M,R) = \left( \frac{1}{3},\frac{4}{3} \right) $.     
 
 Split each file $W_n$, $n = 0,1$,  into $3$ disjoint sub-files of equal size, which are then encoded using a $(4,3)$ MDS code.  
 Each MDS-coded segment contains $ \frac{F}{3}$ bits and due to the property of MDS code,  any $3$ segments can reconstruct the whole file.   
 The 4 MDS-coded segments are denoted by $W_{n,0}, W_{n,1},W_{n,2},W_{n,3}$, and are assigned to segments $w_{n,0}^{(0)}, w_{n,1}^{(0)}, w_{n,0}^{(1)}, w_{n,1}^{(1)} $  based on a randomly generated permutation of $[0:3]$, i.e.,  
 $	w_{n,m}^{(k)} = W_{n, p_{n,m}^{(k)} },    m,k \in \{0,1 \}, $ 
 where  $\boldsymbol{p}_n = \left  (p_{n,0}^{(0)}, p_{n,1}^{(0)}, p_{n,0}^{(1)}, p_{n,1}^{(1)} \right 
 )$ denotes the permutation, which 
 is independently and  uniformly chosen   from the set of all possible permutations of  $[0:3]$.

 In brief, when considering only the MDS-coded segment \( w_{n,0}^{(k)} \), the cache contents and delivery signals under demands \((0,1)\) and \( (1,0)\) follow an \((N,K,M,R) =  \left( 2,2,\frac{1}{2},1\right) \) non-private scheme   stated in \cite[Theorem 1]{Gomez2018}.  
 Meanwhile, \( w_{n,1}^{(k)} \) is delivered directly based on the requirements of decoding.       
 According to the principle of maintaining structural consistency with the established delivery signals, the delivery signals under   demands \( (0,0)\) and \( (1,1)\) are additionally designed. 
 The detailed  scheme  is as follows.  
  
 \emph{Placement Phase.}  
 The cache content of user $k$, i.e., $Z_k$, is  set as
 $	Z_k =   \left(  w_{0,0}^{(k)} \oplus  w_{1,0}^{(k)} ,  P_k \right),  k\in \{0,1\}, $  
where $P_0$, $P_1$ are $2$ i.i.d. random variables  uniformly distributed on $[0:3]$.       
 Since each segment is $\frac{F}{3}$-length and the size of $P_k$ is negligible  
 when $F$ is sufficient large,  we have $M=\frac{1}{3}$. 
 
 \emph{Delivery  Phase.} 
As shown in Table \ref{schAex_X}, for demand $\boldsymbol{D}$, the server delivers  $X_{\boldsymbol{D}} =  \left(X_{\boldsymbol{D}}', J_0, J_1 \right)$, where  
 the  auxiliary variable $J_0$ contains the indices of the segments in  $X_{\boldsymbol{D}}'$, and  $J_1$ contains encrypted indices  of $w_{D_0,0}^{(0)}$ and $w_{D_1,0}^{(1)}$.  
 \begin{table*}
 	\setlength{\abovecaptionskip}{0cm} 
 	\begin{center}  
 		\caption{Delivery Signals of $( N=2,K=2,M= \frac{1}{3},R=\frac{4}{3} )$-private Scheme }  \label{schAex_X} 
 		\begin{tabular}{|c|c|c|c|}
 			\hline
 			$\boldsymbol{D}$   & $X_{\boldsymbol{D}}'$ & $J_0$  & $J_1$ \\
 			\hline
 			$(0,0)$   & $w_{0,1}^{(0)}, w_{0,1}^{(1)},w_{1,0}^{(0)},w_{1,0}^{(1)}$ & $p_{0,1}^{(0)}, p_{0,1}^{(1)},p_{1,0}^{(0)},p_{1,0}^{(1)} $ & $   p_{0,0}^{(0)} \oplus_4 P_0 , p_{0,0}^{(1)} \oplus_4  P_1  $     \\
 			\hline
 			$(0,1)$   & $w_{0,1}^{(0)}, w_{0,0}^{(1)},w_{1,0}^{(0)},w_{1,1}^{(1)}$ & $p_{0,1}^{(0)}, p_{0,0}^{(1)},p_{1,0}^{(0)},p_{1,1}^{(1)}$  & $   p_{0,0}^{(0)}  \oplus_4 P_0 ,  p_{1,0}^{(1)}  \oplus_4 P_1 $    \\
 			\hline
 			$(1,0)$   & $w_{0,0}^{(0)}, w_{0,1}^{(1)},w_{1,1}^{(0)},w_{1,0}^{(1)}$ & $p_{0,0}^{(0)}, p_{0,1}^{(1)},p_{1,1}^{(0)},p_{1,0}^{(1)}$   & $  p_{1,0}^{(0)}  \oplus_4 P_0 ,   p_{0,0}^{(1)}  \oplus_4 P_1 $   \\
 			\hline
 			$(1,1)$   & $w_{0,0}^{(0)}, w_{0,0}^{(1)},w_{1,1}^{(0)},w_{1,1}^{(1)}$ & $p_{0,0}^{(0)}, p_{0,0}^{(1)},p_{1,1}^{(0)},p_{1,1}^{(1)}$  & $   p_{1,0}^{(0)}  \oplus_4 P_0 ,   p_{1,0}^{(1)}  \oplus_4 P_1  $    \\
 			\hline 
 		\end{tabular}
 	\end{center}
 	 \vspace{-1cm}
 \end{table*}
Since $X'_{\boldsymbol{D}}$ contains $4$  $\frac{F}{3}$-length segments and the size of $J_0$ and $J_1$  is negligible when $F$ is sufficiently large, we have $R=\frac{4}{3}$.
 
 \emph{Correctness.} 
Take  $\boldsymbol{D} = (0,1)$ as an example, 
 user $0$ requests file $W_0$ and  directly gets $w_{0,1}^{(0)}$ and  $w_{0,0}^{(1)}$ along with their  indices $p_{0,1}^{(0)}$ and  $p_{0,0}^{(1)}$  from $X_{\boldsymbol{D}}$. 
 Then, user  $0$  decodes $w_{0,0}^{(0)}$ and its index $p_{0,0}^{(0)}$ through  calculating  $ \left( w_{0,0}^{(0)} \oplus  w_{1,0}^{(0)}\right)  \oplus w_{1,0}^{(0)} $ and  $\left(  p_{0,0}^{(0)}  \oplus_4 P_0  \right)  \ominus_4 P_0  $.   
 Thus, according to the property of a $(4,3)$ MDS code, user $0$ can obtain $W_0$ from $w_{0,1}^{(0)}$, $w_{0,0}^{(1)}$ and $w_{0,0}^{(0)}$.
 Similarly, user $1$ gets $w_{1,0}^{(0)}$ and  $w_{1,1}^{(1)}$ directly from $X'_{\boldsymbol{D}}$ and decodes $w_{1,0}^{(1)}$ through  $\left( w_{0,0}^{(1)} \oplus  w_{1,0}^{(1)}\right) \oplus w_{0,0}^{(1)} $.  With corresponding indices, i.e., $p_{1,0}^{(0)}$, $p_{1,1}^{(1)}$ and  $p_{1,0}^{(1)}$,   user $1$ can obtain $W_1$ from 
 $w_{1,0}^{(0)}$, $w_{1,1}^{(1)}$ and  $w_{1,0}^{(1)}$.  
 In the case of other demands, a similar decoding process is used, as detailed in the general proof  in Subsection \ref{secachA}.  
 
 \emph{Privacy.}
	Take $\boldsymbol{D} = (0,0)$ and $\boldsymbol{D} = (0,1)$ as examples, from the perspective of user $0$,  the segments in $X'_{\boldsymbol{D}}$ and auxiliary variables in $J_0$ and $J_1$ are the same. More specifically, in $X'_{\boldsymbol{D}}$, there is one segment of $W_1$  used for decoding, one segment of $W_1$  that can be considered as noise, and two segments of $W_0$ that differ from the segment appearing in $Z_0$.  
 	From $J_0$, $J_1$ and $P_0$, user $0$ can obtain three different indices for the segments of $W_0$ and two different indices for  the segments of $W_1$. Thus, user $0$ cannot infer user $1$’s demand.   
 Similarly, in the case of other demands, no user can obtain any information about the demand of the other user. 
 The proof of privacy in a formal information-theoretic way is detailed in the general proof in Subsection \ref{secachA}.     
 \hfill\mbox{\,$\square$}  
 
  Detailed proofs for the two proposed schemes can be found in Subsections \ref{secachA} and \ref{secachB}, respectively.  
 \hfill \mbox{ {\small$\blacksquare$}} 
  
\subsection{Numerical Evaluations}  \label{main_C}      
  In this subsection, we provide numerical evaluations of the proposed schemes in comparison with other existing demand private schemes.   We show that the  proposed schemes outperform known  achievable schemes for small cache sizes and \( N \le K \), under the privacy constraint \eqref{privacy2}, while the scheme in Theorem \ref{achD}   outperforms known schemes for small cache sizes under the stronger privacy constraint \eqref{privacy1}.

\begin{figure} [htbp]
 	\centering  
 	\subfigcapskip=-5pt 
 	\subfigure[\label{fig1a} For the case $(N,K)=(5,10)$]{
 		\includegraphics[width=0.8\linewidth]{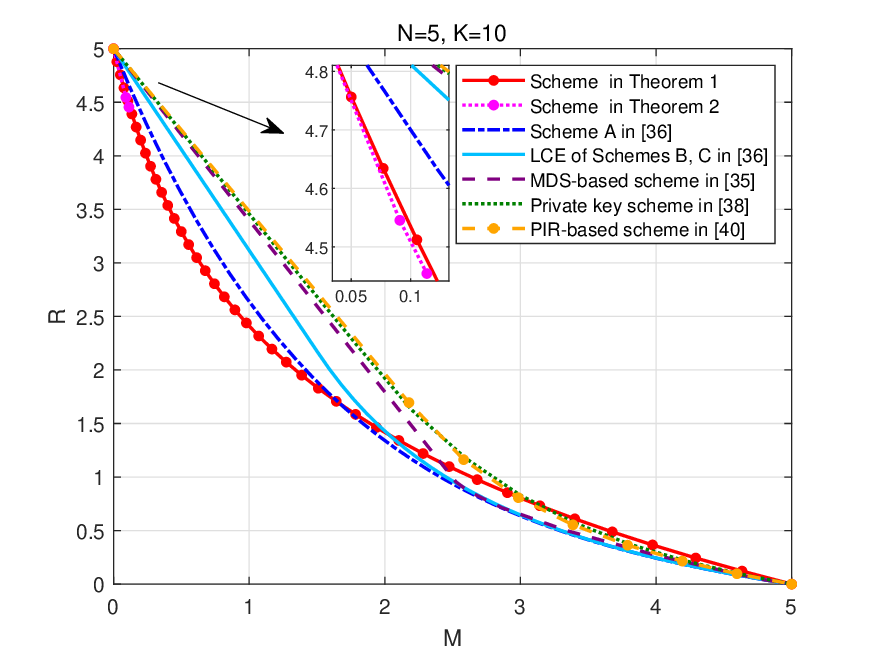}}
 	\subfigure[\label{fig1b} For the case $(N,K)=(10,5)$]{
 		\includegraphics[width=0.8\linewidth]{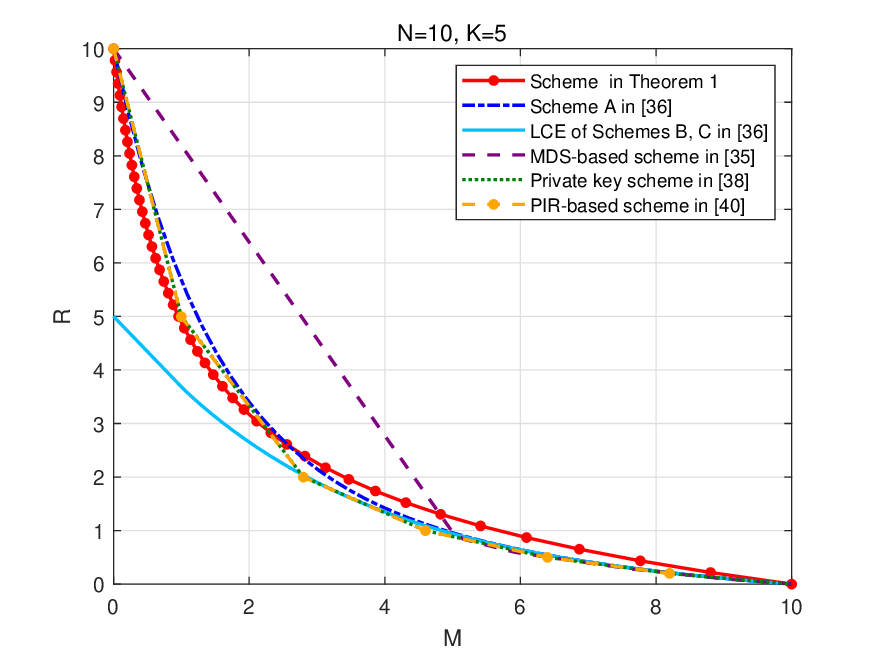}}
 	\caption{Comparison of existing schemes} \label{fig1} 
 \end{figure}

   More specifically, the schemes in Theorems \ref{achD} and \ref{achAB} are illustrated in Fig. \ref{fig1} by the red solid line with dots and the pink dotted line with dots, respectively, with Fig. \ref{fig1a} and Fig. \ref{fig1b} corresponding to the cases \( (N,K) = (5,10) \) and \( (N,K) = (10,5) \), respectively.    
	The other existing demand private schemes shown in Fig. \ref{fig1} are introduced as follows. 
	The virtual user scheme stated in \cite[Theorem 2]{Chinmay2022} outperforms  that in \cite[Theorem 2]{Wan2021} for $M \in [0,N]$ and arbitrary $N$ and $K$, and is denoted by the blue dash-dot line.       
	The lower convex envelope (LCE) of the achievable memory-rate pairs given by Schemes B and C in \cite{Chinmay2022} is  denoted by the sky blue solid line.   
	In addition,  the achievable memory-rate pairs given by the MDS-based scheme in \cite[Theorem 4]{Wan2021}, the private key scheme in  \cite[Theorem 1]{Yan2021}, and the PIR-based scheme  in  \cite{Gholami2023} are denoted by  the purple dashed line, the green dotted line, and the orange dashed line with dots, respectively.     
  
From the numerical results shown  in Fig. \ref{fig1a}, we observe that the LCE of  schemes in Theorems \ref{achD} and \ref{achAB} outperforms known achievable schemes for small cache size and $N \le K $, e.g., $(N,K)=(5,10)$, $M \in [0, 1.642]$.  
Furthermore, when considering the stronger privacy constraint \eqref{privacy1}, we note that Schemes B and C in \cite{Chinmay2022} do not satisfy the stronger privacy constraint \eqref{privacy1}. 
Thus, as shown in Figs. \ref{fig1a} and \ref{fig1b},  under the stronger privacy constraint \eqref{privacy1}, the scheme in Theorem \ref{achD} outperforms known achievable  schemes for small cache size, e.g.,  $(N,K)=(5,10)$, $M \in [0, 1.642]$ and $(N,K)=(10,5)$, $M \in [0, 2.110]$.  

\subsection{Converse and  Optimal Results for General $N$ and $K$}   \label{main_D} 
To further describe the memory-rate tradeoff for the demand private problem, we provide a new converse bound for arbitrary number of files and users in the following theorem.
\begin{Theo} \label{con0}  
	For the $(N,K)$ demand private coded caching problem,
	any $(M, R)$ pair must satisfy  that 
	\begin{align*}
	\left\{ \begin{aligned}
	& (N-1)(K+1)M+(K+1)R  \ge (K+1)N-1, 
	 \quad   \text{when }  N \le K \le 2N-2  \\  
	& \frac{K(K+1)}{2} M + \frac{(K+1)(K+2)}{2N} R \ge  \frac{K(K+3)}{2}, 
	\quad   \text{when }  K > 2N -2   \end{aligned} \right. . 
	\end{align*}
\end{Theo} 
	\emph{\quad Sketch of Proof:} 
	The proof is derived by suitably combining entropy terms containing different cache contents and delivery signals through the submodularity of the entropy function, applying the correctness constraint 	\eqref{decoding}, and exploiting the equality of certain entropy terms caused by the privacy constraint 	\eqref{privacy2}.   
	The content of these entropy terms and the order of combining the terms  for the case $N \le K \le 2N-2$ are inspired by the proof of the converse  bound for the $(2,2)$ demand private coded caching problem in  \cite{Kamath2020},  and the methods of induction and recursion shown in \cite[Lemma 8]{Kumar2023}.    
	In the case where $K > 2N - 2$, some entropy terms and the order of combining them are consistent with those in the case of $N \le K \le 2N - 2$, while others are carefully designed to ensure that the entire  process of combining can determine the maximum number of files.     
	The detailed proof can be found in Section \ref{seccon}. 
	\hfill \mbox{ {\small$\blacksquare$}}   
\begin{remark}[Relations to known converse  results] 
	When $ K = N = 2$, we have $3M + 3R \ge 5$, which is consistent with the converse  result in \cite[Theorem 6]{Chinmay2022}.    
	For arbitrary values of \( N \) and \( K \), where $K \geq \max\{3,N\}$, our converse result provides the only converse bound for the coded caching problem under the  privacy constraint \eqref{privacy2}.  
	Since the privacy constraint in \eqref{privacy1} is stronger than \eqref{privacy2}, the  derived converse result also serves as a bound for the coded caching problem under the constraint    \eqref{privacy1}, and   is tighter than the converse  result in \cite[Theorem 3]{Yan2021} for $N < K$ and certain ranges of $M$. 
   	More specifically, when $N < K \le 2N-2$, our converse  result is tighter   on $M \in \big[\frac{1}{K+1}, \frac{4}{N+1}-\frac{1}{K+1} \big]$, and when $ K > 2N-2$, it is tighter on 	$M \in \big[\frac{1}{K+1},  \min \big\{\frac{3}{N+1},  \frac{(K+1)(K+2) - 2N(N+1)}{(K+1)(N+1)(K+2-2N)} \big\} \big]$. 
   	Furthermore, when $N=K$, our converse is consistent with the  bound in  \cite[Theorem 3]{Yan2021} on $M \in \big[\frac{1}{N+1}, \frac{3}{N+1}\big]$.   
\end{remark} 

By comparing the achievability result in Theorem \ref{achAB} with the converse bound in Theorem \ref{con0} and \cite[Theorem 2]{MaddahAli2014}, we obtain the optimality result presented in the following theorem.   
 \begin{Theo}  
 	For the $(N,K)$ demand private coded caching problem where $ N \le K$, when $M \in \big[0, \frac{1}{K+1}\big] $,   
 	 \begin{subequations}
	\begin{align}   \label{op1}
	R_{N,K}^{*p}(M) = N(1-M). 
	\end{align} 
	Furthermore,  for the case where $N \le K  \le 2N-2$, when $M \in \big[\frac{1}{K+1},  \frac{N}{(K+1)(N-1)}\big] $,  
	\begin{align} \label{op2}
	& R_{N,K}^{*p}(M) = N-\frac{1}{K+1} - (N-1)M. 
	\end{align}
	 \end{subequations}
\end{Theo} 
\begin{IEEEproof}  
 	Setting  $s=N$ in \cite[Theorem 2]{MaddahAli2014}, which is a converse bound for the coded caching problem without privacy constraint,  we get 	$R_{N,K}^{*p}(M) \ge N(1-M)$.  
 	As $\left( \frac{1}{K+1},\frac{KN}{K+1}\right) $ and  $(0,N)$ have been proved to be achievable in  Theorem  \ref{achAB},  \eqref{op1} is thus proved. 
 	When $N \le K  \le 2N-2$, we note that the memory-rate pairs  $\left( \frac{1}{K+1},\frac{KN}{K+1}\right) $ and $\left( \frac{N}{(K+1)(N-1)},\frac{KN-1}{K+1}\right) $, proved to be achievable in  Theorem  \ref{achAB}, meet the converse bound in Theorem \ref{con0}, and  \eqref{op2} is thus proved.    
 \end{IEEEproof}
 
\begin{remark}[Relations to known optimality results]  
	For the case where $ N \le K$, \eqref{op1} contains the optimality result characterized in   \cite[Theorem 4]{Namboodiri2021}
	and  gives the optimal memory-rate tradeoff on $ M \in \big[\frac{1}{K(N-1)+1},\frac{1}{K+1}\big] $ for the first time. 
	Additionally, in the case where $N \le K  \le 2N-2$,   \eqref{op1} and \eqref{op2} together give the optimal  memory-rate tradeoff on $ M \in \big[\frac{1}{K(N-1)+1},\frac{N}{(K+1)(N-1)}\big] $  for the first time.  	  
\end{remark}

\subsection{Converse and Optimal Results for the Case $N=2$}\label{main_E}   
The following lemma states a converse bound for the case of 2 files and arbitrary number of users. 
\begin{Lem}\label{con1}  
	 For the $(2,K)$ demand private coded caching problem,   
	 any $(M, R)$ pair must satisfy that, 
	 	\begin{align*}
	 		& \frac{(K+1)(K+2)}{2N} M + \frac{K(K+1)}{2}  R \ge  \frac{K(K+3)}{2}. 
	 \end{align*}
\end{Lem}
	
\emph{\quad Sketch of Proof:}  The proof follows a similar idea to that of Theorem \ref{con0}. By appropriately swapping the cache contents and delivery signals in the proof of Theorem \ref{con0} for $K > 2N - 2$, we can derive the proof of Lemma \ref{con1}. 	
The detailed proof can be found in Appendix \ref{subseccon3}.  	\hfill \mbox{ {\small$\blacksquare$}}   

 By comparing the achievability and converse
  results mentioned above,  the following theorem states some optimal  memory-rate tradeoff results for  the $(2,K)$ demand private coded caching problem.  
  \begin{Theo}  \label{Theop2}
 	For the $(2,K)$ demand private coded caching problem,  when $M\in \big[0,\frac{2}{K}\big] \cup \big[\frac{2(K-1)}{K+1},2\big]$,
 \begin{subequations} 
 	\begin{align}\label{cor1}
 		R^{*p}_{N,K}(M) = & \max \bigg\{2-2M, \frac{2K(K+3)}{(K+1)(K+2)}-\frac{2K}{K+2}M,
 		1- \frac{1}{2}M,\frac{K+3}{K+1}-\frac{K+2}{2K}M
 		 \bigg\}. 
 	\end{align}  
 	Furthermore, for the $(2,3)$ demand private coded caching problem, the optimal  memory-rate tradeoff is
 	\begin{align}  \label{cor2}
 		& R_{N,K}^{*p}(M)  = \max \bigg \{ 2-2M, \frac{9-6M}{5}, \frac{5-3M}{3}, 
 		 \frac{9-5M}{6} ,  \frac{2-M}{2} \bigg \}. 
 	\end{align} 
\end{subequations} 
 \end{Theo}  

\emph{\quad Sketch of Proof:} 
 	 The theorem can be  proved by combining the results in Theorems \ref{achD} and \ref{con0}, Lemma \ref{con1}, \cite[Theorem 2]{Chinmay2022} and the cut-set bound for the coded caching problem without privacy constraint in \cite{MaddahAli2014}.  
 	The detailed proof can be found in Appendix \ref{pfCoro}. 
 	\hfill \mbox{ {\small$\blacksquare$}}
 \begin{remark}[Relations to known optimality results] 
  Since $\frac{2(K-1)}{K+1} \le 2 \min   \big\{ \frac{2K-1}{2K}, \frac{2^{K-1}}{2^{K-1}+1} \big\}$, in the special case of $N=2$, Theorem \ref{Theop2} contains the optimality  result characterized in \cite[Theorem 6]{Wan2021}. 
 	By setting $K=2$ in  \eqref{cor1}, we obtain the optimal  memory-rate tradeoff of the $(2,2)$ demand private coded caching problem, which was characterized in \cite[Theorem 6]{Chinmay2022}. 
	In addition, \eqref{cor2} provides the first complete description of the optimal memory-rate tradeoff for the case of \( (N,K) = (2,3) \), while \eqref{cor1} presents the  optimal memory-rate tradeoff for \( M \in \big[\frac{1}{K+1}, \frac{2}{K}\big] \cup \big[\frac{2(K-1)}{K+1}, 2 \min   \big\{ \frac{2K-1}{2K}, \frac{2^{K-1}}{2^{K-1}+1} \big\} \big] \) in the case of 2 files and an arbitrary number of users for the first time.      
 \end{remark}

\section{Proof  of  Achievable Schemes}  \label{sec4}  

\subsection{Proof of Lemma \ref{achD}} \label{secach} 
Recalling the definitions of $\mathcal{D_{RS}}$ and $\boldsymbol{d}$ in Subsection \ref{main_A}, we note that
for $k \in [N]$ and $n \in [N]$,  the demand of user $kN+n$, i.e., $\overline{d}_n^{(k)}$, satisfies that  $\overline{d}_n^{(k)} = d_k \oplus_N n$.   
As  any $ \overline {\boldsymbol{d}} \in \mathcal{D_{RS}} $  can be uniquely represented by a vector $ \boldsymbol{d} \in \mathcal{D} $, we use $X_{\boldsymbol{d}}$ to denote the delivery signal under $\overline {\boldsymbol{d}}$ in the $\mathcal{D_{RS}}$-non-private coded caching problem.   

Before we give the proof of Lemma \ref{lemach1} for general $N$ and $K$, we first use an example to illustrate the idea of the scheme.  
\subsubsection*{Example 2} 
We consider the $(N,NK) = (2,6)$  $\mathcal{D_{RS}}$-non-private coded caching problem, and describe the proposed scheme in the case where the parameter  $r = 2$.   
   
The scheme is based on the YMA scheme for $N=2$ files and $NK-K+1=4$ users, where the cache content of user $k$  and the delivery signal under demand $\boldsymbol{g}$ are denoted by  $Z^{\text{YMA}}_{k}$  and $X^{\text{YMA}}_{\boldsymbol{g}}  $, respectively.  
We define $\mathcal{T} = [NK-K+1] = [0:3]$ and   a subset of $ \mathcal{D}$ as  $ \mathcal{D}_0 =  \{(0,0,0),(1,1,1),(0,0,1), (0,1,1) \} $, and  
establish a one-to-one mapping between them  as follows 
\begin{align*} 	
	&  \quad g: \mathcal{T}   \longmapsto \mathcal{D}_{0},   \text{ where} \quad g(0)  = (0,0,0),  
	\\ &  \quad  
	g(1)   = (1,1,1),\quad g(2)  = (0,0,1), \quad  g(3)    = (0,1,1), 
\end{align*}  
and $f: \mathcal{D}_{0}  \longmapsto \mathcal{T}$ is the inverse function of $g$. 
For $t \in \mathcal{T}$, we further define $g_k(t)$ to denote the $ k $-th component of $g(t)$, i.e., $g(t) = \left( g_0(t),g_1(t),g_2(t) \right)$.  

\emph{Placement Phase.}  Split file  $n = 0,1$  into $\binom{NK-K+1}{r} = \binom{4}{2}  = 6 $ equal size subfiles, i.e., $W_n = \left(W_{n, \mathcal{R}}  \right)_{\mathcal{R}\subseteq \mathcal{T}, |\mathcal{R} |=2}$.      
Recalling that each $\boldsymbol{d} \in \mathcal{D} $  represents a vector  $ \overline {\boldsymbol{d}} \in \mathcal{D_{RS}} $,  for each $k \in [0:2]$ and $n   \in [0:1]$, we extract the demand of user $kN+n$  in order from $\mathcal{D}_{0}$, and obtain a $1 \times 4$ demand vector as follows      
\begin{align*} 
	\boldsymbol{g}^{(k,n)} = ( g_k(0)\oplus n, g_k(1)\oplus n, g_k(2)\oplus n, g_k(3)\oplus n ).  
\end{align*}  
Thus, the cache content of  user $kN+n$, i.e., $Z_{kN+n}$,  can be constructed through the delivery signal in the YMA scheme under demand $\boldsymbol{g}^{(k,n)}$, i.e.,   
\begin{align*}
	& Z_{kN+n} =  X^{\text{YMA}}_{\boldsymbol{g}^{(k,n)}} 
	=   \left( Y_{\mathcal{R}^+}^{(k,n)} \right)_{\mathcal{R}^+ \subseteq  \mathcal{T}, |\mathcal{R}^+|=3}, 
	\quad \text{where} \quad Y_{\mathcal{R}^+}^{(k,n)}  =  \bigoplus_{t \in \mathcal{R}^+} W_{g_k(t) \oplus  n, \mathcal{R}^+ \setminus \{t\}}.    
\end{align*}  
Taking the example of $k=1$ and $n=0$, we have  
$\boldsymbol{g}^{(1,0)} = (0,1,0,1)$ and  
$Z_{2} =   \big(  W_{0,\{1,2\}} \oplus W_{1,\{0,2\}} \oplus  W_{0,\{0,1\}},  W_{0,\{1,3\}} \oplus W_{1,\{0,3\}} \oplus  W_{1,\{0,1\}},  W_{0,\{2,3\}} \oplus W_{0,\{0,3\}} \oplus  W_{1,\{0,2\}},  W_{1,\{2,3\}} \oplus W_{0,\{1,3\}} \oplus  W_{1,\{1,2\}}  \big)$.

\emph{Delivery Phase}.    
\begin{table*} 
	\setlength{\abovecaptionskip}{0cm} 
	\begin{center}
		\caption{Delivery Signals of  $( N=2,NK=6,M= \frac{2}{3},R= 1)$  $\mathcal{D_{RS}}$-non-private Scheme}  \label{table1}
		 \setlength{\tabcolsep}{5pt}
		\begin{tabular}{|c|c|c|c|c|c|c|}
			\hline
			$\boldsymbol{d}$   & $\mathcal{V}_{\boldsymbol{d}}$  & $t_{\boldsymbol{d}} $ & $ X^{(n)}_{\boldsymbol{d}, \{0\}}$ &  $ X^{(n)}_{\boldsymbol{d}, \{1\}}$ &   $ X^{(n)}_{\boldsymbol{d}, \{2\}}$ &  $ X^{(n)}_{\boldsymbol{d}, \{3\}}$ \\
			\hline
			$(0,0,0) $   & $\{0\}$ & $ 0 $ & $/$
			& $W_{n,\{0,1\}}$ & $ W_{n,\{0,2\}}$ & 
			$W_{n,\{0,3\}}$   \\
			\hline 
			$(1,1,1)$ &  $\{1\}$  & $ 1 $ & $W_{n,\{0,1\}}$ & $/$  & $W_{n,\{1,2\}}$  & $W_{n,\{1,3\}}$ \\
			\hline 
			$(0,0,1)$ &  $\{2\}$  & $ 2 $ & $W_{n,\{0,2\}}$  &  $W_{n,\{1,2\}}$  &  $/$  &  $W_{n,\{2,3\}}$\\
			\hline 
			$(1,1,0)$ &  $\{0,1,2\}$  & $ 0 $ &  $/$   &  $W_{n,\{0,1\}} \oplus W_{n,\{1,2\}}$  & $W_{n,\{0,2\}}\oplus W_{n,\{1,2\}}$   &  
			 \makecell{$ W_{n,\{0,3\}} \oplus W_{n,\{1,3\}}$  \\ $  \oplus W_{n,\{2,3\}} $ }  \\   
			\hline
			$(0,1,1)$   & $\{3\}$  & $ 3 $ &  $W_{n,\{0,3\}} $ &   $ W_{n,\{1,3\}} $ &   $ W_{n,\{2,3\}}$ & $/$   \\
			\hline 
			$(1,0,0)$ &  $\{0,1,3\}$  &   $ 0 $ &  $/$
			&  $W_{n,\{0,1\}} \oplus W_{n,\{1,3\}}$  
			& \makecell{$W_{n,\{0,2\}} \oplus W_{n,\{1,2\}} $  \\ $ \oplus W_{n,\{2,3\}}$  }    &  $ W_{n,\{0,3\}}\oplus W_{n,\{1,3\}}$  \\
			\hline 
			$(0,1,0)$ &  $\{0,2,3\}$  &   $ 0 $ & $/$
			&   \makecell{$W_{n,\{0,1\}} \oplus W_{n,\{1,2\}} $  \\  $ \oplus W_{n,\{1,3\}}$   } 
			& $W_{n,\{0,2\}} \oplus W_{n,\{2,3\}}$   &  $W_{n,\{0,3\}}\oplus W_{n,\{2,3\}}$\\
			\hline 
			$(1,0,1)$ &  $\{1,2,3\}$  & $ 1 $ &\makecell{$W_{n,\{0,1\}} \oplus W_{n,\{0,2\}}$  \\ $  \oplus W_{n,\{0,3\}}$ }  
			& $/$
			& $W_{n,\{1,2\}} \oplus W_{n,\{2,3\}}$   &  $W_{n,\{1,3\}}\oplus W_{n,\{2,3\}}$\\
			\hline
		\end{tabular}
	\end{center}
	\vspace{-1.2cm}
\end{table*}
For each $\boldsymbol{d} \in \mathcal{D}$, we first construct a set $\mathcal{V_{\boldsymbol{d}}}$ and randomly select  an element from this set, denoted as $t_{\boldsymbol{d}}$,  as detailed in Table \ref{table1}.  Then, based on  $\mathcal{V_{\boldsymbol{d}}}$ and $t_{\boldsymbol{d}}$,  for $\mathcal{S}   = \{0\}, \{1\}, \{2\}, \{3\} $, we define  
\begin{align*} 
	X^{(n)}_{\boldsymbol{d},\mathcal{S}} = &   \bigoplus_{v \in \mathcal{V}_{\boldsymbol{d}} \setminus \mathcal{S} } W_{n,\{v\} \cup \mathcal{S}} , \quad  n = 0,1, 
\end{align*}   
where     
we note that \( X^{(n)}_{\boldsymbol{d},\mathcal{S}} \) is a zero vector of length \( \frac{F}{6} \) for \( \mathcal{V}_{\boldsymbol{d}} \setminus \mathcal{S} = \emptyset \).  
Thus, the delivery signal under $\overline{\boldsymbol{d}}$ is given by
\begin{align*} 
	X_{\boldsymbol{d}} = &  \left( X^{(0)}_{\boldsymbol{d},\mathcal{S}},X^{(1)}_{\boldsymbol{d},\mathcal{S}}  \Big|
	\mathcal{S} \subseteq \mathcal{T} \setminus \{t_{\boldsymbol{d}}  \} 
	,| \mathcal{S} | = 1 \right) ,  
\end{align*}  
where we note  that when $\boldsymbol{d} \in \mathcal{D}_0$,     $X_{\boldsymbol{d}}$ is given by  the cache content  of the YMA scheme for user $f(\boldsymbol{d})$, i.e., $	X_{\boldsymbol{d}} = Z_{ f(\boldsymbol{d}) }^{\text{YMA}} $.  
Table \ref{table1} lists the delivery signals under different $\overline{\boldsymbol{d}}$, where  "/" indicates the absence of that segment in the delivery signal.  

\emph{Decoding}. 
For  user $kN+n$, $k \in [0:2]$, $n   \in [0:1]$, 
the decoding steps for the requested subfile $W_{d_k \oplus  n,  \mathcal{R} }$, $ \mathcal{R} \subseteq  \mathcal{T}, |\mathcal{R}| = 2$,  can be expressed as follows    
\begin{align*}   
	W_{d_k \oplus  n, \mathcal{R}}  
	=  \bigoplus\limits_{t \in \mathcal{V}_{\boldsymbol{d}} \setminus \mathcal{R}} 
	Y^{(k,n)}_{ \{t\} \cup\mathcal{R}}   
	\oplus   \bigoplus\limits_{t \in  \mathcal{R}} 
	X_{\boldsymbol{d},\mathcal{R}\setminus \{t\}} ^{(g_k(t) \oplus n)}. 
\end{align*}   

Taking $\boldsymbol{d} = (0,0,0)$  and $\boldsymbol{d}=(0,1,0)$ as examples, we present the decoding process for user $2$.    
For the case $\boldsymbol{d}=(0,0,0)$, user $2$ obtains $W_{0,\{0,1\}}, W_{0,\{0,2\}}, W_{0,\{0,3\}}$ from  $X_{(0,0,0)}$.  
For $\mathcal{R} = \{1,2\}$, adding $Y_{\{0,1,2\}}^{(1,0)} = W_{0,\{1,2\}} \oplus W_{1,\{0,2\}} \oplus W_{0,\{0,1\}}$, $X_{\boldsymbol{d}, \mathcal{R} \setminus \{1\}}^{(g_1(1))}  = W_{1,\{0,2\}}$ and   
$X_{\boldsymbol{d},\mathcal{R}\setminus \{2\}} ^{(g_1(2))}   =  W_{0,\{0,1\}}$ all together, user $2$ obtains $W_{0,\{1,2\}}$.
Similarly,  user $2$ obtains $W_{0,\{1,3\}}$ by adding $Y_{\{0,1,3\}}^{(1,0)}$, $ X_{\boldsymbol{d}, \{3\}} ^{(1)}$ and $ X_{\boldsymbol{d},  \{1\}} ^{(1)}$ 
and obtains $W_{0,\{2,3\}}$ by adding $Y_{\{0,2,3\}}^{(1,0)}$, $ X_{\boldsymbol{d}, \{3\}} ^{(0)}$ and $ X_{\boldsymbol{d}, \{2\}} ^{(1)}$.	
 
For the case $\boldsymbol{d}=  (0,1,0)$, 
user $2$ first adds $X_{\boldsymbol{d},\{2\}}^{(n)} = W_{n,\{0,2\}} \oplus W_{n,\{2,3\}}$  and $X_{\boldsymbol{d},\{3\}}^{(n)} = W_{n,\{0,3\}}\oplus W_{n,\{2,3\}}$ to obtain $X_{\boldsymbol{d},\{0\}}^{(n)} = W_{n,\{0,2\}} \oplus W_{n,\{0,3\}}$, where $n=0,1$.
For $\mathcal{R} = \{0,1\}$, 
adding  $Y_{\{0,1,2\}}^{(1,0)} = W_{0,\{1,2\}} \oplus W_{1,\{0,2\}}\oplus W_{0,\{0,1\}}$, $Y_{\{0,1,3\}}^{(1,0)} = W_{0,\{1,3\}}\oplus W_{1,\{0,3\}} \oplus W_{1,\{0,1\}} $, $ X_{\boldsymbol{d},  \{1\}} ^{(0)} = W_{0,\{0,1\}} \oplus W_{0,\{1,2\}}  \oplus W_{0,\{1,3\}} $ and  	$ X_{\boldsymbol{d},  \{0\}} ^{(1)} = W_{1,\{0,2\}} \oplus W_{1,\{0,3\}} $  all together, user $2$ obtains $W_{1,\{0,1\}}$. For other $\mathcal{R} $ satisfying $\mathcal{R} \subseteq  \mathcal{T}, |\mathcal{R}|=2$, the decoding of $W_{1,{\mathcal{R} }}$ follows similar decoding steps.  

\emph{Performance.}  
Since each $Z_{kN+n}$ contains $4$ segments $Y_{\mathcal{R}^+}^{(k,n)}$ of length $\frac{F}{6}$, we can determine that $M = \frac{4F}{6F} = \frac{2}{3}$.  As shown in Table \ref{table1}, each $X_{\boldsymbol{d}}$ contains $6$ segments $X_{\boldsymbol{d},\{t\}}^{(n)}$ of length $\frac{F}{6}$, thus, $R = \frac{6F}{6F} = 1$.   \hfill\mbox{\,$\square$}

Now, we generalize \emph{Example 2} to arbitrary  $N$ and  $K$, to present a general achievable scheme for the $(N,NK)$ $\mathcal{D_{RS}}$-non-private coded caching problem. 

Firstly, we define a mapping between demands and users as follows.  
Recalling that $\mathcal{D} =[N]^K$ denotes the set of possible values  for $\boldsymbol{d}$,  we divide it into  $\mathcal{D}_0 $, $\mathcal{D}_1 $ and $ \mathcal{D}_2$, where  
\begin{align*}
	\mathcal{D}_0 &  =  \{(N-1) \boldsymbol{1}_{K} \} 
	\cup  \left\{(a\boldsymbol{1}_{K-k} ,(a\oplus_N 1)\boldsymbol{1}_{k})  \right\}_{k\in[K],a\in[N-1]}, \nonumber \\
	\mathcal{D}_1 &  =    \left\{ a \mathbf{e}'_{k} \right\}_{a\in[1:N-1],k\in [K]} , \text{ and } 
	\mathcal{D}_2    =  \mathcal{D} \setminus (\mathcal{D}_0 \cup \mathcal{D}_1).  
\end{align*}    
We label the $NK-K+1$ demands in set $\mathcal{D}_{0}$ in order as $0,1,\dots,NK-K$, and the one-to-one mapping between labels and demands is defined as follows 
\begin{align*}
	f: \mathcal{D}_{0}  \longmapsto \mathcal{T}, \quad g: \mathcal{T}   \longmapsto \mathcal{D}_{0},\quad  \mathcal{T} = [KN-K+1],
\end{align*}  
where  
\begin{align}  \label{fd} 
	f(\boldsymbol{d}) =   \left\{
	\begin{aligned}
		& f(a\boldsymbol{1}_{K})  =  a ,   \quad  \quad  \quad \quad \quad \quad  \quad  \quad    \quad  \quad \quad \quad \quad \quad  \text{if } a\in[N]  \\
		& f \left( (a\boldsymbol{1}_{K-k} ,(a \oplus_N  1 ) \boldsymbol{1}_{k}  \right)  =  (N-1)k+a+1  , 
		\text{ if } k\in[1:K-1],  a\in[N-1] 
	\end{aligned}
	\right. , 
\end{align}
and $g$ is the inverse function of $f$. 
For $t \in \mathcal{T}$, $g(t)$ is a $K$-length demand vector, and we further define $g_k(t) $ by $  g(t) = (g_0(t),g_1(t) ,\dots,g_{K-1}(t))$.
\begin{remark}   
	Following from \eqref{fd} and the definition of $g_k(t)$, $g_k(t)$ satisfies that  for any $k \in [K]$, $t \in \mathcal{T}$, 
	\begin{align} \label{gt}
		g_k(t)  &  =   \left\{ 
		\begin{aligned} 
			& t, \quad  \quad  \quad  \quad  \quad  \quad  \quad  \quad  \quad  \quad  \quad  \quad  \text{ if } t\in [N]    \\ 
			&    (t-1) \mod(N-1)   ,   
			\quad  \quad   \quad  \text{ if } t \in [N:(K-k)(N-1)]   \\
			& ( (t-1)\mod(N-1) ) +1 , 
			\quad  \text{if } t\in [(K-k)(N-1)+1:KN-K]   
		\end{aligned}
		\right.  . 
	\end{align}  
\end{remark}

\emph{Placement Phase.}  Split each file $W_n$ into $ \binom{NK-K+1}{r} $  subfiles of equal size, i.e.,  
\begin{align*}
	W_n = \left( W_{n,\mathcal{R}} \right)_{\mathcal{R}\subseteq  	\mathcal{T},|\mathcal{R}| = r }, \quad  \mathcal{T} = [NK-K+1].
\end{align*}   
For each $k \in [K]$ and $n \in [N]$, define a vector 
\begin{align*}
	\boldsymbol{g}^{(k,n)}   =   \left(  g_k(0) \oplus_N n, g_k(1) \oplus_N n ,  \dots,   g_k(NK-K)\oplus_N n  \right), 
\end{align*} 
and the cache content of user $kN+n$, i.e., $Z_{kN+n}$, is given by the delivery signal of the YMA scheme under demand $\boldsymbol{g}^{(k,n)}$, i.e.,  
$X_{ \boldsymbol{g}^{(k,n)} }^{\text{YMA}} $.  Thus, for  any $k \in [K]$ and $n \in [N]$, we have 
\begin{align} \label{cache}
	Y^{(k,n)}_{\mathcal{R}^+} 
	= & \bigoplus_{t \in \mathcal{R}^+ } W_{g_k(t) \oplus_N n,\mathcal{R}^+ \setminus \{t\}},  \mathcal{R^+} \subseteq \mathcal{T},|\mathcal{R^+} | = r+1, \nonumber \\
	Z_{kN+n} = &  X_{\boldsymbol{g}^{(k,n)}}^{\text{YMA}} =  \left \{Y^{(k,n)}_{\mathcal{R}^+ } \Big |\mathcal{R}^+ \cap \mathcal{U}_0 \neq \emptyset \right\},    
\end{align}    
where  $\mathcal{U}_0 = [N]$  corresponds to the set of leaders in the YMA scheme. 
The size of each \( Y^{(k,n)}_{\mathcal{R}^+} \) is   \( \frac{F}{\binom{NK-K+1}{r}} \), and the number of sets \( \mathcal{R}^+ \) satisfying the conditions \( \mathcal{R}^+ \cap \mathcal{U}_0 \), \( \mathcal{R}^+ \subseteq \mathcal{T} \), and \( |\mathcal{R}^+| = r+1 \) is  \( \binom{NK-K+1}{r+1} - \binom{NK-K-N+1}{r+1} \). Thus, we have   
\begin{align*}    
	MF = \frac{\tbinom{NK-K+1}{r+1}-\tbinom{NK-K-N+1}{r+1}}{\tbinom{NK-K+1}{r}}F.
\end{align*}

\emph{Delivery Phase.} 
For $\boldsymbol{d} \in  \mathcal{D}_{0} $, the delivery signal under  $\overline {\boldsymbol{d}}$, i.e., $X_{\boldsymbol{d}}$, is given by 
the cache content  of the YMA scheme for user $f(\boldsymbol{d})$, denoted by $Z_{ f(\boldsymbol{d}) }^{\text{YMA}}$, i.e.,   
\begin{align} \label{achxd0}
	X_{\boldsymbol{d}}  
	=  	& Z_{ f(\boldsymbol{d}) }^{\text{YMA}} 
	= \left(  W_{n,\mathcal{R}} | f(\boldsymbol{d}) \in \mathcal{R}, \mathcal{R}\subseteq 	\mathcal{T} , |\mathcal{R}| = r \right)_{n\in[N]}.
\end{align} 

To design $X_{\boldsymbol{d}}$, where $\boldsymbol{d}\in \mathcal{D} \setminus  \mathcal{D}_{0} $, 
we first define an $(NK-K+1)$-length binary vector $V_{\boldsymbol{d}}$ for each $\boldsymbol{d} \in \mathcal{D}$ as follows.
$V_{\boldsymbol{0}_K} = \mathbf{e}_{0} $  and for $\boldsymbol{d} = a \mathbf{e}'_{k} \in \mathcal{D}_1$, 
\begin{align}\label{achVd2} 
	V_{\boldsymbol{d}} =   \left\{
	\begin{aligned} 
		& \mathbf{e}_{0} \oplus \bigoplus_{b\in[1:a]} \left( \mathbf{e}_{b} \oplus \mathbf{e}_{(N-1)(K-1)+b}  \right) , \quad \quad \quad \quad  \quad \quad     \text{if } k=0   \\
		& \mathbf{e}_{0} \oplus \bigoplus_{b\in[1:a]} \left( \mathbf{e}_{(N-1)(K-k)+b} \oplus \mathbf{e}_{(N-1)(K-k-1)+b}  \right), 
		\text{ if } k \in [1:K-2]   \\ 
		&  \mathbf{e}_{0} \bigoplus_{b\in[1:a]} \left( \mathbf{e}_{(N-1)+b} \oplus \mathbf{e}_{b-1}  \right) ,\quad  \quad \quad \quad \quad  \quad \quad \quad    \text{ if } k=K-1    
	\end{aligned}
	\right. . 
\end{align}
For $\boldsymbol{d} = (d_0,d_1,\dots,d_{K-1}) \in \mathcal{D} \setminus (\mathcal{D}_1 \cup \{\boldsymbol{0}_K\}) $,  
\begin{align} 
	V_{\boldsymbol{d}} &  = \mathbf{e}_{0}  \oplus \bigoplus_{i \in[K]} \left( V_{d_i \mathbf{e}'_{i}} \oplus \mathbf{e}_{0}\right).\label{achVd3} 
\end{align} 
Following the definition of $V_{\boldsymbol{d}}$, we define a corresponding set  $\mathcal{V}_{\boldsymbol{d}} \subseteq  \mathcal{T}$, where the $j$-th component of $V_{\boldsymbol{d}}$, denoted by $V_{\boldsymbol{d},j}$,  represents whether element $j$ exists in the set $V_{\boldsymbol{d}}$, i.e.,   
\begin{align*}
	 V_{\boldsymbol{d}} = (V_{\boldsymbol{d},0},V_{\boldsymbol{d},1},\dots,V_{\boldsymbol{d},KN-K} ),\text{ and }    	\mathcal{V}_{\boldsymbol{d}} = \{ j| V_{\boldsymbol{d},j} = 1\}.  
\end{align*}
Thus, for $\mathcal{S} \subseteq \mathcal{T} $ such that $ |\mathcal{S}|= r - 1$, we define  
\begin{align} \label{Xsub1} 
	X^{(n)}_{\boldsymbol{d},\mathcal{S}} = &   \bigoplus_{v \in \mathcal{V}_{\boldsymbol{d}} \setminus \mathcal{S} } W_{n,\{v\} \cup \mathcal{S}} , \quad   n\in[N], 
\end{align}   
where  we note that $X^{(n)}_{\boldsymbol{d},\mathcal{S}}$ equals to a zero vector of length \( \frac{F}{\binom{NK-K+1}{r}} \) when \( \mathcal{V}_{\boldsymbol{d}} \setminus \mathcal{S} = \emptyset \).    
For each $\boldsymbol{d} \in  \mathcal{D} \setminus \mathcal{D}_0$, we  randomly choose an element $t_{\boldsymbol{d}}$ from set $\mathcal{V}_{\boldsymbol{d}}$, and then $	X_{\boldsymbol{d}}$  is set as   
\begin{align} \label{X1}
	X_{\boldsymbol{d}} = &  \left(X^{(n)}_{\boldsymbol{d},\mathcal{S}} \Big |
	\mathcal{S} \subseteq  \mathcal{T} \setminus \{t_{\boldsymbol{d}}  \} 
	,| \mathcal{S} | = r-1\right)_{n\in [N]},  
\end{align}
where $\mathcal{V}_{\boldsymbol{d}} \setminus \mathcal{S}  \neq \emptyset $ since $t_{\boldsymbol{d}} \in  \mathcal{V}_{\boldsymbol{d}} \setminus \mathcal{S}  $.
\begin{remark} \label{remark4}
	Substituting \eqref{achVd2} into \eqref{achVd3},  
	for any $  \boldsymbol{d}  \in \mathcal{D}_{0}$, we have 
	$V_{\boldsymbol{d}} =  \mathbf{e}_{f(\boldsymbol{d})} $, $	\mathcal{V}_{\boldsymbol{d}} =  \{ f(\boldsymbol{d})\} $ and  $t_{\boldsymbol{d}} = f(\boldsymbol{d})$.   
	Thus, the design of $X_{\boldsymbol{d}}$ in $\eqref{achxd0}$ also satisfies \eqref{Xsub1} and \eqref{X1}.
	Since \eqref{Xsub1} and \eqref{X1} are the design of $X_{\boldsymbol{d}}$ for the case where $\boldsymbol{d} \in \mathcal{D} \setminus \mathcal{D}_0$, the design of $X_{\boldsymbol{d}}$ is consistent with \eqref{Xsub1} and \eqref{X1}.  
\end{remark}  

The size of each \( X^{(n)}_{\boldsymbol{d},\mathcal{S}} \) is \( \frac{F}{\binom{NK-K+1}{r}} \), thus, we have 
\begin{align*}  
	RF  =      \frac{NF\tbinom{NK-K}{r-1}}{\tbinom{NK-K+1}{r}}  
	=  \frac{Nr}{NK-K+1}F. 
\end{align*}

\emph{Decoding.}  
The user $kN+n$, $k\in[K]$, $n \in[N]$ obtains its demand file $W_{d_k \oplus_N n}$ by computing 
\begin{align}\label{achVd1} 
	W_{d_k \oplus_N n , \mathcal{R}} 
	=  \bigoplus\limits_{t \in \mathcal{V}_{\boldsymbol{d}} \setminus \mathcal{R}  } 
	Y^{(k,n)}_{ \{t\}\cup\mathcal{R}}    
	\oplus   \bigoplus\limits_{t \in  \mathcal{R}}
	X_{\boldsymbol{d},\mathcal{R}\setminus \{t\}} ^{(g_k(t) \oplus_N n)},
\end{align}      
for all $  \mathcal{R}$ satisfying  $  \mathcal{R} \subseteq \mathcal{T}$ and  $  |\mathcal{R}| = r $. 
The decoding correctness is shown in Appendix  \ref{pfLemdecode}  where we prove that \eqref{achVd1} is correct and all terms on the right-hand side (RHS) of \eqref{achVd1}  can be obtained from $Z_{kN+n}$ and $X_{\boldsymbol{d}}$.    

Thus, the proof of Lemma \ref{lemach1} is complete.

\subsection{Proof of Theorem \ref{achAB},  when $q=N$} \label{secachA}   
In this subsection, we generalize \emph{Example 1}  to arbitrary  $N$ and  $K$, to provide a general proof for Theorem \ref{achAB} in the case of $q=N$. 

\emph{Placement Phase.}
Each file $W_n$, $n \in [N]$,  is  divided into  $K+1$ disjoint sub-files of equal size, which are then encoded using a $(2K,K+1)$ MDS code. The $2K$ MDS-coded segments of file $W_n$ are denoted by  $W_{n,0}, W_{n,1}, \dots ,W_{n,2K-1}$.  
From  the set of all possible permutations of $[2K]$, the server  independently and uniformly chooses $N$ permutations, denoted by 
$\boldsymbol{p}_n =  \big(p^{(k)}_{n,m}\big)_{k\in[K], m \in \{0,1\}} $,   $n \in [N]$.  
The server then shuffles the $2K$ MDS-coded segments based on the permutation $\boldsymbol{p}_n$, and the shuffled segments,  denoted by $w_{n,m}^{(k)}$, satisfy  
\begin{align*} 
	w_{n,m}^{(k)} = W_{n, p_{n,m}^{(k)} }, \quad   k \in [K],m \in \{0,1\}.   
\end{align*}  
The cache content of user $k$, i.e., $Z_k$, is  set as 
\begin{align} \label{schA_z}
Z'_k = &  \bigoplus_{n\in [N]} w_{n,0}^{(k)} ,   \quad 
Z_k =   \left( Z'_k, P_k \right), \quad  k\in[K], 
\end{align}
where $P_k \sim \text{Unif} ([2K])    $, $k \in [K]$,  are $K$ i.i.d. random variables  uniformly distributed on set $[2K]$.    
Since each segments has $\frac{F}{K+1}$ bits and the size of $P_k$ is negligible when $F$ is sufficient large, we have 
\begin{align*}
	M = \frac{F}{(K+1)F} = \frac{1}{K+1}.  
\end{align*} 

\emph{Delivery Phase.} 
To describe the delivery signal, for $k \in [K], n \in [N]$, we define  $X^{(k)}_{\boldsymbol{D},n}$  and  the auxiliary variable $J^{(k)}_{0,n}$  which indicates the index of the shuffled segment in $X^{(k)}_{\boldsymbol{D},n}$, as follows 
\begin{align}   \label{schA_Xn}
	X^{(k)}_{\boldsymbol{D},n} = 	   \left\{
	\begin{aligned}
		& w_{n,0}^{(k)},  &  D_k\neq   n    \\
		& w_{n,1}^{(k)},  &  D_k = n   
	\end{aligned}
	\right. ,  \quad 
	J^{(k)}_{0,n} = 	   \left\{
	\begin{aligned}
		& p_{n,0}^{(k)},  &  D_k\neq   n    \\
		& p_{n,1}^{(k)},  &  D_k = n  
	\end{aligned}
	\right.   . 
\end{align} 
For $k \in [K]$, we define the auxiliary variable \( J_{1,k} \) as follows 
\begin{align}  \label{schA_Jk} 
 	J_{1,k} = p_{D_k,0}^{(k)} \oplus_{2K}   P_k. 
\end{align}
The delivery signal under demand $\boldsymbol{D}$, i.e.,   $X_{\boldsymbol{D} }$, is given by 
\begin{align}\label{schA_x}
	& X_{\boldsymbol{D} } = \left( X'_{\boldsymbol{D} }, J_0, J_1 \right),    \text{ where } X_{\boldsymbol{D}}' =  \big( X^{(k)}_{\boldsymbol{D},n}  \big)_{n \in [N] , k \in [K] },  \nonumber \\
	& J_0  =  \big( J^{(k)}_{0,n}  \big)_{n \in [N] , k \in [K]}, \quad  J_1  =  \left( J_{1,k}  \right)_{k \in [K]}.
\end{align} 
From \eqref{schA_Xn},  \eqref{schA_Jk} and \eqref{schA_x}, we note that 
$X'_{\boldsymbol{D}}$ contains $NK$ $\frac{F}{K+1}$-length segments, and the sizes of $J_0$ and $J_1$ are negligible when $F$ is sufficiently large.  Thus,  we have 
 \begin{align*}
 	R = \frac{KNF}{(K+1)F} = \frac{KN}{K+1}.  
 \end{align*}

\emph{Proof of Correctness.}
We show the process for each user $k$ to decode  file $W_{D_k}$. 
From $X_{\boldsymbol{D}}$, user $k$ directly obtains $K$ MDS-coded segments of $W_{D_k}$ and their corresponding indices in $X^{(i)}_{\boldsymbol{D},D_k}$ and $J^{(i)}_{0, D_k }$, $i \in [K]$.  
In addition, user $k$ decodes $w_{D_k,0}^{(k)}$ and its corresponding index $p_{D_k,0}^{(k)}$ by   
\begin{align*} 
w_{D_k,0}^{(k)}  &  \overset{(a)} =   Z'_k  \oplus   \bigoplus_{n\in [N] \setminus \{D_k\}}  X^{(k)}_{\boldsymbol{D},n}, \quad    
p_{D_k,0}^{(k)}    \overset{(b)} =   J_{1,k}   \ominus_{2K} P_k,    
\end{align*} 
where $(a)$ follows from the design of cache content and delivery signal, i.e., \eqref{schA_z} and \eqref{schA_Xn}, 
and  $(b)$  follows from \eqref{schA_Jk}. 
Following from \eqref{schA_Xn}, we note that   $w_{D_k,0}^{(k)}$ and $X^{(i)}_{\boldsymbol{D},D_k}$, $i \in [K]$,  are all different.  
Thus, user $k$ gets $K+1$  different MDS-coded segments of $W_{D_k}$  and can therefore recover $W_{D_k}$.
This is because  the property of a $(2K,K+1)$ MDS code allows any $K+1$ of the $2K$ MDS-coded segments of $W_{D_k}$ to decode $W_{D_k}$, with the indices of the $K+1$ MDS-coded segments given by $\big( J_{0,D_k}^{(i)} \big)_{i\in[K]}$ and $p_{D_k,0}^{(k)}$.

\emph{Proof of Privacy.}
First, for any fixed  value of $\left(J_0,J_1,P_k,D_{[K]}\right)$,  we have 
\begin{align}  \label{schA_p1}
	H\left(X'_{\boldsymbol{D}}, Z'_k\right) &  \overset{(a)}   = H \left(W_{D_k},  \big(X^{(i)}_{\boldsymbol{D},n} \big)_{n\in [N],i\in[K]}, \bigoplus_{n\in [N]} w_{n,0}^{(k)}  \right) \nonumber \\ 
	& \ge H \left(W_{D_k}, \big(X^{(i)}_{\boldsymbol{D},n} \big)_{n\in [N] \setminus \{D_k\}, i\in[K] } \right)  \nonumber \\  
	&  \overset{(b)}  = F + \frac{K(N-1)}{K+1} F =   \frac{KN+1}{K+1} F  ,  
\end{align}   
where $(a)$ follows from user $k$'s decoding process shown in the proof of correctness, and 
$(b)$ follows from the fact that $ X^{(0)}_{\boldsymbol{D},n}, X^{(1)}_{\boldsymbol{D},n}, \dots, X^{(K-1)}_{\boldsymbol{D},n} $ are $K$ different $\frac{F}{K+1}$-length MDS-coded segments of  $W_n$, and the assumption that the files are independent, uniformly distributed and $F$ bits each. 
Since  the lengths of $X'_{\boldsymbol{D}}$ and $Z_k'$ add up to $ \frac{KN+1}{K+1} F$,
\eqref{schA_p1} holds only when \((X'_{\boldsymbol{D}}, Z'_k)\) follows a uniform distribution. 
In addition, since $H \left(X'_{\boldsymbol{D}}, Z'_k \big| W_{[N]} \right) = 0$, and  $W_{[N]}$ is independent of  $\left(J_0,J_1,P_k,D_{[K]} \right)$, for any fixed  $\left(J_0,J_1,P_k,D_{[K]} \right)$, we have     
 \begin{align} \label{schA_p2}
	\left( X'_{\boldsymbol{D}}, Z'_k \big|J_0,J_1,P_k, D_{[K]}  \right)  \sim \text{Unif} \left( \{0,1\}^{ \frac{KN+1}{K+1} F }\right) .    
\end{align} 

Next, considering the auxiliary variables,  we have
\begin{align} \label{schA_p3}
	 & \Pr \left( \big(J_{0,n}^{(i)}  \big)_{n\in[N], i\in[K] },  p_{D_k,0}^{(k)}   \Big |D_{[K]}\right) \nonumber   \\
	  \overset{(c)}  =  &  \Pr \left( J_{0,D_k}^{(0)},J_{0,D_k}^{(1)}, \dots, J_{0,D_k}^{(K-1)},  p_{D_k,0}^{(k)}   \Big |D_{[K]} \right) 
	  \times \prod_{n \in [N] \setminus \{D_k\}} \Pr \left(J_{0,n}^{(0)},J_{0,n}^{(1)}, \dots,J_{0,n}^{(K-1)} \Big |D_{[K]}\right)  \nonumber \\
	  	\overset{(d)}  = &  \frac{(K-1)!}{(2K)!}  \left(  \frac{K!}{(2K)!}  \right)^{N-1},   
\end{align}
where $(c)$ follows from the fact that the permutations $ \boldsymbol{p}_0,  \boldsymbol{p}_1,  \dots,  \boldsymbol{p}_{N-1} $ are mutually independent, 
and $(d)$ follows from  \eqref{schA_Xn} and the fact that $ \boldsymbol{p}_n $ is uniformly chosen from  all possible permutations of $[2K]$. 
Following \eqref{schA_p3}, we can conclude that $\left(  \big(  J_{0,n}^{(i)} \big)_{n\in[N], i\in[K] },  p_{D_k,0}^{(k)}\right)$  is independent of $D_{[K]}$. 
Therefore, we can further obtain
\begin{align}	 \label{schA_p4}
	I\left( D_{[K] \setminus \{k\} }; \big( J_{0,n}^{(i)}  \big)_{n\in[N], i\in[K] },  p_{D_k,0}^{(k)} \Big|  D_k \right)    =   0.  
\end{align} 

Finally, for any $k\in[K]$, we have 
	\begin{align}
	& I \left(D_{[K]\setminus \{k\}}; X_{\boldsymbol{D}} , Z_k, D_k \right)  
	    =  
	I \left(D_{[K]\setminus \{k\}}; X_{\boldsymbol{D}} \big |Z_k, D_k \right)  + I  \left(D_{[K]\setminus \{k\}}; Z_k, D_k \right)   \nonumber  \\
	 \overset{(e)}   =  &   
	I \left(D_{[K]\setminus \{k\}}; X_{\boldsymbol{D}} \big |Z_k, D_k \right) 
	= 
	I \left(D_{[K]\setminus \{k\}};   X'_{\boldsymbol{D}},J_0,J_1 \big|P_k,Z'_k, D_k \right) \nonumber \\
	 \overset{(f)} =  &    
	I \left( D_{[K] \setminus \{k\} }; \big(J_{0,n}^{(i)}  \big)_{n\in[N], i\in[K] },  \left(J_{1,i}\right)_{ i\in[K] }  \Big | P_k, D_k \right)  \nonumber \\
	 \overset{(g)} =   &   
	I\left( D_{[K] \setminus \{k\} }; \big( J_{0,n}^{(i)}  \big)_{n\in[N], i\in[K] },  p_{D_k,0}^{(k)} \Big|  D_k \right)    \overset{(h)}  =   0, \nonumber 
	\end{align} 
where $(e)$ follows from the fact that $D_{[K]}$ are all i.i.d. random variables and $Z_k$ is independent of $D_{[K]}$,  
$(f)$ follows from  \eqref{schA_p2}, 
$(g)$ follows from  the fact that $p_{D_i,0}^{(i)}$, $i \in [K]$, is encrypted by a one-time pad $P_i$, which is only available to user $i$,  
and $(h)$  follows from \eqref{schA_p4}.  
The proof of privacy is thus complete.
 
The proof of Theorem \ref{achAB} when $q=N$ is thus complete.

\subsection{Proof of Theorem \ref{achAB}, when $q=N-1$ and $N\ge3$}  \label{secachB}  
Before presenting the proof of Theorem \ref{achAB} when $q=N-1$ and $N\ge3$, we first recall some definitions  used  in the previous work \cite{Qian2018,Gomez2018,Chinmay2022}  as follows.   
\begin{Def}[\cite{Qian2018,Gomez2018}]  
	\label{def1}
	For a demand vector  $ \boldsymbol{D}$,
	let $\mathcal{N}_e( \boldsymbol{D})$ denote the set containing the requested files. 
 For each requested file, the server arbitrarily selects one user from those requesting the file, denoted as \( u_n, n \in \mathcal{N}_e(\boldsymbol{D}) \). These $|\mathcal{N}_e( \boldsymbol{D}) |$ users are referred to as \emph{leaders}, and the set composed of them is  denoted by $\mathcal{U} =  \{u_n\}_{n \in \mathcal{N}_e( \boldsymbol{D})}$.     
\end{Def}      
\begin{Def}[\cite{Chinmay2022}] 
For a positive integer $l$, if the permutation  $\pi = \left(\pi(0), \pi(1), \dots, \pi(l-1) \right)$  is a permutation of $[l]$, then for $Y = (Y_0, Y_1, \dots, Y_{l-1})$, we define $\pi(Y) =  \left(Y_{\pi^{-1}(i)} \right)_{i \in [l]}$, which will be used to shuffle the segments in the delivery signal.    
\end{Def}  

Next, we use a simple example to illustrate the idea of the   proposed scheme.    
\subsubsection*{Example 3}  
We describe a demand private scheme for $N = K = 3$ which achieves the memory-rate pair $(M,R) =  \left( \frac{3}{8}, 2 \right) $.   

\emph{Placement Phase.}  
Similar to the operation that generates   segment $w_{n,m}^{(k)}$ in  Subsection \ref{secachA},  
each file $W_n$, $n \in [0:2]$, is cut into $8$ disjoint sub-files of equal size, and then encoded using a $(12,8)$ MDS code. 
The $12$ MDS-coded segments are denoted as $W_{n,0}, W_{n,1}, \dots,  W_{n,11}$ and each contains $\frac{F}{8}$ bits.  
From the set of all possible permutations of  $[0:11]$,  the server independently and uniformly  chooses $N=3$ permutations, denoted by $\boldsymbol{p}_n = \big( p_{n,m}^{(k)}\big)_{k\in [0:2],m \in [0:3]}$, and  the segment $w_{n,m}^{(k)}$ is generated by  $ w_{n,m}^{(k)} = W_{n,p^{(k)}_{n,m}},   n \in [0:2], k \in [0:2],  m  \in [0:3]$.     
Each user $k \in  [0:2]$ caches $Z_k' =  \left(w^{(k)}_{1,0} \oplus  w^{(k)}_{2,0}, w^{(k)}_{0,1} \oplus w^{(k)}_{2,1},w^{(k)}_{0,2} \oplus  w^{(k)}_{1,2} \right)$, and some additional random variables of negligible size.  
Since $Z'_k$ consists of $3$ $\frac{F}{8}$-length segments, we have $ M   =  \frac{3}{8}$.  

Next, without loss of generality, we give the construction of delivery signals  under two specific demands, i.e., $(0,1,2)$ and $(0,1,1)$,  and demonstrate the correctness and privacy of the scheme based on this.

\emph{Delivery Phase.} 
From the set of all possible permutations of  $[0:2]$,  the server independently and uniformly  chooses $N=3$ permutations, denoted as $\pi_n = (\pi_{n,0}, \pi_{n,1},\pi_{n,2})$, $n \in [0:2]$.   
For $\boldsymbol{D} = (0,1,2)$, we set $\mathcal{U} = \{u_0, u_1,u_2\} = \{ 0,1,2\}$,  and 
\begin{align*}
	& Y_{\boldsymbol{D},0} =  w^{(0)}_{0,0}, \quad  Y_{\boldsymbol{D},1} = w^{(1)}_{1,1},  \quad  Y_{\boldsymbol{D},2} =  w^{(2)}_{2,2},  \\
	& X_{\boldsymbol{D},0} =  \left(  w^{(1)}_{0,2},    w^{(2)}_{0,1} ,
	\pi_0 \left(  w^{(0)}_{0,3}  , 
	Y_{\boldsymbol{D},0}  \oplus w^{(1)}_{0,1},    Y_{\boldsymbol{D},0} \oplus w^{(2)}_{0,2} \right) \right) ,  	\\ 
	& X_{\boldsymbol{D},1}=   \left(
	w^{(1)}_{1,3}, w^{(2)}_{1,0},
	\pi_1 \left(  
	Y_{\boldsymbol{D},1} \oplus  w^{(0)}_{1,0}, 	w^{(0)}_{1,2}, Y_{\boldsymbol{D},1} \oplus  w^{(2)}_{1,2} \right)  \right) ,  	\\
	& X_{\boldsymbol{D},2} =    \left(  w^{(1)}_{2,0}, w^{(2)}_{2,3},
	\pi_2 \left(    
	Y_{\boldsymbol{D},2}  \oplus w^{(0)}_{2,0},  Y_{\boldsymbol{D},2}  \oplus w^{(1)}_{2,1}, w^{(0)}_{2,1}  \right) \right).    
\end{align*}
For $\boldsymbol{D} = (0,1,1)$, we set $\mathcal{U} = \{u_0, u_1\} = \{ 0,1\}$,   and  
\begin{align*}
	& Y_{\boldsymbol{D},0} =  w^{(0)}_{0,0}, \quad  Y_{\boldsymbol{D},1} = w^{(1)}_{1,1},  \quad  Y_{\boldsymbol{D},2} =   w^{(1)}_{2,1} \oplus  w^{(0)}_{2,1}, \\
	& X_{\boldsymbol{D},0} =  \left(  w^{(1)}_{0,2},    w^{(2)}_{0,2} ,
	\pi_0 \left(  w^{(0)}_{0,3}  , 
	Y_{\boldsymbol{D},0} \oplus w^{(1)}_{0,1}, Y_{\boldsymbol{D},0} \oplus w^{(2)}_{0,1} \right) \right) ,  	\\
	& X_{\boldsymbol{D},1}=   \left( 
	w^{(1)}_{1,3}, w^{(2)}_{1,3},
	\pi_1  \left(  
	Y_{\boldsymbol{D},1} \oplus  w^{(0)}_{1,0}, 	w^{(0)}_{1,2}, Y_{\boldsymbol{D},1}  \oplus  w^{(2)}_{1,1} \right) \right) ,  	\\
	& X_{\boldsymbol{D},2} = \left(w^{(1)}_{2,0}, w^{(2)}_{2,0},
	\pi_2  \left(   
	Y_{\boldsymbol{D},2} \oplus w^{(0)}_{2,0},   w^{(0)}_{2,1}, Y_{\boldsymbol{D},2} \oplus  w^{(2)}_{2,1}\right) \right).   	
\end{align*} 
Hence, for demand $\boldsymbol{D}$, the server delivers 
$	X_{\boldsymbol{D}}' = \left(    \left(  X_{\boldsymbol{D},n}\right)_{n \in [0:2]}, \bigoplus_{ n\in[0:2] }  Y_{\boldsymbol{D},n} \right)  $, and some auxiliary variables of negligible size, denoted by $J = (J_0,J_1,J_2,J_3)$.  
The specific roles of these auxiliary variables are outlined as follows:  
\begin{itemize} 
	\item   
	$J_0  =  (J_{0,0},J_{0,1},J_{0,2})$, where $J_{0,n}$ indicates to all users the indices of the MDS-coded segments of  $W_n$ that are contained in $X_{\boldsymbol{D},n}$ and different  from  $Y_{\boldsymbol{D},n}$.  
	\item  $J_1$ and $J_2$ can be further divided into $NK = 9$ parts, i.e., $J_i  = \big(J_{i,n}^{(k)} \big)_{k\in [0:2], n\in [0:2] }$, $i = 1,2$.  
	Each $J_{i,n}^{(k)}$  is encrypted using a one-time random variable  cached by user $k$, ensuring that the file index or position information contained in $J_{i,n}^{(k)}$, denoted by $T_{i,n}^{(k)}$,  can only be decoded by user $k$.  	 
	For $n \neq D_k$,  
	$T_{1,n}^{(k)} = p_{D_k,n}^{(k)}$ indicates the index of the segment of $W_{D_k}$ cached by user $k$, and 
	$T_{2,n}^{(k)} = \pi_{n, k}$ indicates which position in the shuffled part of $X_{\boldsymbol{D},n}$ will be used to decode $Y_{\boldsymbol{D},D_k}$.  
	For $n = D_k$, $T_{1,n}^{(k)} = p_{D_k,n}^{(u_n)}$ indicates the index of the segment of $W_{D_k}$ in $Y_{\boldsymbol{D},D_k}$, and $T_{2,n}^{(k)} = \pi_{D_k, u_n}$ indicates the position in the shuffled part of  $X_{\boldsymbol{D},D_k}$ whose content is an MDS-coded segment. 	
	\item 	 
	$J_3 = (J_{3,n})_{n \in [0:2] \setminus \{D_0\}}$, 
	where each $J_{3,n}$ is encrypted using a one-time random variable  cached by user $0$,  ensuring that the position information contained in $J_{3,n}$, denoted by $T_{3,n}$,  can only be decoded  by user $0$. 
	More specifically, for $\boldsymbol{D} = (0,1,2)$, we set $(T_{3,1},T_{3,2}) = (\pi_{1,1},\pi_{2,2})$, and for $\boldsymbol{D} = (0,1,1)$,  we set $(T_{3,1},T_{3,2}) = (\pi_{1,1},\pi_{2,1})$.   
\end{itemize}
Since  $X'_{\boldsymbol{D}}$ contains $16$ $\frac{F}{8}$-length segments  and the size of $J$ is negligible when $F$ is sufficiently large, we have $R = 2$. 

\emph{Correctness.}    
Take $\boldsymbol{D} = (0,1,2)$  as an example, user $0$  requests file $W_0$ and decodes $w_{0,0}^{(0)}$,  $w_{0,1}^{(0)}$ and  $w_{0,2}^{(0)}$ through $ 	w_{0,0}^{(0)}  = Y_{\boldsymbol{D},0}   = \left( w^{(0)}_{1,0} \oplus  w^{(0)}_{2,0}\right)  \oplus 	\left(Y_{\boldsymbol{D},1} \oplus  w^{(0)}_{1,0} \right) 
\oplus \left( Y_{\boldsymbol{D},2} \oplus  w^{(0)}_{2,0}\right)  \oplus\left(  \bigoplus_{n \in [0:2]} Y_{\boldsymbol{D},n}  \right)  $, $ 
w_{0,1}^{(0)}  = \left( w^{(0)}_{0,1} \oplus w^{(0)}_{2,1}\right)  \oplus w^{(0)}_{2,1}$,  and $ 
w_{0,2}^{(0)}   = \left(  w^{(0)}_{0,2} \oplus  w^{(0)}_{1,2}\right)  \oplus w^{(0)}_{1,2}$.  Additionally, user $0$   
directly obtains  $ w^{(1)}_{0,2}$, $w^{(2)}_{0,1}$ and $w^{(0)}_{0,3}$ from  $X_{\boldsymbol{D},0}$,  and further decodes $  w^{(1)}_{0,1} $ and $w^{(2)}_{0,2}$ by adding $w_{0,0}^{(0)}$ to $ Y_{\boldsymbol{D},0} \oplus w^{(1)}_{0,1}$ and $Y_{\boldsymbol{D},0}  \oplus w^{(2)}_{0,2} $, separately.    
Similarly, user $1$   decodes  $w^{(1)}_{1,0}$, $w^{(1)}_{1,1}= Y_{\boldsymbol{D},1}$ and  $ w^{(1)}_{1,2}$, directly obtains  $w^{(1)}_{1,3}$, $w^{(2)}_{1,0}$ and $w^{(0)}_{1,2}$ from $ X_{\boldsymbol{D},1}$, and further decodes $ w^{(0)}_{1,0}$ and $w^{(2)}_{1,2}$ from $ X_{\boldsymbol{D},1}$. 
User $2$   decodes  $w^{(2)}_{2,0}$, $w^{(2)}_{2,1}$ and  $ w^{(2)}_{2,2} = Y_{\boldsymbol{D},2} $, directly obtains $ w^{(1)}_{2,0}$, $w^{(2)}_{2,3}$ and $w^{(0)}_{2,1}$ from $ X_{\boldsymbol{D},2}$, and further decodes $ 	 w^{(0)}_{2,0},$ and $w^{(1)}_{2,1}$ from $ X_{\boldsymbol{D},2}$. 
Note that the positions of the segments required for decoding are either not shuffled by permutations or are given by $J_2$ and $J_3$. Additionally, the indices of the $8$ different  MDS-coded segments obtained above are provided by $J_{0}$ and $J_1$.  Thus, according to the property of a $(12,8)$ MDS code, user  $k$  can obtain $W_{D_k}$. 
In the case of other demands, a similar decoding process is used. 

\emph{Privacy.} 
By observing the delivery signals under $\boldsymbol{D} = (0,1,2)$ and $\boldsymbol{D} = (0,1,1)$,  we have 
\begin{align*}
	& H \left(X'_{(0,1,2)},Z'_0\right) \overset{(a)}{=}  H(W_0)  + H\left(X'_{(0,1,2)},Z'_0 \Big| W_0 \right)   \\  
	\overset{(b)}{\ge}   &  H\left(W_0\right) + H \left(w_{1,1}^{(1)} \oplus w_{2,2}^{(2)}, w^{(1)}_{1,3},  w^{(2)}_{1,0}, 	w^{(0)}_{1,2},  w^{(0)}_{1,0},w^{(2)}_{1,2},  
	 w^{(1)}_{2,0}, w^{(2)}_{2,3}, w^{(0)}_{2,1},w^{(0)}_{2,0},  w^{(1)}_{2,1} \right) \\   \overset{(c)}{=} &   F + \frac{11F}{8} = \frac{19F}{8}, \text{ and} \\ 
	& H\left(X'_{(0,1,1)},Z'_0 \right) \overset{(a)}{=}   H(W_0)  + H\left(  X'_{(0,1,1)},Z'_0 \Big| W_0 \right)  \\
	\overset{(b)}{\ge}  &  H(W_0) + H \left( w^{(1)}_{1,1}  \oplus    w^{(1)}_{2,1}, w^{(1)}_{1,3}, w^{(2)}_{1,3}, w^{(0)}_{1,2}  , w^{(0)}_{1,0},w^{(2)}_{1,1}, 
	w^{(1)}_{2,0}, w^{(2)}_{2,0}, w^{(0)}_{2,1}, w^{(0)}_{2,0},  w^{(2)}_{2,1}  \right) \\ \overset{(c)}{=}  &  F + \frac{11F}{8} = \frac{19F}{8},   
\end{align*} 
where $(a)$ follows from user $0$'s decoding steps for $W_{0}$, $(b)$ follows from the design of $Z'_0$, $X'_{(0,1,2)}$ and $X'_{(0,1,1)}$,  
and $(c)$ follows from the independence of the files and the fact that at most $8$ MDS-coded segments of $W_n$ are mutually independent, uniformly distributed, and each consists of $\frac{F}{8}$ bits.        
Since the length of $\left( X'_{\boldsymbol{D}},Z'_k\right)$ is $\frac{19}{8}F$,  we have $\left(X'_{(0,1,2)},Z'_0 \right)  \sim \left(X'_{(0,1,1)},Z'_0  \right  
) \sim \text{Unif}  \left(\{0,1\}^{ \frac{19}{8}F} \right)$.   
Similarly, for each user $k$ and any demand $\boldsymbol{D} \in \mathcal{D}$, the proposed scheme consistently satisfies $(X'_{\boldsymbol{D}},Z'_k)  \sim  \text{Unif}\left( \{0,1\}^{ \frac{19}{8}F}\right) $. This implies that, from the perspective of user $k$, $X'_{\boldsymbol{D}}$ is always uniformly distributed for any $\boldsymbol{D} \in \mathcal{D}$.  

When considering the auxiliary variables accessible to user $k$, the   indices of the same file or the position information within the same permutation $\pi_{n}$ are all  different. 
Since the permutations that determine the file indices and  the positions of the segments in $X_{\boldsymbol{D},n}$ are uniformly and independently chosen from all possible permutations and are unknown to any user, the auxiliary variables also do not reveal any information about the demands of other users to user $k$.   

In summary, user \(k\) cannot obtain any information about the demands of other users. The proof of privacy in a formal information-theoretic way is detailed in Appendix \ref{schB_pfP}. \hfill\mbox{\,$\square$}

Now, we generalize \emph{Example 3}  to arbitrary  $N$ and  $K$, where $K \ge N  \ge 3$, to provide a scheme achieving the memory-rate pair  $ \left(\frac{N}{(K+1)(N-1)},\frac{KN-1}{K+1} \right)$. 
 
\emph{Placement Phase.} 
We divide each $W_n$ into $(K+1)(N-1)$ equal size sub-files, and then encode them into $2K(N-1)$ MDS-coded segments, denoted by  $W_{n,0}, W_{n,1}, \dots, W_{n,2K(N-1)-1} $,  using a  $(2K(N-1),(K+1)(N-1))$ MDS code. 
The segment $w_{n,m}^{(k)}$ is generated by  
\begin{align*}
	w_{n,m}^{(k)} = W_{n, p^{(k)}_{n,m }}, 
	\quad     m\in [2N-2], k\in [K], 
\end{align*} 
where  $ \boldsymbol{p}_n = \big(p^{(k)}_{n,m}\big)_{k\in[K], m\in [2N-2]} $  is a permutation chosen independently and uniformly from the set of all possible permutations of $[2K(N-1)]$.     
For $k \in [K]$, let  
\begin{align} \label{schBZkn}
Z'_k = (Z_{k,n})_{n\in [N]}, \text{ where } Z_{k,n} =    \bigoplus_{m \in [N] \setminus \{n\}} w_{m,n}^{(k)}.
\end{align} 
Consider  $KN$ i.i.d. random variables uniformly distributed on  $[2K(N-1)]$, denoted by $P_{k,n}$ for $k\in[K]$ and $n\in[N]$, and  $(K+1)N$ i.i.d. random variables uniformly distributed on  $[K]$, denoted by $S_{k,n}$ for $k\in[K+1]$ and $n\in[N]$.  Let $P_k = (P_{k,n})_{n\in [N]}$ and $S_k = (S_{k,n})_{n\in [N]}$.  
Thus, the cache content of user $k$, i.e., $Z_k$, is given by   
\begin{align*}
	Z_k = 	
	& \left\{
	\begin{aligned}
		&  (Z'_{k}, P_{k}, S_k, S_K),     \\
		& (Z'_{k}, P_{k}, S_k),   
	\end{aligned}
	\quad 
	\begin{aligned}
		&   \text{if } k = 0   \\ 
		&    \text{if } k \neq 0   
	\end{aligned}
	\right.  .
\end{align*}

Since $Z'_k$ contains $N$ segments, each with a size of $\frac{F}{(K+1)(N-1)}$, and  the sizes of $P_k$ and $S_k$ can be neglected when $F$ is sufficiently large, we have 
\begin{align*}
M  = \frac{NF}{(K+1)(N-1)F} =  \frac{N}{(K+1)(N-1)}.
\end{align*}

\emph{Delivery Phase.} 
To describe the delivery signal, recalling Definition \ref{def1}, for $ n \in \mathcal{N}_e(\boldsymbol{D}) $, $ u_n $ is the leader of users requesting file $ W_n $, and    
for $ n \notin \mathcal{N}_e(\boldsymbol{D}) $, we further define $ u_n \equiv 1 $.  
When $N \ge 3$, for each file index $n \in [N]$, define an arbitrary one-to-one mapping function $h_{n}(m)$, which  maps each file index $m \in [N] \setminus \{n\}$ to another file index $h_{n} (m) \in [N] \setminus \{n,m\}  $, such that $h_{n}(m_1) \neq  h_{n}(m_2) $ if $m_1 \neq m_2 $.  For example, if $n = N-1$, we can take $h_{n}(m) = m \ominus_{N-1} 1 $.  

For the demand $\boldsymbol{D}$, the server delivers $ X'_{\boldsymbol{D}}$, which consists of $ N + 1 $ parts, i.e.,   
$ X'_{\boldsymbol{D}} = (	X_{\boldsymbol{D},n})_{n\in [N+1]} $. 
Here, $X_{\boldsymbol{D},n}$, $n \in [N]$, only contains the content of file $W_n$, while $X_{\boldsymbol{D},N}$ is composed of the content of all $N$ files and is given by 
\begin{align}  \label{schB_XN}
	X_{\boldsymbol{D},N}  = 	& \bigoplus_{n \in [N]} 	Y_{\boldsymbol{D},n}, \quad \text{and} \\
	\label{schB_Yn}
	Y_{\boldsymbol{D},n} = 	 
	& \left\{
	\begin{aligned}
		& w^{(u_n)}_{n,n},      \\
		& w^{(1)}_{n,D_1} \oplus w^{(0)}_{n,h_{n} (D_0)},   
	\end{aligned}
	\quad 
	\begin{aligned}
		&   \text{if } n \in \mathcal{N}_e(\boldsymbol{D})   \\ 
		&    \text{if } n \notin \mathcal{N}_e(\boldsymbol{D})    
	\end{aligned}
	\right. .
\end{align} 
Next,  to describe $X_{\boldsymbol{D},n}$, $n \in [N]$,   for $k \in [1:K-1]$, set
\begin{align} \label{schB_Y11}
	X_{\boldsymbol{D},n}^{(k)}  = 	 & \left\{
	\begin{aligned}
		&  \big(w_{n,m}^{(k)} \big)_{ m \in [N] \setminus \{n,D_k \} },  	\quad    \text{if }  n \neq D_k \\
		&  \big( w_{n,m}^{(k)}\big)_{ m \in [N:2N-3]}  ,      	\quad    \text{   }   \text{ if }   n = D_k
	\end{aligned} 
	\right.  , 
\end{align}
for  user $k = 0$,  set
\begin{align} \label{schB_Y12}
 X_{\boldsymbol{D},n}^{(k)}   =  & \left\{
	\begin{aligned}
		& \big(w_{n,m}^{(k)} \big)_{ m \in [N] \setminus \{n,D_0, h_{n}  (D_0) \} },   \text{   }    \text{ if }  n \neq D_0 \\
		& \big(w_{n,m}^{(k)} \big)_{ m \in [N+1:2N-3]}  ,  	\quad  	\quad  \quad      \text{if }  n = D_0      
	\end{aligned}
	\right.  , 
\end{align}
 and for  $k\in[K]$, set 
\begin{align}  \label{schB_Y2}
	V_{\boldsymbol{D},n}^{(k)}  = 	 & \left\{
	\begin{aligned}
	& w_{n,h_{n}  (D_0)}^{(0)},   \quad  \quad \quad  \text{if }  k = u_n, n \neq D_0   \\ 
	& w_{n,N}^{(0)},   \quad \quad \quad \quad \quad \text{if }  k = u_n, n = D_0 \\ 
	& 	Y_{\boldsymbol{D},n} \oplus w_{n,D_k}^{(k)}, \quad  \text{ if }  k \neq u_n  
	\end{aligned} 
	\right.  . 
\end{align}
For each \( n \in [N] \), the server uniformly and independently selects \( N \) permutations of \( [K] \) from all possible permutations of \( [K] \), denoted by \( \pi_{n} = (\pi_{n,0}, \pi_{n,1}, \dots, \pi_{n,K-1}) \).  
Thus, \( X_{\boldsymbol{D}, n} \),   \( n \in [N] \), is given by 
\begin{align}  \label{schB_Xn} 
	X_{\boldsymbol{D},n} = \left( \big( 		X_{\boldsymbol{D},n}^{(k)}  \big)_{k\in [K]}, \pi_{n} \left( \big(	V_{\boldsymbol{D},n}^{(k)}\big)_{k \in [K] } \right) \right).    
\end{align}  
 
Besides $ X'_{\boldsymbol{D}} $, the server also delivers some auxiliary variables to users, denoted by  $J = (J_0,J_1,J_2,  J_3)$.   
 $J_0 = (J_{0,n})_{n \in [N]}$ indicates to all users  the indices of the MDS-coded segments in \( X_{\boldsymbol{D},n} \), \( n \in [N] \).    
Thus, in accordance with \eqref{schB_Y11}-\eqref{schB_Xn},  
$J_{0,n}$ is given by 
 \begin{align*}
	 J_{0,n} = \left(\big(J_{0,n,0}^{(k)}\big)_{k \in [K]}, \big(J_{0,n,1}^{(k)}\big)_{k \in [K]} \right), 
\end{align*} 
where for  $k \in [1:K-1]$,  
\begin{align}  \label{schB_J011} 
	J_{0,n,0}^{ (k)}  = 	 & \left\{
	\begin{aligned}
		&  \big( p_{n,m}^{(k)} \big)_{ m \in [N] \setminus \{n,D_k \} }, \quad   \text{if }  n \neq D_k \\
		& \big( p_{n,m}^{(k)} \big)_{ m \in [N:2N-3]}  ,    \quad   \text{ }  \text{ if }   n = D_k 
	\end{aligned}
	\right.  , 
\end{align}
for $k = 0$, 
\begin{align} \label{schB_J010}
	J_{0,n,0}^{ (k)}  = 	 & \left\{
	\begin{aligned}
		& \big( p_{n,m}^{(k)} \big)_{ m \in [N] \setminus \{n,D_0, h_{n}  (D_0) \} },    \text{ if }  n \neq D_0  \\
		& \big( p_{n,m}^{(k)} \big)_{ m \in [N+1:2N-3]}  ,       \quad   \quad   \quad     \text{if }   n = D_0     
	\end{aligned} 
	\right.  ,   
\end{align} 
and for  $k  \in [K]$, 
\begin{align} \label{schB_J32}
	J_{0,n,1}^{(\pi_{n,k})}  = 	 & \left\{
	\begin{aligned}
		& p_{n,h_{n}  (D_0)}^{(0)},  \quad  \text{if }   k = u_n, n \neq D_0   \\
		& p_{n,N}^{(0)},  \quad \quad \quad  \text{if }  k = u_n, n = D_0   \\
		& p_{n,D_{k}}^{(k)},    \quad \quad \text{ if }        k \neq u_n 
	\end{aligned} . 
	\right.    
\end{align} 
$J_1$ and $J_2$ can be further divided into $NK$ parts, i.e., $J_i  =  \big( J_{i,n}^{(k)}  \big)_{k\in [K], n\in [N]}$, $i \in \{1,2\}$, where  
\begin{align} \label{schB_T}
	& \left( T_{1,n}^{ (k)}, T_{2,n}^{ (k)} \right)  = 	  \left\{
	\begin{aligned}
		& \left( p_{D_k,n}^{(k)},  \pi_{n, k }  \right), \quad  \text{ if }  n \neq D_k  \\
		& \left( p_{D_k,n}^{(u_{n})} ,\pi_{n, u_{n}}  \right), \quad  \text{if }   n = D_k   
	\end{aligned} 
	\right.,    \quad \text{and }  \nonumber \\
	& J_{1,n}^{ (k)} =   P_{k,n} \oplus_{2K(N-1)} T_{1,n}^{(k)}, \quad J_{2,n}^{ (k)} =   S_{k,n} \oplus_{K}  T_{2,n}^{(k)}.
\end{align} 
The auxiliary variable $J_3$ is given by  $ J_3 =  (J_{3,n})_{n \in [N] \setminus \{ D_0\}} $, where   
\begin{align}  \label{schB_T2} 
	T_{3,n} = \pi_{n, u_{n} },   \quad \text{and } 
	J_{3,n}  =   S_{K,n} \oplus_{K}T_{3,n}  .  
\end{align}  

In summary,  the  delivery signal is given by 
\begin{align*}
	X_{\boldsymbol{D}} = (	X'_{\boldsymbol{D}},J ). 
\end{align*} 
From \eqref{schB_Y11}-\eqref{schB_Y2}, we observe that $\big(X_{\boldsymbol{D},n}^{(k)}\big)_{k\in [K]}$ contains $(N-2)K-1$ MDS-coded segments of $W_n$, and   $\big(V_{\boldsymbol{D},n}^{(k)}\big)_{k \in [K] }$ contains $K$ segments. As each segment mentioned above is  of length  \(\frac{F}{(K+1)(N-1)}\), and   $X_{\boldsymbol{D},N}$ has  $\frac{F}{(K+1)(N-1)}$ bits, we have  
\begin{align*}  
	R = \frac{N (K(N-2)+K-1)+1}{(K+1)(N-1)F} F  = \frac{KN-1}{K+1}. 
\end{align*} 

\emph{Proof of Correctness.}
Now we describe the steps for user $k$, $k \in [K]$, to decode $W_{D_k}$.  
 
First, user $k$ decodes 
$w_{D_k,m}^{(k)}$, $ m \in [N] \setminus \{D_k\}$, by 
\begin{align} \label{schBdecode1}
	w_{D_k,m}^{(k)}  = Z_{k,m} \oplus  \bigoplus_{n \in [N] \setminus \{m, D_k\}} w_{n,m}^{(k)},  
\end{align} 
where \eqref{schBdecode1} follows from the design of cache content, i.e., \eqref{schBZkn}, and all \( w_{n,m}^{(k)} \) used, except for \( w_{n, h_{n}(D_0)}^{(0)}\), can be found in \( 	X_{\boldsymbol{D},n}^{(k)}\), $n \in [N] \setminus \{D_k\}$.  
The position of \(	V_{\boldsymbol{D},n}^{(u_n)}   =  w_{n, h_{n}(D_0)}^{(0)} \), $n \in [N] \setminus  \{D_0\}$,  in \( \pi_n \left( \big(	V_{\boldsymbol{D},n}^{(i)}  \big)_{i \in [K] } \right)   \) is known to user $0$ 
through  $J_{3,n}  \ominus_{K}  S_{K,n}  = T_{3,n} =  \pi_{n, u_{n} } $.  
User $k$ further decodes  $w_{D_k,D_k}^{(u_{D_k})}$ by 
\begin{align} \label{schBdecode2}
 w_{D_k,D_k}^{(u_{D_k})}  \overset{(a)}  =  &  \bigoplus_{n \in [N] \setminus \{D_k\}} w_{n,D_k}^{(k)} \oplus    \bigoplus_{n \in [N]} 	Y_{\boldsymbol{D},n} 
 \oplus  \bigoplus_{n \in [N] \setminus \{D_k\}} 	\left( Y_{\boldsymbol{D},n} \oplus w_{n,D_k}^{(k)}   \right)    \nonumber \\ \overset{(b)}  =  & Z_{k,D_k} \oplus   X_{\boldsymbol{D},N}  \oplus   \bigoplus_{n \in [N] \setminus \{D_k\}} 	  V_{\boldsymbol{D},n}^{(k)},         
\end{align}     
where $(a)$ follows from \eqref{schB_Yn}, and $(b)$ follows from  \eqref{schBZkn}, \eqref{schB_XN} and \eqref{schB_Y2}. 
The position of 
$V_{\boldsymbol{D},n}^{(k)}$, $n \in [K] \setminus \{D_k\}$, in \( \pi_n \left( \big( 	V_{\boldsymbol{D},n}^{(i)}  \big)_{i \in [K] } \right)\) is known to user $k$ through  $J_{2,n}^{(k)}  \ominus_{K}  S_{k,n}  = T_{2,n}^{(k)} = \pi_{n, k} $.  
Following from \eqref{schB_T}, the indices of the $N$ MDS-coded segments decoded from \eqref{schBdecode1} and \eqref{schBdecode2} are given by $T_{1,m}^{(k)} = J_{1,m}^{(k)} \ominus_{2K(N-1)} P_{k,m}$, $m \in [N]$, which are known to user $k$.  

Next, user \( k \) sequentially obtains \( (N-1)K-1 \)  different MDS-coded segments from   
\(  X_{\boldsymbol{D},D_k} \), with the corresponding indices provided by \( J_{0,D_k} \).   
More specifically, from $ \big(X_{\boldsymbol{D},D_k}^{(i)}\big)_{i \in [K]}$ as defined in \eqref{schB_Y11} and \eqref{schB_Y12}, user $k$ directly obtains  $(N-2)K-1$ different  MDS-coded segments, whose indices are provided in order by  $\big(J_{0,D_k,0}^{(i)} \big)_{i\in[K]}$.    
User $k$ directly obtains $w_{D_k,N}^{(0)}$ or $w_{D_k,h_{D_k}  (D_0)}^{(0)}$ from $V_{\boldsymbol{D},D_k}^{(u_{D_k})}$ as  defined in \eqref{schB_Y2}, and the position of $  V_{\boldsymbol{D},D_k}^{(u_{D_k})} $ in $ \pi_{D_k} \left( \big( V_{\boldsymbol{D},D_k}^{(i)} \big)_{i\in[K]} \right)$ is specified by $J_{2,D_k}^{(k)} \ominus_{K} S_{k,D_k} =  T_{2,D_k}^{(k)} = \pi_{D_k, u_{D_k}}$.      
Using the terms in $ \pi_{D_k} \left( \big( V_{\boldsymbol{D},D_k}^{(i)} \big)_{i\in[K]} \right)$  excluding $ V_{\boldsymbol{D},D_k}^{(u_{D_k})} $ in order, user $k$ decodes $w_{D_k,D_i}^{(i)}$,  $i \in [K] \setminus \{u_{D_k }\}$, by 
\begin{align} \label{schBdecode3} 
	w_{D_k,D_i}^{(i)} =   w_{D_k,D_k}^{(u_{D_k})} \oplus 	V_{\boldsymbol{D},D_k}^{(i)},  
\end{align}
where \eqref{schBdecode3} follows from  \eqref{schB_Yn} and \eqref{schB_Y2}. 
The indices of the above $K$ segments, whether directly obtained or through decoding, are sequentially given by $ \big(J_{0,D_k,1}^{(i)} \big)_{i\in[K]}$. 

In summary,  from the above process, user $ k $ can obtain $(K+1)(N-1)$ different MDS-coded segments of file $W_{D_k}$ and their corresponding indices. 
Due to the property of a $(2K(N-1),(K+1)(N-1))$ MDS code, user $k$ can decode the file $W_{D_k}$ using these segments and indices. 
  
\emph{Proof of Privacy.} 
The proof of privacy for the above scheme is similar to that in Subsection \ref{secachA}. More specifically,  we begin by proving that for any fixed   $\left(D_{[K]},\boldsymbol{p}_{[N]}, \pi_{[N]}, P_{[K]},S_{[K+1]} \right)$, 
\begin{align*}
	  & \left( X'_{\boldsymbol{D} }, Z'_k  \Big| D_{[K]},\boldsymbol{p}_{[N]}, \pi_{[N]}, P_{[K]},S_{[K+1]}    \right) 
	  \sim \text{Unif}\left( \{0,1\}^{ \frac{KN(N-1)+1}{(K+1)(N-1)}F}\right) , 
\end{align*}
which can be proved by carefully examining the independence between each term contained in $ Z'_{k}$ and $ X'_{\boldsymbol{D} }$. 
Next, we prove that when $D_k$ is fixed, the auxiliary variables  follow the same distribution under different demands from the perspective of user $k$.  
Finally, following from the two conclusions above, \eqref{privacy2} is proved to be correct.  
The proof of privacy in a formal information-theoretic way is detailed in Appendix \ref{schB_pfP}.

The proof of Theorem \ref{achAB} with $q=N-1$ and $N\ge3$ is thus complete.  
 
\section{Proof of Theorem \ref{con0}} \label{seccon}
In this section, to prove our converse, we first provide some preliminary results and simple corollaries, which are used repeatedly in the paper. 

Recalling the definitions of \textit{user-index-symmetric} and \textit{file-index-symmetric} in \cite{Tian2016}, let $\overline{\pi}( \cdot )$ be a permutation function on the user index set $[K]$, $\hat{\pi}( \cdot )$ be a permutation function on the file index set $[N]$, and $\overline{\pi}^{-1}( \cdot )$ and $\hat{\pi}^{-1}( \cdot )$ be their inverses.  
For subsets $ \mathcal{W} \subseteq W_{[N]}, \mathcal{Z} \subseteq Z_{[K]}, \mathcal{X} \subseteq X_{\mathcal{D}} $, we further define 
\begin{align*} 
& \overline{\pi}( \mathcal{Z})  \triangleq  \{ \overline{\pi}(Z_k ): Z_k \in \mathcal{Z} \},  	&       \overline{\pi}( \mathcal{X}) \triangleq  \{ \overline{\pi}(X_{\boldsymbol{d}} ): X_{\boldsymbol{d}} \in  \mathcal{X} \},  \\
& \hat{\pi}( \mathcal{W})  \triangleq  \{  \hat{\pi}(W_n ): W_n \in \mathcal{W} \},  	&         \hat{\pi}( \mathcal{X}) \triangleq  \{  \hat{\pi}(X_{\boldsymbol{d}} ): X_{\boldsymbol{d}} \in  \mathcal{X} \},  
\end{align*}  
where  $ \overline{\pi}(Z_k ) \triangleq Z_{\overline{\pi}(k)} $, $\hat{\pi}(W_n ) \triangleq W_{\hat{\pi}(n)}$, and for $ \boldsymbol{d} = (d_0,d_1, \dots, d_{K-1}) \in \mathcal{D} $, 
\begin{align*}
&  \overline{\pi}(X_{\boldsymbol{d}})   \triangleq  X_{(d_{\overline{\pi}^{-1}(0)} ,d_{\overline{\pi}^{-1}(1)}, \dots, d_{\overline{\pi}^{-1}(K-1)})},   
&   \hat{\pi}(X_{\boldsymbol{d}})  \triangleq  X_{(\hat{\pi} (d_0) ,\hat{\pi} (d_1), \dots, \hat{\pi} (d_{K-1}) )}. 
\end{align*}  
A demand private coded caching scheme is called \textit{user-index-symmetric} and \textit{file-index-symmetric} if for any subsets $ \mathcal{W} \subseteq W_{[N]}, \mathcal{Z} \subseteq Z_{[K]}, \mathcal{X} \subseteq X_{\mathcal{D}} $, and any permutations $\overline{\pi}$ and $\hat{\pi}$, the following relations hold:
\begin{subequations} 
	\begin{align}
	& H  \left(\mathcal{W}, \mathcal{Z}, \mathcal{X}  \right) = H  \left(\mathcal{W}, \overline{\pi} (\mathcal{Z} ) ,\overline{\pi} (\mathcal{X} ) \right), \label{symmetryuser} \\ 
	& H  \left( \mathcal{W}, \mathcal{Z}, \mathcal{X} \right) = H  \left(\hat{\pi} (\mathcal{W} ) ,\mathcal{Z},\hat{\pi} (\mathcal{X} ) \right). \label{symmetryfile}
	\end{align}
\end{subequations}
Similar to \cite{Tian2016}, the following proposition demonstrates the symmetry in the demand private coded caching problem.
\begin{Proposition} \label{symmetry} 
	For any demand private coded caching scheme, there always exists a  corresponding \textit{user-index-symmetric} and \textit{file-index-symmetric} scheme, which has the same or smaller cache size and delivery rate. 
\end{Proposition}
\begin{IEEEproof}
	The proof follows the same argument as \cite[Proposition 1]{Tian2016}  and is thus omitted. 
\end{IEEEproof}

The following lemma is derived from \cite[Lemma 8]{Kumar2023},  and is a converse result for the traditional $(N, K)$ coded caching problem, where there is no privacy constraint.  Note that a converse for the traditional coded caching problem also  serves as a converse for the demand private coded caching problem.   
\begin{Lem} \label{conLemma3}
	Consider a sequence of demands $\boldsymbol{d}_i$,   where each of the $N$ files is requested at least once in $\boldsymbol{d}_i$, and file $W_0$ is requested exclusively by user $\gamma_i$. 
	Then, for a sequence of user sets $\mathcal{G}_l \subseteq [K] $ and  $\mathcal{G}_i \subseteq [K] \setminus \{\gamma_i \}$, $ l <i  \le j $,  such that $\mathcal{G}_i  = \mathcal{G}_{i+1} \cup  \{ \gamma_{i+1} \}$ for $ l \le i < j$,   
	any coded caching scheme satisfies the following inequality, 
	\begin{align*}
	& H(W_{[1:N-1]},Z_{\mathcal{G}_{l}}) + \sum_{i=l+1}^j H(W_{[1:N-1]},Z_{\mathcal{G}_{i}},X_{\boldsymbol{d}_i}) 
	  \ge (j-l)NF + H(W_{[1:N-1]},Z_{\mathcal{G}_{j}}). 
	\end{align*}
\end{Lem} 
\begin{IEEEproof}
	The proof follows the same steps as \cite[Lemma 8]{Kumar2023}  and is thus omitted. 
\end{IEEEproof}   

The following lemma shows that from the perspective of   user $k$, the information it can obtain, i.e., $X_{ \boldsymbol{D}}$, $Z_k$ and $W_{D_k}$,  is identically distributed when the demands of other users change.  
\begin{Lem} \label{Lemconp} 
	For all $d_0,d_1,\dots, d_{K-1},  d'_0,d'_1, \dots, d'_{K-1} \in [N]$ and user $k$, $k \in [K]$, any demand private coded caching scheme satisfies the following distribution equation
	\begin{align*}   
	& \left(X_{\left(d_0,d_1,\dots,d_{k-1},d_k,d_{k+1}, \dots, d_{K-1}\right)},Z_{k},W_{d_k} \right)   
	 \sim     \left(X_{ (d_0',d'_1,\dots,d'_{k-1},d_k, d'_{k+1},\dots, d'_{K-1})},Z_{k},W_{d_k} \right).
	\end{align*}
\end{Lem}  
\begin{IEEEproof}
	Lemma \ref{Lemconp} extends the result of  \cite[Lemma 4]{Chinmay2022}, which works for $K=2$, to arbitrary $K$. The proof  is similar to that of \cite[Lemma 4]{Chinmay2022} and is thus omitted.   
\end{IEEEproof} 

\subsection{Proof of Theorem \ref{con0},  when $N \le K \le 2N-2 $ }   \label{secconA} 
	In this subsection, we first rewrite the converse of the case  $ K = N = 2 $ in \cite{Kamath2020}. By observing this example and building on the induction and recursion in Lemma \ref{conLemma3}, we extend the proof to arbitrary $ N$ and $K $, where $ N \leq K \leq 2N-2 $. 

\subsubsection*{Example 4} 
We consider the $(N,K)=(2,2)$  demand private coded caching problem,  and reformulate the converse proof of $ 3M + 3R \ge 5$ in \cite{Kamath2020} in the form of using induction and recursion described in Lemma \ref{conLemma3}. 	
	In the rewritten proof,  first, each delivery signal is combined with \( (N-1) \) cache contents, which not only allows the proof to extend from two files to an arbitrary number of files but also lays the foundation for the subsequent use of Lemma \ref{conLemma3}.   In addition, the application of Lemma \ref{conLemma3} allows the extension of the proof from two users to an arbitrary number of users. The detailed proof is presented as follows.   
	First, following from the correctness constraint \eqref{decoding}, it is easy to obtain
	\begin{align}\label{conex1_2}
	&  (M+R)F      \ge   H(Z_1) + H\big( X_{(0,1)}\big) 
	\ge 
	H \big(Z_1,X_{(0,1)}\big) = H\big(W_1, Z_1,X_{(0,1)}\big). 
	\end{align}  
	Next,  we have, 
	\begin{align}\label{conex1_1}
	&  (2M+2R)F \ge  H(Z_0) + H(Z_1) + 2H\big(X_{(1,1)}) 
	\ge 
	H\big(Z_0,X_{(1,1)}\big) + H\big(Z_1, X_{(1,1)}\big) \nonumber \\
	\overset{(a)}{=}  & H\big(W_1, Z_0,X_{(1,1)}\big) + H\big(W_1,  Z_1, X_{(1,1)}\big) 
	\overset{(b)}{\ge}  
	H\big(W_1, Z_0,Z_1, X_{(1,1)}\big) + H\big(W_1, X_{(1,1)}\big) \nonumber \\
	\overset{(c)}{=} & H\big(W_1, Z_0,Z_1, X_{(1,1)}\big) + H\big(W_1, X_{(1,0)}\big),   
	\end{align}
	where $(a)$  follows from the correctness constraint \eqref{decoding}, 
	and $(b)$ follows from the submodularity of the entropy function \cite{book}, i.e., let $\mathcal{X}$ represent a set of random variables, for any subsets $\mathcal{X}_1, \mathcal{X}_2 \subseteq \mathcal{X}$, we have
	\begin{align*} 
	H(\mathcal{X}_1) + H(\mathcal{X}_2) \geq H(\mathcal{X}_1 \cup \mathcal{X}_2) + H(\mathcal{X}_1 \cap \mathcal{X}_2).
	\end{align*}
	$(c)$ follows from the  privacy constraint, more specifically, Lemma \ref{Lemconp}.  
		Adding   \eqref{conex1_2}  and \eqref{conex1_1} together, we have
	\begin{align*}
	&  (3M+3R)F   
	\ge 
	H(W_1, Z_0,Z_1)  +   H \big(W_1, Z_1,X_{(0,1)}\big)  + H\big(W_1, X_{(1,0)}\big)    
	\overset{(d)} {\ge} 
	4F + H(W_1) 
	\overset{(e)} {=}     5F   ,  
	\end{align*} 
	where $(d)$  follows from Lemma \ref{conLemma3} with $l=0$, $j=2$, $\mathcal{G}_0 = \{0,1\}$,   $\mathcal{G}_1 = \{1\}$,    $\mathcal{G}_2 = \emptyset$, $\boldsymbol{d}_1 = (0,1)$ and $\boldsymbol{d}_2 = (1,0) $,   
	and  $(e)$ follows from the assumption that the files are independent, uniformly distributed  and  $F$ bits each. 
	\hfill\mbox{\,$\square$}

In the proof for arbitrary $N$ and $K$, we consider a series of user sets and demand vectors defined as follows. 
Let $\boldsymbol{a}_0 =  (1,2, \dots, N-1, 0, 1, 2, \dots, K-N)$.
Following the definition of $\boldsymbol{a}_0$, we define $\boldsymbol{a}_i$ for $i \in [1:K]$, where $\boldsymbol{a}_i$ is obtained by left cyclically shifting the vector $\boldsymbol{a}_0$ by $i$ positions, e.g., 
in the case of $N=4$, $K=5$, and $i=2$, we have $\boldsymbol{a}_2 = (3,0,1,1,2)$.
Since $\boldsymbol{a}_0$ is a  $K$-length vector, we note that  $\boldsymbol{a}_0=\boldsymbol{a}_K$.
Thus, in demand $\boldsymbol{a}_i $, the user requires  $W_0$, denoted by $\gamma_i$,  satisfies 
\begin{align} \label{gammai} 
\gamma_i =  \left\{
\begin{aligned}
& N-i-1 ,       \\
& K+N-i-1,    
\end{aligned}
\quad 
\begin{aligned}
&   \text{if } 0 \le i \le N-1  \\
&   \text{if }   N \le i \le K 
\end{aligned}
\right. .  
\end{align}  
Another series of demand vectors $\boldsymbol{b}_i$,  for $i \in [K-N+1]$, are obtained by setting the $(N-1)$-th position of the vector $\boldsymbol{a}_i$, i.e., the demand of user $N-1$, to $K-N+1$. 
For example, when $N=4$ and $K=5$, we have $\boldsymbol{b}_0 = (1,2,3,2,1)$ and $\boldsymbol{b}_1 = (2,3,0,2,1)$. 

As shown in Table \ref{conset1}, we define the user sets $\mathcal{A}_i$ and $\mathcal{B}_i$ for $i \in [N]$, and the user sets $\mathcal{C}_i$ and $\mathcal{P}_i$ for $i \in [K-N+1]$.    
\begin{table*}   [h!]
	\begin{center}  
		\caption{ }  
		\label{conset1}
		  \vspace{-3mm}
		\begin{threeparttable}
			\begin{tabular}{|c|c|c|} %
				\hline  
				User Set & Set Size  
				& Files Requested  in  $\boldsymbol{a}_i$ \tnote{a}   \\
				\hline
				$\mathcal{A}_i = [0:N-i-2]$ & $N-i-1$    & $W_{i+1},W_{i+2},\dots,W_{N-1}$ \\
				\hline
				$\mathcal{B}_i = [K-i: K-1]  $ & $i$   & $W_1,W_2,\dots,W_{i}$ \\
				\hline
				$\mathcal{C}_i = [K-N-i+1:N-i-2] $ & $2N-K-2 $     & $W_{K-N+2},W_{K-N+3}, \dots, W_{N-1}$ \\
				\hline
				$\mathcal{P}_i = \big[N:K-i-1 \big] $ & $K-N-i$      & $ W_{i+1}, W_{i+2}, \dots, W_{K-N}  $ \\
				\hline
			\end{tabular}
			\begin{tablenotes}  
				\footnotesize
				\item[a] Since $\boldsymbol{b}_i$ differs from $\boldsymbol{a}_i$ only at the demand of user  $N-1$, the results in this column hold true for $\boldsymbol{b}_i$  when $i \in [K-N+1]$.
			\end{tablenotes}
			 \vspace{-12mm} 
		\end{threeparttable}
	\end{center}
\end{table*}
In Table \ref{conset1}, we note that
\begin{align} 
& \mathcal{A}_{N-1} = \mathcal{B}_{0} =  \mathcal{P}_{K-N} =  \emptyset, \quad 
 \mathcal{P}_i  \cup \mathcal{B}_i =  [N:K-1], \quad  \forall  i \in [K-N+1].   \label{defE}
\end{align} 
Following \eqref{defE}, we further define the user set $\mathcal{E} = \{N-1\} \cup \mathcal{P}_i \cup \mathcal{B}_i = [N-1:K-1]$. For demand $\boldsymbol{b}_i$,   $i \in [K-N+1]$, we  note that  users in the set $\mathcal{E}$ request $W_1, W_2, \dots, W_{K-N+1}$.  
Due to \( |\mathcal{A}_i \cup \mathcal{B}_i| = N-1 \) and the correctness constraint, i.e., \eqref{decoding},  
 for $i \in [K-N+1: N-1]$, we have
\begin{align} \label{con1_1}
& (N-1)MF + RF 
\ge  
 H(Z_{\mathcal{A}_i \cup \mathcal{B}_i } ,X_{\boldsymbol{a}_i }) =  H(W_{[1:N-1]}, Z_{\mathcal{A}_i \cup \mathcal{B}_i } ,X_{\boldsymbol{a}_i }).  
\end{align} 
In relation to Example 4, \eqref{con1_1} can be seen as a generalization of the bound in \eqref{conex1_2}.   

Next, the following lemma is generalized from the bound in \eqref{conex1_1}.   
For $ i \in [K-N+1]$, Lemma \ref{Lemcon11} lower bounds $2(N-1)RF + 2MF$ by terms involving the joint entropy of the files $W_{[1:N-1]}$, the cache contents of certain users, and $X_{\boldsymbol{a}_i}$ or $X_{\boldsymbol{b}_i}$.    
\begin{Lem} \label{Lemcon11} 
	For the $(N,K)$ demand private coded caching problem and $i \in [K-N+1]$, the sets $\mathcal{A}_i$, $\mathcal{B}_i$, and $\mathcal{E}$, and the demands $\boldsymbol{a}_i$ and $\boldsymbol{b}_i$, defined above, satisfy 
	\begin{align*} 
	& 2(N-1)MF+2RF 
	\ge H\left( W_{[1:N-1]},Z_{\mathcal{A}_i \cup    \mathcal{E}}, X_{\boldsymbol{b}_{i}}\right)  
	+  H\left(W_{[1:N-1]},Z_{\mathcal{B}_i},X_{\boldsymbol{a}_{i}}\right).
	\end{align*} 
\end{Lem}
\begin{IEEEproof}	
	For $i \in [K-N+1]$, we have  
	\begin{align*} 
	& 2(N-1)MF + 2RF 
	\overset {(a)}  \ge 
	\sum_{j \in  \mathcal{C}_i \cup    \mathcal{E}}  H(Z_j)  +   H(X_{\boldsymbol{a}_i} )        
	+ \sum_{j \in  \mathcal{A}_i \cup  \mathcal{B}_i } H(Z_j) +   H(X_{\boldsymbol{b}_i} )  \nonumber \\ 
	\ge & \sum_{j \in   \mathcal{C}_i \cup \mathcal{E} \setminus \mathcal{B}_i    }  H\left(Z_j,X_{\boldsymbol{b}_i}\right) 
	- |\mathcal{C}_i \cup \mathcal{E} \setminus \mathcal{B}_i  | H(X_{\boldsymbol{b}_i} ) 
	+ H\left( Z_{\mathcal{A}_i \cup \mathcal{B}_i} , X_{\boldsymbol{b}_i}\right)   
	+ H(Z_{ \mathcal{B}_i}, X_{\boldsymbol{a}_i})  
	\nonumber \\ 
	\overset {(b)} { = }		
	&   \sum_{j \in  \mathcal{C}_i}  H \left(W_{j+i+1}, Z_j,X_{\boldsymbol{b}_i}\right) +
	\sum_{j \in   \mathcal{P}_i   }  H \left(W_{j-N+i+1}, Z_j,X_{\boldsymbol{b}_i} \right)    
	 + H\left( W_{K-N+1}, Z_{N-1},X_{\boldsymbol{b}_i} \right) \nonumber \\    & - |\mathcal{C}_i \cup \mathcal{E} \setminus \mathcal{ B}_i | H\left( X_{\boldsymbol{b}_i} \right)
	+ H \left(W_{[1:N-1]}, Z_{\mathcal{A}_i \cup \mathcal{B}_i} , X_{\boldsymbol{b}_i}\right)  + H\left(W_{[1:i]}, Z_{ \mathcal{B}_i}, X_{\boldsymbol{a}_i}\right)    
	\nonumber \\
	\overset {(c)} { \ge }		
	&   \sum_{n = i+1}^{N-1}  H \left( W_{n}, X_{\boldsymbol{b}_i}\right) 
	+ H \left( W_{[1:N-1]}, Z_{\mathcal{A}_i \cup \mathcal{E} } , X_{\boldsymbol{b}_i}\right)  
	+ H\left( W_{[1:i]}, Z_{ \mathcal{B}_i}, X_{\boldsymbol{a}_i}\right) - |\mathcal{C}_i \cup \mathcal{E} \setminus \mathcal{ B}_i | H\left( X_{\boldsymbol{b}_i} \right)   
	\nonumber \\
	\overset {(d)} { = }		
	& \sum_{n = i+1}^{N-1}  H \left( W_{n}, X_{\boldsymbol{a}_i}\right)
	+ H\left( W_{[1:N-1]}, Z_{\mathcal{A}_i \cup \mathcal{E} } , X_{\boldsymbol{b}_i}\right)  
	+ H\left(  W_{[1:i]}, Z_{ \mathcal{B}_i}, X_{\boldsymbol{a}_i}\right) - |\mathcal{C}_i \cup \mathcal{E} \setminus \mathcal{ B}_i | H\left( X_{\boldsymbol{a}_i}\right)  
	\nonumber \\
	\overset {(e)} { \ge }		
	&  H \left(   W_{[1:N-1]}, Z_{\mathcal{A}_i \cup \mathcal{E} } , X_{\boldsymbol{b}_i} \right) 
	+ H  \left( W_{[1:N-1]}, Z_{ \mathcal{B}_i}, X_{\boldsymbol{a}_i}\right) ,   
	\end{align*} 
 where $(a)$ follows from the facts that  $H(Z_j) \leq MF$, $H(X_{\boldsymbol{D}}) \leq RF$  for $\forall j \in [K], \forall \boldsymbol{D} \in \mathcal{D}$, and $|\mathcal{C}_i \cup    \mathcal{E}| = |\mathcal{A}_i \cup  \mathcal{B}_i |= N-1$ for $\forall i \in [K-N+1]$,  
	$(b)$ follows from the correctness constraint \eqref{decoding} and  $\mathcal{E} = \{N-1\} \cup  \mathcal{P}_i \cup \mathcal{B}_i $, 
	$(c)$ follows from the submodularity of the entropy function with  $ Z_{\mathcal{A}_i  \cup  \mathcal{B}_i } \cup   Z_{\mathcal{P}_i}  \cup \{Z_{N-1}\} = Z_{\mathcal{A}_i  \cup  \mathcal{E} }$,  
	$(d)$ follows from the privacy constraint, more  specifically, Lemma \ref{Lemconp},  
	and $(e)$ follows from the submodularity of the entropy function with  $W_{[i+1:N-1]} \cup W_{[1:i]} = W_{[1:N-1]}$.    
\end{IEEEproof}
  
Finally, armed with the above lemmas, we have,  
\begin{align*} 
	& (K+1)(N-1)MF  + (K+1) RF  
	=  
	 \left( 2(K-N+1)  + 2N-K-1  \right) \left((N-1)MF+RF  \right) \nonumber \\
	\overset{(a)}{\ge}  &
	\sum_{i = 0  }^{K-N} \left( H \left( W_{[1:N-1]}, Z_{\mathcal{A}_i \cup \mathcal{E}} , X_{\boldsymbol{b}_{i}}\right)  +  H \left(  W_{[1:N-1]}, Z_{\mathcal{B}_i}, X_{\boldsymbol{a}_{i}}\right)   \right)  \nonumber \\ & 
	 + \sum_{i= K-N+1}^{N-1}
	H \left(   W_{[1:N-1]},  Z_{\mathcal{A}_i \cup  \mathcal{B}_i }, X_{\boldsymbol{a}_i} \right)   \nonumber \\
	\overset{(b)}{\ge}  & 
	(K-N)NF +  H \left( W_{[1:N-1]}, Z_{\mathcal{A}_{K-N} \cup \mathcal{E}} \right)  \nonumber \\ & 
	+ \sum_{i= K-N+1}^{N-1}
	H \left( W_{[1:N-1]}, Z_{\mathcal{A}_i \cup \mathcal{E} }, X_{\boldsymbol{a}_i} \right)     + \sum_{i = 0 }^{K-N}  H \left(  W_{[1:N-1]}, Z_{\mathcal{B}_i},  X_{\boldsymbol{a}_{i+N}}\right)  \nonumber \\
	\overset{(c)}{\ge} &
	(N-1)NF      + H   \left(  W_{[1:N-1]}, Z_{ \mathcal{E}}  \right)      + \sum_{i = 0 }^{K-N}  H  \left(   W_{[1:N-1]}, Z_{\mathcal{B}_i},  X_{\boldsymbol{a}_{i+N}}  \right)   \nonumber \\
	\overset{(d)}{\ge}    &    
	KNF     +   H \left( W_{[1:N-1]} \right)    
	\overset{(e)}{=}   \left(  (K+1)N-1 \right) F,    
\end{align*}    
where $(a)$  follows from Lemma \ref{Lemcon11} and \eqref{con1_1},   
the first two terms of $(b)$ follow from Lemma \ref{conLemma3} with $\mathcal{G}_i = \mathcal{A}_i \cup \mathcal{E}$, $\boldsymbol{d}_i = \boldsymbol{b}_{i}$, $l = 0$, $j = K-N$, and $\mathcal{G}_i = \mathcal{G}_{i+1} \cup \{N-i-2\} = \mathcal{G}_{i+1} \cup \{\gamma_{i+1}\}$ for $i \in [K-N]$,   
the third term of $(b)$ follows from $\mathcal{E} \subseteq \mathcal{B}_i$ for $i \in [K-N+1 : K-1]$,  
the forth term of $(b)$ follows from Proposition \ref{symmetry}, more specifically, \eqref{symmetryuser}, 
$(c)$ follows from Lemma \ref{conLemma3} with $\mathcal{G}_i = \mathcal{A}_i \cup \mathcal{E}$,  $\boldsymbol{d}_i = \boldsymbol{a}_{i}$, $l = K-N$, $j = N-1$,  $\mathcal{G}_i = \mathcal{G}_{i+1} \cup \{N-i-2\} = \mathcal{G}_{i+1} \cup  \{\gamma_{i+1} \} $ for $i \in [K-N:N-2]$,  and  $\mathcal{G}_{N-1} = \mathcal{E}$,  
$(d)$ follows from Lemma \ref{conLemma3} with $\mathcal{G}_i = \mathcal{B}_{K-i} = [i:K-1]$,  $\boldsymbol{d}_i =  \boldsymbol{a}_{K+N-i}$, $l = N-1$, $j = K$, $\mathcal{G}_i = \mathcal{G}_{i+1} \cup \{i\} = \mathcal{G}_{i+1} \cup  \{\gamma_{i+1} \} $ for $i \in [N-1:K-1]$, and ${\mathcal{G}_{K}} = \emptyset $, and
$(e)$ follows from the assumption that the files are independent, uniformly distributed and $F$ bits each. 

The proof of Theorem \ref{con0} when $N \le K \le 2N-2 $  is thus complete.

\subsection{Proof of Theorem \ref{con0}, when $ K > 2N -2 $ } \label{subseccon2}
In this subsection, we first provide a new converse for the case where $K = 3$ and $N = 2 $, and then  extend this proof to arbitrary $N$ and $K$, where $K > 2N - 2 $, using methods similar to those in Section \ref{secconA}. 
\subsubsection*{Example 5} 
We now provide the converse proof of $6M + 5R \geq 9$ for the $(N,K) = (2,3)$ demand private coded caching problem as follows.  
	First, 	we have 
	\begin{align} \label{conex2_1}
	& H \left( W_1, Z_0,Z_1, X_{(1,1,1)} \right)   +  H\left( Z_2\right)  +  \frac{1}{2} H\left( X_{(1,1,1)}\right)  \nonumber  \\
	\ge   & H \big(W_1, Z_{ [0:1]}, X_{(1,1,1)} \big)    +  H \big( Z_2,X_{(1,1,1)}\big) - \frac{1}{2} H \big( X_{(1,1,1)} \big)    \nonumber   \\
	\overset{(a)} {\ge }  & H \big( W_1, Z_{[0:2]}, X_{(1,1,1)}\big)     +  H \big(W_1,X_{(1,1,1)}\big)  - \frac{1}{2} H\big(X_{(1,1,1)}\big)   \nonumber   \\
	\overset{(b)} {=}  & H\big(W_1, Z_{[0:2]}, X_{(1,1,1)}\big)  
	 + \frac{1}{2}  \left(H\big(W_1, X_{(1,0,1)}\big) + H\big(W_0, X_{(1,0,1)}\big) - H\big(X_{(1,0,1)}\big)   \right)   \nonumber  \\  
	\overset{(c)} {\ge}  & H\big(W_1, Z_{[0:2]}, X_{(1,1,1)}\big)  + \frac{1}{2}  H(W_0,W_1)    
	\overset{(d)} {=}  
	H\big(W_1, Z_{[0:2]}, X_{(1,1,1)}\big) + F,        
	\end{align}
	where $(a)$ follows from the submodularity of the entropy function and the correctness constraint \eqref{decoding},  
 	$(b)$ follows from the privacy constraint, i.e., Lemma  \ref{Lemconp},  and  Proposition 1,  more specifically, 
	\begin{align*}
	H \big(W_1,X_{(1,1,1)} \big)   & =  H\big(W_1,X_{(1,0,1)} \big),  \text{ and} \\ 
	H\big(W_1,X_{(1,1,1)} \big)  &   =  H\big(W_0,X_{(0,0,0)} \big)  =  H\big(W_0,X_{(1,0,1)} \big),  
	\end{align*} 
$(c)$  follows from the submodularity of the entropy function, and  $(d)$  follows from the assumption that the files are independent, uniformly distributed  and  $F$ bits each. 
	Thus, we have  
	\begin{align} \label{conex2_2}
	&   3MF + \frac{5}{2}RF =   (2M+2R)F +  MF + \frac{1}{2}RF    \nonumber  \\
	\overset{(e)} {\ge }   &   H \big(W_1, Z_0,Z_1, X_{(1,1,1)}\big)   + H \big(W_1, X_{(1,0,1)}\big) 
	+  H(Z_2)  +  \frac{1}{2} H\big(X_{(1,1,1)}\big)  \nonumber  \\ 
	\overset{(f)} {\ge}  & H\big(W_1, Z_{[0:2]}, X_{(1,1,1)}\big) + H\big(W_1, X_{(1,0,1)}\big)  + F,       
	\end{align}  
	where the first two terms of $(e)$ can be obtained by replacing $X_{(1,1)}$ and $X_{(1,0)}$ in the proof of \eqref{conex1_1} with $X_{(1,1,1)}$ and $X_{(1,0,1)}$, respectively, and $(f)$ follows from \eqref{conex2_1}.
	
	Similar to \eqref{conex1_2}, \eqref{conex2_1}, and \eqref{conex2_2}, we have
	\begin{align} \label{conex2_3}  
	&   3MF + \frac{5}{2}RF    =   2(M+R)F +  MF + \frac{1}{2}RF   \nonumber  \\
	\ge   & 
	2 H\big(W_1, Z_{2}, X_{(0,1,1)}\big) + H(Z_1) + \frac{1}{2}H\big(X_{(0,1,1)}\big)   \nonumber  \\
	\ge  &  H\big(W_1,Z_{1}, Z_{2}, X_{(0,1,1)}\big) +  H\big(W_1, Z_{2}, X_{(0,1,1)}\big) + F.   
	\end{align} 
	Adding  \eqref{conex2_2} and  \eqref{conex2_3} together,  we have 
	\begin{align*}
	&  (6M+5R)F   \\
	\ge  &  H \big(W_1, Z_{[0:2]}, X_{(1,1,1)}\big) +   H\big(W_1,Z_{1}, Z_{2}, X_{(0,1,1)}\big)    \\  
	&  +  H\big(W_1, Z_{2}, X_{(0,1,1)}\big) + H\big(W_1, X_{(1,0,1)}\big) + 2F \\  
	\overset{(a)} {\ge }  %
	&  H(W_1, Z_{[0:2]}) +   H\big(W_1,Z_{1}, Z_{2}, X_{(0,1,1)}\big)   
	+  H\big(W_1, Z_{2}, X_{(1,0,1)}\big) + H\big(W_1, X_{(1,1,0)}\big) + 2F \\ 
	\overset{(b)} {\ge }  %
	&  6F  + H(W_1) + 2F 	
	\overset{(c)} { = }  9F, 
	\end{align*} 
	where $(a)$ follows from \eqref{symmetryuser}, 
	$(b)$  follows from Lemma \ref{conLemma3} with $l=0$, $j=3$, $\mathcal{G}_0 = \{0,1,2\}$,   $\mathcal{G}_1 = \{1,2\}$,    $\mathcal{G}_2 = \{2\}$, $\mathcal{G}_3 = \emptyset$,  $\boldsymbol{d}_1 = (0,1,1)$, $\boldsymbol{d}_2 = (1,0,1)$ and $\boldsymbol{d}_3 = (1,1,0) $, 
	and $(c)$ follows from the assumption that the files are independent, uniformly distributed  and  $F$ bits each.
		\hfill\mbox{\,$\square$}    

In the proof for arbitrary $N$ and $K$,  let \( \boldsymbol{a}_0 \) be a \( K \)-length vector that ends with \( K - 2N + 2 \) copies of \( N-1 \), and is given by \( \boldsymbol{a}_0 = (1, 2, \dots, N-1, 0, 1, 2, \dots, N-2, N-1, N-1, \dots, N-1) \).  Similar to the proof in Section \ref{secconA}, 
for $i \in [1:K]$, define $\boldsymbol{a}_i$ by left cyclically shifting the vector $\boldsymbol{a}_0$ by $i$ positions.   
For $i \in [N]$, we further define $\boldsymbol{b}_i$ by setting the $(N-1)$-th position of  $\boldsymbol{a}_i$ as $N-1$.  
Note that in demand $\boldsymbol{a}_i$,
the user requesting $W_0$, denoted by $\gamma_i$, also satisfies \eqref{gammai}. 
\begin{table*} 
	\begin{center}  
		\caption{ } 
		\label{conset2}
		 \vspace{-2.5mm}
		\begin{threeparttable}
			\begin{tabular}{|c|c|c|} 
				\hline  
				User Set & Set Size  
				& Files Requested  in  $\boldsymbol{a}_i$  \tnote{a}     \\
				\hline
				$\mathcal{A}_i = [0:N-i-2]$ & $N-i-1$    & $W_{i+1},W_{i+2},\dots,W_{N-1}$ \\
				\hline
				$\mathcal{B}_i = [K-i: K-1]  $ & $i$   & $W_1,W_2,\dots,W_{i}$ \\
				\hline
				$\mathcal{P}_i = \big[N:2N-i-3 \big] $& $N-i-2$      & $ W_{i+1}, W_{i+2}, \dots, W_{N-2}  $ \\
				\hline
				$\mathcal{O}_i = [2N-i-2:K-i-1] $ & $K-2N+2$    & $W_{N-1}$ \\ 
				\hline
			\end{tabular}
			\begin{tablenotes} 
				\footnotesize
				\item[a] For $\mathcal{A}_i$, $\mathcal{B}_i$ and $\mathcal{O}_i$, $ i \in [N]$ is considered while for $\mathcal{P}_i$,  $i \in [N-1]$ is considered.  
			\end{tablenotes}
			\vspace{-10mm}
		\end{threeparttable} 
	\end{center}
\end{table*}

As shown in Table \ref{conset2}, define the user sets   $\mathcal{A}_i$  and $\mathcal{O}_i$ for $i \in [N]$,   the user set $ \mathcal{P}_i$ for $i \in [N-1]$, and  the user set  $\mathcal{B}_i$  for  $i \in [K-N+1]$.  In addition, we  define $\mathcal{E}  = [N-1:K-1]$, and note that  $\mathcal{E} = \{N-1\} \cup \mathcal{P}_i  \cup \mathcal{O}_i \cup \mathcal{B}_i$ for $i \in [N-1]$.  

Following the above definitions, we generalize \eqref{conex2_2} and \eqref{conex2_3} to arbitrary $N$ and $K$.   
For $ i \in [N]$, Lemma \ref{con2Lem1} is generalized from the bounds in  \eqref{conex2_2} and  \eqref{conex2_3}, and  lower bounds $\left( KM + \frac{K+2}{N} R \right) F$   by terms involving the joint entropy of the files $W_{[1:N-1]}$, the cache contents of certain users, and $X_{\boldsymbol{a}_i}$ or $X_{\boldsymbol{b}_i}$, and the entropy of files.  
\begin{Lem} \label{con2Lem1} 
	For the $(N,K)$ demand private coded caching problem and $i \in [N]$, the sets $\mathcal{A}_i$, $\mathcal{B}_i$, and $\mathcal{E}$, and the demands $\boldsymbol{a}_i$ and $\boldsymbol{b}_i$, defined above, satisfy   
	\begin{align*}
	& KMF + \frac{K+2}{N} RF   \ge H \big(W_{[1:N-1]}, Z_{ \mathcal{A}_i \cup \mathcal{E} }  , X_{\boldsymbol{b}_i}\big)    
	+ H\big(W_{[1:N-1]}, Z_{ \mathcal{B}_i }, X_{\boldsymbol{a}_{i}} \big) +  (K-2N+2)F.
	\end{align*}  
\end{Lem}
\begin{IEEEproof} 
	In the proof of Lemma \ref{Lemcon11}, we substitute $\mathcal{E}$ with $\mathcal{E} \setminus \mathcal{O}_i$, and set $\mathcal{C}_i = \emptyset$, $K=2N-2$. It is observed that for $i \in [N-1]$, each step in the proof of Lemma \ref{Lemcon11} holds for the demands $\boldsymbol{a}_{i}$, $\boldsymbol{b}_{i}$, and the  sets $\mathcal{A}_i$, $\mathcal{B}_i$, $\mathcal{P}_i$, $\mathcal{E}$ defined in this subsection. Thus, for $i \in [N-1]$, we have
	\begin{align} 
	\label{con2Lem2}   
	&   2(N-1)MF+2RF  \ge   
	H \big(W_{[1:N-1]}, Z_{\mathcal{A}_i \cup(\mathcal{E} \setminus \mathcal{O}_i ) }, X_{\boldsymbol{b}_i}\big)  
	+ H\big( W_{[1:N-1]}, Z_{ \mathcal{B}_i}, X_{\boldsymbol{a}_i} \big). 
	\end{align}
	For $i = N-1$, the result of \eqref{con2Lem2} also holds, i.e.,    
	\begin{align} 	
	\label{con2Lem3} 
	& 2(N-1)MF+2RF  
	\overset {(a)}  \ge  2 \sum_{j \in \mathcal{B}_i} H(Z_j) +  2 H(X_{\boldsymbol{a}_i} )  
	\ge 
	2 H(Z_{ \mathcal{B}_i }, X_{\boldsymbol{a}_i} )   \nonumber \\ 
	 \overset {(b)}   =  & 2 H \big(W_{[1:N-1]}, Z_{ \mathcal{B}_i }, X_{\boldsymbol{a}_i} \big) 
	\overset {(c)}  = 
	H \big(W_{[1:N-1]},Z_{ \mathcal{A}_i \cup (\mathcal{E} \setminus \mathcal{O}_i) }, X_{\boldsymbol{b}_{i}}\big)
	+ H \big(W_{[1:N-1]},Z_{\mathcal{B}_i}, X_{\boldsymbol{a}_{i}}\big),   
	\end{align}
	 where $(a)$ follows from the facts that  $H(Z_j) \leq MF$, $H(X_{\boldsymbol{D}}) \leq RF$   for $\forall j \in [K], \forall \boldsymbol{D} \in \mathcal{D}$, and $|\mathcal{B}_{N-1}| = N-1$,  
	$(b)$ follows from the correctness constraint \eqref{decoding},  
	and $(c)$ follows from $\boldsymbol{a}_{N-1} = \boldsymbol{b}_{N-1}$, $\mathcal{A}_{N-1} = \emptyset$, and $\mathcal{E} \setminus \mathcal{O}_{N-1} = \mathcal{B}_{N-1} = [K-N+1:K-1]$. 
	
	For $i \in [N]$, following from \eqref{con2Lem2}, \eqref{con2Lem3}, and the fact that $|\mathcal{O}_i| = K - 2N + 2$, we have 
	\begin{align} \label{con2Lem4} 
	& KMF + \frac{K+2}{N} RF 
	= 
	 2(N-1)MF+2RF + (K-2N+2)  \left(M+ \frac{1}{N} R \right)F \nonumber \\ 
	\ge &  H \big(W_{[1:N-1]},Z_{ \mathcal{A}_i \cup (\mathcal{E} \setminus \mathcal{O}_i) }, X_{\boldsymbol{b}_{i}} \big) + H\big(W_{[1:N-1]},Z_{\mathcal{B}_i}, X_{\boldsymbol{a}_{i}}\big)
 	\nonumber \\ 	& 
	  + \sum_{j \in \mathcal{O}_i }   \left( H\left( Z_j ,X_{\boldsymbol{b}_i} \right)  - \frac{N-1}{N} H\left( X_{\boldsymbol{b}_i}\right)  \right)     \nonumber \\ 
	\overset {(a)} = &  H \big(W_{[1:N-1]},Z_{ \mathcal{A}_i \cup (\mathcal{E} \setminus \mathcal{O}_i) }, X_{\boldsymbol{b}_{i}} \big) + H \big(W_{[1:N-1]},Z_{\mathcal{B}_i}, X_{\boldsymbol{a}_{i}}\big)
	\nonumber \\ 
	&   + \sum_{j \in \mathcal{O}_i }   \left(    H\left( W_{N-1}, Z_j ,X_{\boldsymbol{b}_i} \right)  - \frac{N-1}{N} H \left(  X_{\boldsymbol{b}_i}\right)  \right) \nonumber \\ 
	\overset {(b)}  \ge &  H\big(W_{[1:N-1]},Z_{ \mathcal{A}_i \cup \mathcal{E}  }, X_{\boldsymbol{b}_{i}}\big) + H\big(W_{[1:N-1]},Z_{\mathcal{B}_i}, X_{\boldsymbol{a}_{i}}\big)
	\nonumber \\ 
	&  
	 + \sum_{j \in \mathcal{O}_i }   \left(   H  \left(W_{N-1},X_{\boldsymbol{b}_i} \right)   - \frac{N-1}{N} H \left( X_{\boldsymbol{b}_i}\right)   \right) \nonumber \\   
	\overset {(c)}  = &  H\big(W_{[1:N-1]},Z_{ \mathcal{A}_i \cup \mathcal{E}  }, X_{\boldsymbol{b}_{i}}\big) + H\big(W_{[1:N-1]},Z_{\mathcal{B}_i}, X_{\boldsymbol{a}_{i}}\big)
	\nonumber \\ 
	&  + \frac{|\mathcal{O}_i|}{N}  \left(  \sum_{n=0}^{N-1}  H \left(W_n ,X_{\boldsymbol{a}_i} \right)    - (N-1) H \left(X_{\boldsymbol{a}_i}\right)    \right)  
	\nonumber \\
	\overset {(d)}  \ge &  H\big(W_{[1:N-1]},Z_{ \mathcal{A}_i \cup \mathcal{E}  }, X_{\boldsymbol{b}_{i}}\big) + H\big(W_{[1:N-1]},Z_{\mathcal{B}_i}, X_{\boldsymbol{a}_{i}}\big)
	 + (K-2N+2)F,    
	\end{align} 
	where $(a)$ follows from the correctness constraint \eqref{decoding},  
	$(b)$ follows from the submodularity of the entropy function, 
	$(c)$ follows from Lemma \ref{Lemconp} and Proposition 1, 
	and $(d)$  follows from the submodularity of the entropy function and the assumption that the files are independent, uniformly distributed  and  $F$ bits each.  
\end{IEEEproof}

Similar to the proof of Lemma \ref{con2Lem1}, we can  derive the following lemma, which lower bounds $ \left( iM  + \frac{i+1}{N} R \right) F $  by terms involving the entropy of files, and the joint entropy of  $W_{[1:N-1]}$, the cache contents of certain users, and $X_{\boldsymbol{a}_{i+N}}$.   
\begin{Lem}  \label{con2Lem5} 
	For the $(N,K)$ demand private coded caching problem and $i \in [N:K-N]$, the set $\mathcal{B}_i$ and the demand $\boldsymbol{a}_{i+N}$, defined above, satisfy 
	\begin{align*} 
	& iMF   + \frac{i+1}{N} RF  
	\ge  H \big(W_{[1:N-1]}, Z_{\mathcal{B}_i},X_{\boldsymbol{a}_{i+N} } \big) + (i-N+1)F.
	\end{align*}   
\end{Lem}
\begin{IEEEproof}
	Let $ \mathcal{Q}_i = [K-i+N-1:K-1]$ for $i \in [N:K-N]$. 
	Recalling the definitions of $\boldsymbol{a}_{i+N}$ and $ \mathcal{B}_i$, we note that for the demand $\boldsymbol{a}_{i+N}$, users in $ \mathcal{Q}_i $ request file $ W_{N-1}$, while users in $ \mathcal{B}_i \setminus \mathcal{Q}_i = [K-i:K-i+N-2]$ respectively request $ W_{1}, W_{2}, \dots, W_{N-1} $. Thus, for  $i \in [N:K-N]$, we have  
	\begin{align}
	&  iMF   + \frac{i+1}{N} RF    
	= 
	(i-N+1)  \left( MF+\frac{1}{N} RF \right)  +  (N-1)MF + RF    \nonumber \\ 
	\ge   &  
	 \sum_{j\in \mathcal{Q}_i  } \left( H \left( Z_j,X_{\boldsymbol{a}_{i+N}}\right) - \frac{N-1}{N} H\left(X_{\boldsymbol{a}_{i+N}}\right)    \right)  
	+ 	H  \left( Z_{\mathcal{B}_i \setminus \mathcal{Q}_i}, X_{\boldsymbol{a}_{i+N}} \right)  \nonumber \\ 
	\overset {(a)} { = } &  
	\sum_{j\in \mathcal{Q}_i  } \left(  H \left( W_{N-1},  Z_j,X_{\boldsymbol{a}_{i+N}}\right)        - \frac{N-1}{N} H \left(X_{\boldsymbol{a}_{i+N}}\right)  \right)   
	+ 	H \left(W_{[1:N-1]}, Z_{\mathcal{B}_i \setminus \mathcal{Q}_i}, X_{\boldsymbol{a}_{i+N}}\right)  \nonumber \\ 
	\overset {(b)} { \ge } &   
	\frac{|\mathcal{Q}_i|}{N}
	\left(  N H \left( W_{N-1},X_{\boldsymbol{a}_{i+N}}\right) 
	-  (N-1)H \left(X_{\boldsymbol{a}_{i+N}}\right)  \right)   
	+ 	H \left(W_{[1:N-1]}, Z_{\mathcal{B}_i}, X_{\boldsymbol{a}_{i+N}} \right)  \nonumber \\
	\overset {(c)} { = } &   
	\frac{|\mathcal{Q}_i|}{N}
	\left( \sum_{n = 0}^{N-1}  H \left( W_{n},X_{\boldsymbol{a}_{i+N}}\right) 
	-  (N-1)H \left(X_{\boldsymbol{a}_{i+N}}\right)  \right) 
	 + 	H \left(W_{[1:N-1]}, Z_{\mathcal{B}_i}, X_{\boldsymbol{a}_{i+N}} \right)  \nonumber \\
	\overset {(d)} { \ge } &   
	(i-N+1) F + 	H \left(W_{[1:N-1]}, Z_{\mathcal{B}_i}, X_{\boldsymbol{a}_{i+N}} \right) ,   \nonumber 
	\end{align} 
	where $(a)$ to $(d)$ follow from reasons consistent with those in \eqref{con2Lem4}. 
\end{IEEEproof}

Armed with Lemmas \ref{con2Lem1} and \ref{con2Lem5}, we have 
\begin{align*}
& \frac{K(K+1)}{2} M F + \frac{(K+1)(K+2)}{2N} RF  \nonumber \\  
=  & \sum_{i=N}^{K-N} \left(iMF  + \frac{i+1}{N} RF \right) +  N\left( KM + \frac{K+2}{N} R\right)F    \nonumber \\ 
	\overset {(a)} { \ge }  & 
  \sum_{i=N}^{K-N} \left(   H \left(W_{[1:N-1]}, Z_{\mathcal{B}_i}, X_{\boldsymbol{a}_{i+N}}\right) + (i-N+1)F \right)  \nonumber \\
& + \sum_{i=0}^{N-1}  \left(   (K-2N+2)F + H \left(W_{[1:N-1]}, Z_{\mathcal{A}_i \cup \mathcal{E}}, X_{\boldsymbol{b}_i}\right)     +  H \left(W_{[1:N-1]}, Z_{\mathcal{B}_i}, X_{\boldsymbol{a}_i}\right)   \right)  \nonumber \\
\overset {(b)} { \ge } 
 &        \sum_{i=0}^{N-1}    H \left(W_{[1:N-1]}, Z_{\mathcal{A}_i \cup \mathcal{E}}, X_{\boldsymbol{b}_i}\right)     +  \sum_{i=0}^{K-N}    H \left(W_{[1:N-1]}, Z_{\mathcal{B}_i}, X_{\boldsymbol{a}_{i+N}}\right)  
     + \frac{(K-2N+2)(K+1)}{2} F   \nonumber \\
\overset {(c)} { \ge } 
& (N-1)NF    + H \left(W_{[1:N-1]}, Z_{ \mathcal{E}}\right)    +  \sum_{i=0}^{K-N}    H \left(W_{[1:N-1]}, Z_{\mathcal{B}_i}, X_{\boldsymbol{a}_{i+N}}\right)     +  \frac{(K-2N+2)(K+1)}{2} F     \\   
\overset {(d)} { \ge } & 
   (N-1)NF    +  (K-N+1)NF 
   + H\left(W_{[1:N-1]}\right)  + \frac{(K-2N+2)(K+1)}{2} F  
\overset {(e)} {=} 
\frac{K(K+3)}{2} F,    
\end{align*} 
where $(a)$ follows from Lemmas \ref{con2Lem1} and  \ref{con2Lem5}, 
$(b)$ follows from  Proposition \ref{symmetry}, more specifically, \eqref{symmetryuser},
$(c)$ follows from Lemma \ref{conLemma3} with $\mathcal{G}_i = \mathcal{A}_i \cup \mathcal{E}$,  $\boldsymbol{d}_i = \boldsymbol{b}_{i}$, $l = 0$, $j = N-1$, $\mathcal{G}_i = \mathcal{G}_{i+1} \cup \{N-i-2\} = \mathcal{G}_{i+1} \cup  \{\gamma_{i+1} \} $ for $i \in [N-1]$, and ${\mathcal{G}_{N-1}} = \mathcal{A}_{N-1} \cup \mathcal{E} = \mathcal{E}$,  
$(d)$ follows from Lemma \ref{conLemma3} with $\mathcal{G}_i = \mathcal{B}_{K-i} = [i:K-1]$,  $\boldsymbol{d}_i =  \boldsymbol{a}_{K+N-i}$, $l = N-1$, $j = K$,  $\mathcal{G}_i = \mathcal{G}_{i+1} \cup \{i\} = \mathcal{G}_{i+1} \cup  \{\gamma_{i+1} \} $ for $i \in [N-1:K-1]$, and ${\mathcal{G}_{K}} = \emptyset $,  and 
$(e)$ follows from the assumption that the files are independent, uniformly distributed and $F$ bits each.

The proof of Theorem 1 when $ K > 2N -2 $ is thus complete.

\section{Conclusions} \label{sec7}
   In this paper, some progress has been made towards characterizing the optimal  memory-rate tradeoff  for the demand private coded caching problem.  	More specifically, we present a new virtual-user-based achievable scheme  and two MDS-code-based achievable schemes. 
  Then, we derive a new converse bound for the case \( N \le K \), along with an additional new converse bound for the case \( N = 2 \). 
 As a result,  we obtain the optimal  memory-rate tradeoff of the demand private coded caching problem for $M \in \big[0, \frac{N}{(K+1)(N-1)} \big] $ where   $N \le K \le 2N-2$, and  for  $M \in \big[0, \frac{1}{K+1} \big] $ where $ K > 2N-2$.   
 Finally, we characterize the entire optimal  memory-rate tradeoff curve for any cache size in the case of 2 files and 3 users. 
 As for the case of 2 files and arbitrary number of users, the optimal  memory-rate tradeoff is characterized  for $M\in \big[0,\frac{2}{K} \big] \cup \big[\frac{2(K-1)}{K+1},2 \big]$.  
\appendices{}   

\section{Proof of Decoding Correctness for  Lemma  \ref{achD}} \label{pfLemdecode} 
In this section, we prove the correctness of decoding for the scheme described in Section \ref{secach}. 
More specifically, we need to show that \eqref{achVd1} is correct and that all terms on the  RHS of \eqref{achVd1}  can be obtained from $Z_{kN+n}$ and $X_{\boldsymbol{d}}$.   
Before giving the proof of decoding correctness, we first provide the following two lemmas. 

The first lemma  serves as an intermediate step in the proof of the correctness of  \eqref{achVd1}, and  shows that  the file requested by user \(kN+n\) under  demand  \(\boldsymbol{d}\) is equal to the sum of the files requested by user \(kN+n\) under the demands in the set \(\mathcal{V}_{\boldsymbol{d}}\).   
\begin{Lem} \label{Lemachx2} 
	For any $\mathcal{R} \subseteq \mathcal{T}, |\mathcal{R}| = r , \boldsymbol{d} \in  \mathcal{D},   k \in [K]$, and $ n \in [N] $,   we have
	\begin{align}   \label{Lemachx3} 
	 W_{d_k \oplus_N n ,\mathcal{R}} = &    \bigoplus_{t \in \mathcal{V}_{\boldsymbol{d}} } W_{g_k(t) \oplus_N n ,\mathcal{R}}. 
	\end{align} 
\end{Lem}
\begin{IEEEproof} The proof can be found in Appendix \ref{pfLemachx2}.
\end{IEEEproof}

The following lemma shows that,  although some symbols $X^{(n)}_{\boldsymbol{d},\mathcal{S}}$ defined in \eqref{Xsub1} are not directly delivered in the delivery signal, these symbols can still be recovered from the delivery signal. 
\begin{Lem} \label{Lemachx1}
	For $ \forall t_{\boldsymbol{d}} \in \mathcal{V}_{\boldsymbol{d}} $, 
	$\forall \boldsymbol{d} \in  \mathcal{D} \setminus \mathcal{D}_{0}$, $ \forall \mathcal{S}\subseteq \mathcal{T}, |\mathcal{S}| = r-1$ and $ \forall n \in [N] $,  $X^{(n)}_{\boldsymbol{d},\mathcal{S}}$ can be recovered from $X_{\boldsymbol{d}} $. 
\end{Lem}
\begin{IEEEproof}  The proof of this lemma  is similar to the proof of \cite[Lemma 2]{Chinmay2022}. For completeness, we provide the proof in Appendix \ref{pfLemachx1}.
\end{IEEEproof}  
 
Next, we show the decodability  for the case $\boldsymbol{d}  \in \mathcal{D}_{0}$. If $f(\boldsymbol{d}) \in \mathcal{R} $, for any user $kN+n$,  $k\in[K], n \in[N]$, we have      
\begin{align}  \label{D0} 
	& \bigoplus\limits_{t \in \mathcal{V}_{\boldsymbol{d}}  \setminus \mathcal{R}  } 
	Y^{(k,n)}_{ \{t\}\cup\mathcal{R}}   
	\oplus   \bigoplus\limits_{t \in  \mathcal{R}}
	X_{\boldsymbol{d},\mathcal{R}\setminus \{t\}} ^{(g_k(t) \oplus_N n )} 
	\overset{(a)}{=} 
	 X_{\boldsymbol{d},\mathcal{R}\setminus \{f(\boldsymbol{d})\}} ^{(g_k(f(\boldsymbol{d})) \oplus_N n ) }
	\overset{(b)}{=} X_{\boldsymbol{d},\mathcal{R}\setminus \{f(\boldsymbol{d})\}} ^{(d_k \oplus_N n)} 
	\overset{(c)}{=} 	W_{d_k \oplus_N n , \mathcal{R}},      
\end{align}  
where $(a)$  follows from 
$ \mathcal{V}_{\boldsymbol{d}} = \{f(\boldsymbol{d}) \}$, $  \mathcal{V}_{\boldsymbol{d}}  \setminus \mathcal{R}  = \emptyset $,  and the fact that  $ X_{\boldsymbol{d},\mathcal{R}\setminus \{t\}} ^{(j)}$ equals to a $ \frac{F}{\tbinom{NK-K+1}{r}} $-length zero vector    for all $t\in \mathcal{R} \setminus \{ f(\boldsymbol{d})\} $ and $j \in [N]$,   
$(b)$ follows from the  definitions of the functions $f(\cdot)$ and $g_k(\cdot)$, and  $(c)$ follows from  \eqref{Xsub1}.   
From \eqref{achxd0}  and Remark \ref{remark4}, we note that all terms on the left-hand side (LHS) of \eqref{D0} are composed of some zero vectors and $W_{d_k \oplus_N n , \mathcal{R}}$, which is contained in  $X_{\boldsymbol{d}}$.   
  
If $f(\boldsymbol{d}) \notin \mathcal{R}$, for any user $kN+n$,  $k\in[K], n \in[N]$, we have  
\begin{align} \label{D1} 
	& \bigoplus\limits_{t \in \mathcal{V}_{\boldsymbol{d}} \setminus \mathcal{R}  } 
	Y^{(k,n)}_{ \{t\}\cup\mathcal{R}}    
	\oplus   \bigoplus\limits_{t \in  \mathcal{R}}
	X_{\boldsymbol{d},\mathcal{R}\setminus \{t\}} ^{(g_k(t) \oplus_N n)}  
	\overset{(a)}{=}  
	Y^{(k,n)}_{ \{f(\boldsymbol{d})\}\cup\mathcal{R}}   
	\oplus   \bigoplus \limits_{t \in  \mathcal{R}}
	X_{\boldsymbol{d},\mathcal{R}\setminus \{t\}} ^{(g_k(t) \oplus_N n)}  \nonumber \\ 
	\overset{(b)}{=}  &  
	\bigoplus_{t \in  \{f(\boldsymbol{d})\}\cup\mathcal{R} } W_{g_k(t) \oplus_N n, \{f(\boldsymbol{d})\}\cup\mathcal{R}  \setminus \{t\}}  
	\oplus \bigoplus \limits_{t \in  \mathcal{R}}
	W_{g_k(t) \oplus_N n, \{f(\boldsymbol{d}  )\} \cup \mathcal{R}\setminus \{t\}}   \nonumber \\ 
	=     &  
	W_{g_k(f(\boldsymbol{d} )) \oplus_N n,   \mathcal{R} } 
	\overset{(c)}{=}  W_{d_k  \oplus_N n , \mathcal{R}},  
\end{align}
where $(a)$  follows from $\mathcal{V}_{\boldsymbol{d}}  \setminus \mathcal{R}    =  \mathcal{V}_{\boldsymbol{d}} = \{ f(\boldsymbol{d})\}$,
$(b)$ follows from the design of the cache content and the delivery signal, i.e., \eqref{cache} and  \eqref{Xsub1},  
and $(c)$ follows from the  definitions of the functions $f(\cdot)$ and $g_k(\cdot)$. 
Since from the YMA scheme\cite{Qian2018}, all symbols $Y^{(k,n)}_{\mathcal{R}^+}$  such that $\mathcal{R}^+ \subseteq \mathcal{T} $ and  $|\mathcal{R}^+ | =r+1 $ can be obtained from $X_{\boldsymbol{g}^{(k,n)}}^{\text{YMA}}$, the first term on the LHS of \eqref{D1} can be obtained from cache content, i.e., $Z_{kN+n}$.  From  \eqref{achxd0} and Remark \ref{remark4}, we note that  the second term on the LHS of \eqref{D1} can be obtained from $X_{\boldsymbol{d}} $. 
Thus, for the case  $\boldsymbol{d}  \in \mathcal{D}_{0}$, \eqref{achVd1} is proved to be correct, and all terms on the  RHS of \eqref{achVd1}  can be obtained from $Z_{kN+n}$ and $X_{\boldsymbol{d}}$.      
 
For any  demand $\boldsymbol{d} \in \mathcal{D} \setminus \mathcal{D}_0$, any user $kN+n$,  $k\in[K], n \in[N]$, and  any $  \mathcal{R}$  satisfying  $  \mathcal{R} \subseteq \mathcal{T}$ and  $  |\mathcal{R}| = r $, we have
\begin{align} 
	 W_{d_k \oplus_{N} n, \mathcal{R}}  
	\overset{(a)}{=}  &   \bigoplus_{t \in \mathcal{V}_{\boldsymbol{d}} } W_{ g_k (t) \oplus_N n,\mathcal{R}}   \nonumber \\
    	\overset{(b)}{=}  & \bigoplus_{t \in \mathcal{V}_{\boldsymbol{d} }\cap \mathcal{R} } W_{ g_k (t) \oplus_N n,\mathcal{R}}  \oplus  \bigoplus_{t \in \mathcal{V}_{\boldsymbol{d} } \setminus \mathcal{R}   }  \left(  Y^{(k,n)}_{ \{t\}\cup\mathcal{R}} \oplus  \bigoplus\limits_{u \in \mathcal{R}} W_{ g_k (u) \oplus_N n, \{t\} \cup \mathcal{R} \setminus  \{u\} } \right) \nonumber  \\ 
    = & \bigoplus\limits_{t \in \mathcal{V}_{\boldsymbol{d}} \setminus  \mathcal{R}  } 
    Y^{(k,n)}_{ \{t\}\cup\mathcal{R}} 
    \oplus   
    \bigoplus_{t \in \mathcal{V}_{\boldsymbol{d} }\cap \mathcal{R} }   W_{ g_k (t) \oplus_N n,\mathcal{R}}    \oplus
    \bigoplus\limits_{t \in \mathcal{R}} \bigoplus_{u \in \mathcal{V}_{\boldsymbol{d} } \setminus \mathcal{R}   }     W_{ g_k (t) \oplus_N n, \{u\} \cup \mathcal{R} \setminus  \{t\} }   \nonumber  \\ 
    = & \bigoplus\limits_{t \in \mathcal{V}_{\boldsymbol{d}} \setminus  \mathcal{R}  } 
    Y^{(k,n)}_{ \{t\}\cup\mathcal{R}} 
    \oplus   
    \bigoplus_{t \in \mathcal{V}_{\boldsymbol{d} }\cap \mathcal{R} }  \left(W_{ g_k (t) \oplus_N n,\mathcal{R}}    \oplus \bigoplus_{u \in \mathcal{V}_{\boldsymbol{d} } \setminus \mathcal{R}   }     W_{ g_k (t) \oplus_N n, \{u\} \cup \mathcal{R} \setminus  \{t\} }   \right)  \nonumber \\
    &    \oplus
    \bigoplus\limits_{t \in \mathcal{R}  \setminus\mathcal{V}_{\boldsymbol{d} }   } \bigoplus_{u \in \mathcal{V}_{\boldsymbol{d} } \setminus \mathcal{R}   }     W_{ g_k (t) \oplus_N n, \{u\} \cup \mathcal{R} \setminus  \{t\} }   \nonumber \\
	= & \bigoplus\limits_{t \in \mathcal{V}_{\boldsymbol{d}} \setminus  \mathcal{R}  }  
	Y^{(k,n)}_{ \{t\}\cup\mathcal{R}}    \oplus   \bigoplus\limits_{t \in   \mathcal{R}  }
	\bigoplus\limits_{u \in  \mathcal{V}_{\boldsymbol{d}} \setminus (\mathcal{R} \setminus \{t\})} 
	W_{ g_k (t) \oplus_N n,\{u\} \cup \mathcal{R}\setminus \{t\}}     \nonumber   \\
	\label{pfach_c}
	\overset{(c)}{=} & \bigoplus\limits_{t \in \mathcal{V}_{\boldsymbol{d}} \setminus \mathcal{R}  } 
	Y^{(k,n)}_{ \{t\}\cup\mathcal{R}}   
	\oplus   \bigoplus\limits_{t \in \mathcal{R} }
	X_{\boldsymbol{d},\mathcal{R}\setminus \{t\}} ^{(g_k(t) \oplus_N n ) },  
\end{align}   
where $(a)$ follows from Lemma \ref{Lemachx2}, 
$(b)$ follows from   \eqref{cache}, 
and $(c)$  follows from  the design of the delivery signal, i.e., \eqref{Xsub1}. 
Following the same reasoning used to obtain the first term on the LHS of \eqref{D1} from \( Z_{kN+n} \), and following from Lemma \ref{Lemachx1}, the first term on the RHS of \eqref{pfach_c} can be obtained from \( Z_{kN+n} \), and the second term on the RHS of \eqref{pfach_c} can be obtained from \( X_{\boldsymbol{d}} \).

Hence, the proof of decoding correctness is complete, provided that Lemmas \ref{Lemachx2} and \ref{Lemachx1} are correct.  These will be proved next. 
 
\subsection{Proof of Lemma \ref{Lemachx2}} \label{pfLemachx2} 
For $\boldsymbol{d}\in \mathcal{D}_{0} $, we have  
\begin{align*}  
	\bigoplus_{t \in \mathcal{V}_{\boldsymbol{d}} } W_{g_k(t) \oplus_N n ,\mathcal{R}} 	\overset{(a)}{=}     W_{g_k( f(\boldsymbol{d} ) ) \oplus_N n ,\mathcal{R}}  	\overset{(b)}{=}    W_{d_k \oplus_N n  ,\mathcal{R}},  
\end{align*}   
where $(a)$ follows  from Remark \ref{remark4}, i.e., $ \mathcal{V}_{\boldsymbol{d}}  = \{ f( \boldsymbol{d} )\}$, and 
$(b)$ follows from the  definitions of the functions $f(\cdot)$ and $g_k(\cdot)$.   
 
For  $\boldsymbol{d} = a \mathbf{e}'_0 $, $a \in [1:N-1]$, following from the design of $V_{\boldsymbol{d}}$ and  $\mathcal{V}_{\boldsymbol{d}}$, i.e., \eqref{achVd2},  we have   
\begin{align}  \label{pfL81}
	&  	\bigoplus_{t \in \mathcal{V}_{\boldsymbol{d}} } W_{g_k(t) \oplus_N n ,\mathcal{R}} 
	=     W_{g_k(0) \oplus_N n ,\mathcal{R}}  \oplus  \bigoplus_{b \in [1:a]} \left( W_{g_k(b) \oplus_N n ,\mathcal{R}} 
	  \oplus  W_{g_k((N-1)(K-1)+b) \oplus_N n ,\mathcal{R}}     \right).      
\end{align}  
Substituting the result of  \eqref{gt}  into  \eqref{pfL81}, we have, if  $k = 0$,   
\begin{align} \label{pfL82}   
	&  \text{LHS of  \eqref{pfL81}}    
	  =   W_{n ,\mathcal{R}}  \oplus  \bigoplus_{b \in [1:a]} \left( W_{b \oplus_N n ,\mathcal{R}}  \oplus  W_{(b-1) \oplus_N n ,\mathcal{R}}     \right) 
	= 
	W_{a \oplus_N n ,\mathcal{R}}     \overset{(a)}{=} W_{d_k \oplus_N n,\mathcal{R}},      
\end{align} 
and if  $k \in [1:K-1] $,  
\begin{align} 
	& \text{LHS of  \eqref{pfL81}}  
	=    W_{n ,\mathcal{R}}  \oplus  \bigoplus_{b \in [1:a]} \left( W_{b \oplus_N n ,\mathcal{R}}  \oplus  W_{b \oplus_N n ,\mathcal{R}}     \right) 
	 = 
	 W_{n,\mathcal{R}} \overset{(a)}{=}  W_{d_k \oplus_N n,\mathcal{R}},    
\end{align}  
where  $(a)$ follows  from $\boldsymbol{d} = a \mathbf{e}'_0$.   
Similarly, for  $\boldsymbol{d} = a \mathbf{e}'_i$, $i \in [1:K-2]$, $a \in [1:N-1]$, following from  \eqref{achVd2},   we have
\begin{align}   \label{pfL83} 
	&  	\bigoplus_{t \in \mathcal{V}_{\boldsymbol{d}} } W_{g_k(t) \oplus_N n ,\mathcal{R}} \nonumber \\ 
	 = &    
	W_{g_k(0) \oplus_N n ,\mathcal{R}}  
	\oplus  \bigoplus_{b \in [1:a]} \left( W_{g_k((N-1)(K-i)+b )   \oplus_N n ,\mathcal{R}} 
	\oplus  W_{g_k((N-1)(K-i-1)+b) \oplus_N n ,\mathcal{R}}     \right).   
\end{align}  
Substituting the result of  \eqref{gt}  into  \eqref{pfL83}, we have, if   $k \in [0:i-1]$, 
\begin{align}  \label{pfL84} 
	&  \text{LHS of  \eqref{pfL83}}  
	=   W_{n ,\mathcal{R}}  \oplus  \bigoplus_{b \in [1:a]} \big( W_{(b-1)  \oplus_N n ,\mathcal{R}} 
	 \oplus  W_{(b-1) \oplus_N n ,\mathcal{R}}   \big) =  W_{n,\mathcal{R}} = W_{d_k \oplus_N n,\mathcal{R}},     
\end{align}  
if $k=i$,  
\begin{align} 
	&    \text{LHS of  \eqref{pfL83}}  
	=    W_{n ,\mathcal{R}}  \oplus  \bigoplus_{b \in [1:a]} \left( W_{b \oplus_N n ,\mathcal{R}}  \oplus  W_{(b-1) \oplus_N n ,\mathcal{R}}     \right) 
	  = 
	   W_{ a \oplus_N n,\mathcal{R}}   =   W_{d_k \oplus_N n,\mathcal{R}},    
\end{align}  
and if $k \in [i+1:K-1] $, 
\begin{align} 
		&    \text{LHS of  \eqref{pfL83}}  
	=    W_{n ,\mathcal{R}}  \oplus  \bigoplus_{b \in [1:a]} \left( W_{b \oplus_N n ,\mathcal{R}}  \oplus  W_{b \oplus_N n ,\mathcal{R}}     \right)  
	=  
	W_{n,\mathcal{R}} =   W_{d_k \oplus_N n,\mathcal{R}}.   
\end{align} 
For  $\boldsymbol{d} = a \mathbf{e}'_{K-1}$, $a \in [1:N-1]$, following from  \eqref{achVd2},   we have    
\begin{align} \label{pfL85} 
	&  	\bigoplus_{t \in \mathcal{V}_{\boldsymbol{d}} } W_{g_k(t) \oplus_N n ,\mathcal{R}}   
	=      W_{g_k(0) \oplus_N n ,\mathcal{R}}  
	\oplus  \bigoplus_{b \in [1:a]} \left( W_{g_k(N-1+b) \oplus_N n ,\mathcal{R}}   \oplus  W_{g_k(b-1) \oplus_N n ,\mathcal{R}}     \right).     
\end{align}     
Substituting the result of  \eqref{gt}  into  \eqref{pfL81}, we have,   if $k = K-1$,
\begin{align}
	&  \text{LHS of  \eqref{pfL85}}    
	=   W_{n ,\mathcal{R}}  \oplus  \bigoplus_{b \in [1:a]} \big( W_{b \oplus_N n ,\mathcal{R}}  \oplus  W_{(b-1) \oplus_N n ,\mathcal{R}}     \big)  
	= 
	W_{a \oplus_N n ,\mathcal{R}}  = W_{d_k \oplus_N n,\mathcal{R}}, 
\end{align}
and if $k \in [0:K-2] $,  
\begin{align} \label{pfL86}   
	&    \text{LHS of  \eqref{pfL85}}    
	=     W_{n ,\mathcal{R}}  \oplus  \bigoplus_{b \in [1:a]} \big( W_{(b-1) \oplus_N n ,\mathcal{R}}   
	\oplus  W_{(b-1) \oplus_N n ,\mathcal{R}}     \big)     =      W_{n,\mathcal{R}} =  W_{d_k \oplus_N n,\mathcal{R}}.    
\end{align}  
Combining the results in \eqref{pfL81}-\eqref{pfL86}, we find that   \eqref{Lemachx3} holds for $\boldsymbol{d} \in \mathcal{D}_1$.  

Finally, considering \(\boldsymbol{d} \in \mathcal{D}_2\), we have  
\begin{align*}  
&  \bigoplus_{t \in \mathcal{V}_{\boldsymbol{d}} } W_{g_k(t) \oplus_N n ,\mathcal{R}} 
\overset{(a)}{=}  
 W_{n,\mathcal{R}}  \oplus  (W_{d_k \oplus_N n,\mathcal{R}}  \oplus W_{n,\mathcal{R}}  ) \oplus 
\bigoplus_{i \in[K] \setminus \{k\}}  (   W_{n,\mathcal{R}} \oplus  W_{n,\mathcal{R}} )   
= 
 W_{d_k \oplus_N n,\mathcal{R}} ,
\end{align*}    
where $(a)$ follows from the design of $V_{\boldsymbol{d}}$, i.e.,  \eqref{achVd3},  and the correctness of  \eqref{Lemachx3} for $\boldsymbol{d}\in \mathcal{D}_{0}  \cup  \mathcal{D}_1$.  
 
The proof of Lemma \ref{Lemachx2}  is thus complete.

\subsection{Proof of Lemma \ref{Lemachx1}} \label{pfLemachx1}  
When the randomly chosen $t_{\boldsymbol{d}} \notin \mathcal{S}$, $X^{(n)}_{\boldsymbol{d},\mathcal{S}}$  already exists in $X_{\boldsymbol{d}}$.   
We only need to prove that   any $ X^{(n)}_{\boldsymbol{d},\mathcal{S}}$, $  \mathcal{S} \subseteq   \mathcal{T}, t_{\boldsymbol{d}}  \in \mathcal{S}  $ can be recovered from $ X_{\boldsymbol{d}}$. 
Let $\mathcal{A} = \mathcal{S} \setminus \{t_{\boldsymbol{d} } \}$,   and we have
\begin{align}
X^{(n)}_{\boldsymbol{d},\mathcal{S}} \overset{(a)}{=}  &   
\bigoplus_{t\in \mathcal{V}_{\boldsymbol{d}} \setminus \mathcal{S}} 
W_{n, \mathcal{S}  \cup \{t\} }  
\overset{(b)}{=}   \bigoplus_{t \in \mathcal{V}_{\boldsymbol{d}} \setminus  \mathcal{S}   }  
W_{n, \mathcal{A} \cup \{t,t_{\boldsymbol{d} }\} }   \nonumber  \\ 
\overset{(c)}{=}  &   \bigoplus_{t\in \mathcal{V}_{\boldsymbol{d}} \setminus \mathcal{S}}  
\bigg( 
\bigoplus_{v\in \mathcal{V}_{\boldsymbol{d}} \setminus (\mathcal{A}\cup \{t\}) } 
W_{n,\mathcal{A}  \cup \{t,v\} } 
 \oplus   
\bigoplus_{v \in \mathcal{V}_{\boldsymbol{d}} \setminus (\mathcal{S}\cup \{ t\})}  
W_{n, \mathcal{A} \cup \{t,v\} }  \bigg) 
 \nonumber \\ 
\overset{(d)}{=}  &  \bigoplus_{ t\in \mathcal{V}_{  \boldsymbol{d}} \setminus \mathcal{S} }  \bigg(
X^{(n)}_{\boldsymbol{d},\mathcal{A}\cup \{ t \}}    \oplus
\bigoplus_{v \in \mathcal{V}_{\boldsymbol{d}} \setminus (\mathcal{S}\cup \{ t\})   }  
W_{n, \mathcal{A} \cup \{t,v\} }  \bigg)
 \nonumber \\ 
= &   \bigoplus_{t\in \mathcal{V}_{\boldsymbol{d}} \setminus \mathcal{S} }  
X^{(n)}_{\boldsymbol{d},\mathcal{A}\cup \{ t \}}    \oplus 
\bigoplus_{t\in \mathcal{V}_{\boldsymbol{d}} \setminus \mathcal{S}}  
\bigoplus_{v \in \mathcal{V}_{\boldsymbol{d}} \setminus (\mathcal{S}\cup \{ t\}) }   
W_{n, \mathcal{A} \cup \{t,v\} }  \nonumber \\
\overset{(e)}{=} &   \bigoplus_{t\in \mathcal{V}_{\boldsymbol{d}} \setminus  \mathcal{S}  }  
X^{(n)}_{\boldsymbol{d},\mathcal{A}\cup \{ t \}},  \nonumber 
\end{align}
where  $(a)$ and $(d)$ follow  from the definition of $	X^{(n)}_{\boldsymbol{d},\mathcal{S}} $, i.e.,  \eqref{Xsub1},
$(b)$ follows from the fact that for all $t \in \mathcal{V}_{\boldsymbol{d}} \setminus \mathcal{S}$ and $t_{\boldsymbol{d}} \in \mathcal{S}$, $ t \neq t_{\boldsymbol{d}}$,   
$(c)$ follows from the fact that $    \mathcal{V}_{\boldsymbol{d}} \setminus  (\mathcal{S}\cup \{ t\} )   \cup \{t_{\boldsymbol{d} } \} =   \mathcal{V}_{\boldsymbol{d}} \setminus  (\mathcal{A}\cup \{t\})  $, 
and $(e)$  follows from the fact that every term $W_{n, \mathcal{A} \cup \{t,v\} }  $ appears twice in the binary addition.

The proof of Lemma \ref{Lemachx1}  is thus complete. 

\section{Proof of Privacy for  Scheme in Subsection \ref{secachB} }  \label{schB_pfP}  
In this section, for a demand vector  $ \boldsymbol{D}$ and $n \in [N]$,  we use $\mathcal{K}_n$ to denote  the set of users requesting file $W_n$.   
With this new defined notation, we now present the proof of privacy for the scheme proposed  in Subsection \ref{secachB}.   
 
First, considering the part related to the file content within the cache content and the delivery signal, i.e., \( Z'_k \) and \( X'_{\boldsymbol{D}} \), for any fixed $ \left  (D_{[K]},\boldsymbol{p}_{[N]}, \pi_{[N]}, P_{[K]},S_{[K+1]}  \right ) $ and  $k \in [K]$,  we have    
	\begin{align} \label{schB_privacy1}
		& H \left(Z'_k, X'_{\boldsymbol{D}} \right) 
		\overset{(a)}  \ge   H \left( W_{D_k}, Z_{k,D_k}, \big(X_{\boldsymbol{D},n}\big)_{n \in  [N]   \setminus \{D_k\}} \right) \nonumber \\
		\overset{(b)}  =  &  H ( W_{D_k}) + H ( Z_{k,D_k}) + H \left(  
		\big(X_{\boldsymbol{D},n} \big)_{n \in  [N] \setminus \{D_k\}}  \Big| Z_{k,D_k} \right) \nonumber \\ 
		\overset{(c)}\ge  &   F + \frac{1}{(K+1)(N-1)}F  
		+ \sum_{n \in  [N] \setminus \{D_k\}  } H \left(X_{\boldsymbol{D},n}  \Big|w_{n,D_k}^{(k)} \right) \nonumber \\   
		\overset{(d)} = &    
		\frac{KN(N-1)+1}{(K+1)(N-1)}F,  
	\end{align} 
	where $(a)$ follows from user $k$'s decoding steps for $W_{D_k}$,   
	$(b)$ follows from the fact that $  \Big(  Z_{k,D_k}$, $ 
	 \big(X_{\boldsymbol{D},n}\big)_{n \in [N] \setminus \{D_k\}} \Big) $ is independent of $W_{D_k}$,      
	$(c)$ follows from \eqref{schBZkn},    
	$H \left( X_{\boldsymbol{D},n}|W_n \right) = 0$, and the fact that $W_0, W_1, \dots, W_{N-1}$ are independent and each has $F$ bits, and $(d)$ follows from the fact that, for $n \neq D_k$, $ H  \left( X_{\boldsymbol{D},n} \Big|w_{n,D_k}^{(k)} \right) = \frac{K(N-1)-1}{(K+1)(N-1)}F$,  which will be proved in the following equations, i.e., \eqref{schB_privacy2}  and \eqref{schB_privacy3}.     
	\begin{itemize} 
		\item 
		For $k \in [K]$ and $ n \in  \mathcal{N}_e(\boldsymbol{D})  \setminus \{D_k\} $, following	from the design of $X_{\boldsymbol{D},n}$, i.e., \eqref{schB_Y11}-\eqref{schB_Xn},    
		we have \begin{align}    \label{schB_privacy2} 
			& H  \left( X_{\boldsymbol{D},n} \Big|w_{n,D_k}^{(k)} \right)   
			=  
			H \left(w_{n,n}^{(u_n)},  \big(w_{n,n}^{(u_n)}  \oplus w_{n,D_i}^{(i)} \big)_{i \in [K] \setminus \{u_n,k \}}, 
			V_{\boldsymbol{D},n}^{(u_n)}, \big(X_{\boldsymbol{D},n}^{(i)} \big)_{i \in [K] }    \Big| w_{n,D_k}^{(k)}  \right) \nonumber  \\ 
			=  &  H \left( \big(w_{n,D_i}^{(i)} \big)_{i \in [K] \setminus \{k\}}, \big(w_{n,m}^{(i)} \big)_{i \in  \mathcal{K}_n ,m\in [N:2N-3] },  
			 \big(w_{n,m}^{(i)}\big)_{ i \in [K] \setminus \mathcal{K}_n, m\in [N] \setminus \{n,D_i\} }    \Big| w_{n,D_k}^{(k)}  \right) \nonumber  \\  
			=  &  \frac{K(N-1)-1}{(K+1)(N-1)}F, 
		\end{align}  
		where the last step follows from the fact that 	
		$\big(w_{n,D_i}^{(i)}\big)_{i \in [K]}$, $ \big(w_{n,m}^{(i)}\big)_{i \in  \mathcal{K}_n ,m\in [N:2N-3] }$   and  $\big(w_{n,m}^{(i)}\big)_{ i \in [K] \setminus \mathcal{K}_n, m\in [N] \setminus \{n,D_i\} }$  are   $K(N-1)$ different $\frac{F}{(K+1)(N-1)}$-length  MDS-coded segments of file $W_n$.      
	
		\item For $k \in [K]$ and  $n \in [K] \setminus  \mathcal{N}_e(\boldsymbol{D})$,  similarly,  we have 
		\begin{align}    \label{schB_privacy3}  
			& H \left(  X_{\boldsymbol{D},n} \Big| w_{n,D_k}^{(k)}  \right)  \nonumber \\ 
			=  &  H \left(   \big(w_{n,m}^{(i)}\big)_{ i \in [K], m\in [N] \setminus \{n,D_i\} }, 
			 \big(w_{n,D_1}^{(1)} \oplus w_{n,h_n(D_0)}^{(0)} \oplus w_{n,D_i}^{(i)} \big)_{i \in [K] \setminus \{1\}}   \Big| w_{n,D_k}^{(k)}  \right) \nonumber  \\   
			= &  H \left(    \big(w_{n,m}^{(i)} \big)_{ i \in [K], m\in [N] \setminus \{n\} }   \right) - H \left(   w_{n,D_k}^{(k)} \right)  \nonumber  \\  
			= &  \frac{K(N-1)-1}{(K+1)(N-1)}F.    
		\end{align}
	\end{itemize}	  
	Following from \eqref{schB_privacy1}, since   the lengths of $X'_{\boldsymbol{D}}$ and $Z_k'$ add up to $\frac{KN(N-1)+1}{(K+1)(N-1)}F$, $H \left( X'_{\boldsymbol{D}}, Z'_k \big| W_{[N]}\right) = 0$, and  $W_{[N]}$ is independent of  $ \left(  D_{[K]},\boldsymbol{p}_{[N]}, \pi_{[N]}, P_{[K]},S_{[K+1]} \right) $, for any fixed   $ \big( D_{[K]},\boldsymbol{p}_{[N]}, \pi_{[N]},$  $ P_{[K]},S_{[K+1]} \big) $, we have,  
	\begin{align} \label{schB_privacy4} 
		& \left( X'_{\boldsymbol{D} }, Z'_k  \Big| D_{[K]},\boldsymbol{p}_{[N]}, \pi_{[N]}, P_{[K]},S_{[K+1]}    \right)  
		\sim \text{Unif} \left( \{0,1\}^{ \frac{KN(N-1)+1}{(K+1)(N-1)}F}\right).  
	\end{align}  
 
Next, we consider the auxiliary variables.  Following from the definition of $J_{0,n}$, i.e., \eqref{schB_J011}, \eqref{schB_J010} and  \eqref{schB_J32}, we can observe that $J_{0,n}$ consists of  $K(N-1)-1$ different  indices of file $n$, i.e., $\big(p_{n,D_i}^{(i) }\big)_{i \in [K] \setminus \{u_n\}}$,   $ \big(p_{n,m}^{(i)}\big)_{i \in  \mathcal{K}_n ,m\in [N:2N-3] }$,   and $\big(p_{n,m}^{(i)}\big)_{ i \in [K] \setminus \mathcal{K}_n, m\in [N] \setminus \{n,D_i\} }$.     
Furthermore, following from \eqref{schB_T}, 
$ \Big(	J_{0,D_k}, \big(T^{(k)}_{1,n}\big)_{n \in [N]}   \Big)$ consists of  $(K+1)(N-1)$ different    indices of file $D_k$,  
i.e., $\big(p_{D_k,n}^{(k)}\big)_{n \in  [N]\setminus \{D_k\}}$, $ \big(p_{D_k,n}^{(i)}\big)_{i \in [K] \setminus \mathcal{K}_{D_k}, n\in [N] \setminus \{D_i\}}$, and  $ \big( p_{D_k,n}^{(i)} \big)_{i \in  \mathcal{K}_{D_k},n\in [N:2N-3] \cup \{D_k\}} $.

Based on the above observations,  for $k \in [1:K-1]$, we have 
\begin{align} \label{schB_privacy5}   
	& \Pr \left(	J_0, \big(T^{(k)}_{1,n} \big)_{n\in [N]},\big(T^{(k)}_{2,n}\big)_{n\in [N]} \Big |D_{[K]}\right) \nonumber   \\
	\overset{(a)} =  & \Pr \left(J_{0,D_k},(T^{(k)}_{1,n})_{n\in [N]} \Big|D_{[K]}\right)  \Pr \left(  \pi_{D_k, u_{D_k} } \Big|D_{[K]} \right) 
	 \times \prod_{n \in [N] \setminus \{D_k\}}  \left(\Pr  \left(J_{0,n}  \big|D_{[K]} \right) \Pr   \left( \pi_{n,k}   |D_{[K]}  \right)    \right)  \nonumber \\
	\overset{(b)} =  & \frac{1}{K^N}  \frac{ ((K-1)(N-1) )!}{(2K(N-1))!}  \left(  \frac{  (KN-K+1)!}{(2K(N-1))!}  \right)  ^{N-1},     
\end{align} 
and for $k=0$, we have  
\begin{align}  \label{schB_privacy6} 
	& \Pr \left(	J_0,  \big(T^{(0)}_{1,n} \big)_{n\in [N]},\big(T^{(0)}_{2,n}\big)_{n\in [N]}, 
	  \left( T_{3,n}\right)_{n\in[N] \setminus \{D_0\}} \Big |D_{[K]}\right) \nonumber   \\ 
	\overset{(a)} =  & \Pr \left(J_{0,D_0}, \big(T^{(0)}_{1,n}\big)_{n\in [N]} \Big|D_{[K]}\right)  \Pr \left(  \pi_{D_0, u_{D_0} } \big|D_{[K]} \right) \nonumber  \\ 
	& \times \prod_{n \in [N] \setminus \{D_0\}}  \left( \Pr \left( J_{0,n}   \big|D_{[K]} \right)  \Pr  \left( \pi_{n,0}, \pi_{n,u_n} \big|D_{[K]} \right)    \right)  \nonumber \\
	\overset{(b)} = &    \frac{1}{K^N(K-1)^{N-1}}  
	\frac{((K-1)(N-1))!}{(2K(N-1))!} 
	 \times \left(    \frac{(KN-K+1)!}{(2K(N-1))!}  \right)^{N-1},   
\end{align} 
where $(a)$ follows from \eqref{schB_T}, \eqref{schB_T2}, and the fact that the permutations $ \boldsymbol{p}_0,  \boldsymbol{p}_1,  \dots,  \boldsymbol{p}_{N-1} ,  \pi_{0}, \allowbreak \pi_{1}, \dots,  \pi_{N-1}$ are mutually independent, 
and $(b)$ follows from the above observations regarding $J_0$ and $\big(T^{(k)}_{1,n} \big)_{n\in [N]}$,  and the fact that $ \boldsymbol{p}_n$ and $\pi_n$ are uniformly selected from all possible permutations of  $[2K(N-1)]$ and $[K]$, respectively.    
Following from \eqref{schB_privacy5} and  \eqref{schB_privacy6}, we can conclude that 
$ \left( J_0, \big(T^{(k)}_{1,n}\big)_{n\in [N]},\big(T^{(k)}_{2,n}\big)_{n\in [N]}\right)$, $k \in [1:K-1]$, and    $ \left( J_0, \big(T^{(0)}_{1,n} \big)_{n\in [N]},   \big(T^{(0)}_{2,n} \big)_{n\in [N]},  \left(   T_{3,n} \right)_{n\in[N] \setminus \{D_0\}} \right)$  are  each independent of $D_{[K]}$.   
   
Finally, for user $k \in [1:K-1]$, we have 
\begin{align*}
	&   
	I \left(  D_{[K]\setminus \{k\}}; X_{\boldsymbol{D}}  \big|Z_k, D_k  \right)   
	=  
	I   \left( D_{[K]\setminus \{k\}}; X'_{\boldsymbol{D}}, J  \big|Z'_k, P_k, S_k, D_k\right)    \nonumber \\
	\overset{(a)} = &  
	I  \left(D_{[K]\setminus \{k\}};  J_0,J_1,J_2,J_3 \big| S_k, P_k, D_k \right)  \\
	\overset{(b)} = &   
	I  \left(D_{[K]\setminus \{k\}}; J_0, \big(T^{(k)}_{1,n}\big)_{n\in [N]},\big(T^{(k)}_{2,n}\big)_{n\in [N]}  \Big| D_k  \right) 	\overset{(c)} = 0, 
\end{align*}
and for $k=0$, we have 
\begin{align*}
	&  I \left( D_{[K]\setminus \{k\}}; X_{\boldsymbol{D}}  \big|Z_k, D_k  \right)  
	=  
	I \left( D_{[K]\setminus \{0\}}; X'_{\boldsymbol{D}}, J  \big|Z'_0, P_0, S_0, D_0,S_K \right)  \nonumber \\
	\overset{(d)} = &  
	I   \left(D_{[K]\setminus \{0\}}; J_0,J_1,J_2,J_3  \big | P_0, S_0, D_0, S_K  \right)  \\ 
	\overset{(b)} = &  
	I\left( D_{[K] \setminus \{0\} }; J_0, \big(T^{(0)}_{1,n} \big)_{n\in [N]},   \big(T^{(0)}_{2,n} \big)_{n\in [N]}, 
	\left( T_{3,n} \right)_{n\in[N] \setminus \{D_0\}} \Big |  D_0 \right)   \overset{(e)} =   0.    
\end{align*}    
where $(a)$  follows from  \eqref{schB_privacy4} and the fact that  $ \left(J,S_k,P_k,D_{[K]} \right)$ can be determined by  $\big( D_{[K]},  \boldsymbol{p}_{[N]}, \allowbreak  \pi_{[N]}, P_{[K]},S_{[K+1]}\big)$,  
$(b)$ follows from the fact that $T_{1,n}^{(i)}$ and $T_{2,n}^{(i)}$,  $n \in [N]$, $i \in [K]$, are encrypted using one-time pads $P_{i,n}$ and $S_{i,n}$, which are available only to user $i$, and the fact that $T_{3,n}$, $n \in [N] \setminus \{D_0\}  $,  is encrypted using a one-time pad $S_{K,n}$, which is available only to user $0$,  
$(c)$  follows from the fact that $ \left( J_0, \big(T^{(k)}_{1,n}\big)_{n\in [N]},\big(T^{(k)}_{2,n}\big)_{n\in [N]}\right)$, $k \in [1:K-1]$, is independent of $D_{[K]}$,  
$(d)$ follows from  \eqref{schB_privacy4} and the fact that  $ \left(J,S_0,P_0,S_K,D_{[K]} \right)$ can be determined by  $\left( D_{[K]},\boldsymbol{p}_{[N]}, \pi_{[N]}, P_{[K]},S_{[K+1]}\right)$,  
and $(e)$ follows from the fact that $ \Big( J_0, \big(T^{(0)}_{1,n} \big)_{n\in [N]},   \big(T^{(0)}_{2,n} \big)_{n\in [N]}, \allowbreak  \left(  T_{3,n}\right)_{n\in[N] \setminus \{D_0\}} \Big)$ is independent of $D_{[K]}$.  

The proof of privacy for the scheme proposed in Subsection \ref{secachB} is thus complete.

\section{Proof of Lemma \ref{con1} }   \label{subseccon3}   
Recalling the definitions of  $\boldsymbol{a}_i$ and $\boldsymbol{b}_i$  in Subsection \ref{subseccon2},  
for $N=2$ and $i \in [K]$,  we further define  $\boldsymbol{c}_i = \boldsymbol{a}_{i+2}$, $i \in [K-1]$, and $\boldsymbol{c}_{K-1} = \boldsymbol{b}_{0}=(1,1,\dots,1)$.  We note that $\boldsymbol{a}_{1} = (0,1,1,\dots,1)$, and in the demand  $\boldsymbol{c}_i$, where $i \in [K-1]$, user $K-i-1$ requests $W_0$, while all other users request $W_1$. 
By replacing all $Z_i$ with $\boldsymbol{c}_i$, all $\boldsymbol{a}_{i+2}$ with $Z_i$, $\boldsymbol{a}_{0}$ and  $\boldsymbol{b}_{0}$  with $Z_0$, and   $\boldsymbol{a}_{1}$ and $\boldsymbol{b}_{1}$ with $Z_{1}$ or $Z_{K-1}$,  the proof of Lemma \ref{con1} follows the same steps as the proof of Theorem \ref{con0} when $N=2$ and $K>2N-2$, except for a few steps which use the privacy constraints. The detailed proof is presented as follows.

First, recalling that $\mathcal{B}_{i} = [K-i:K-1]$,  and following from a similar recursive approach as presented in the proof of Lemma \ref{conLemma3}, we have  
\begin{align}  \label{conLemma4} 
	&  H \left(W_{1}, \left( X_{\boldsymbol{c}_j} \right)_{j \in \mathcal{B}_K } \right)   
	+ \sum_{i=1}^{K-1}  H\left( W_1, Z_{i},\left(  X_{\boldsymbol{c}_j} \right)_{j \in \mathcal{B}_i }\right) \nonumber \\     
	\overset {(a)} \ge &     H\left( W_1,Z_{K-1},  \left( X_{\boldsymbol{c}_j} \right)_{j \in \mathcal{B}_K }  \right)  +   H \left(W_{1}, \left( X_{\boldsymbol{c}_j} \right)_{j \in \mathcal{B}_{K-1} } \right)  
	+ \sum_{i=1}^{K-2}  H\left( W_1, Z_{i},\left(  X_{\boldsymbol{c}_j} \right)_{j \in \mathcal{B}_i }\right)  \nonumber \\  
	\overset {(b)} \ge &     H(W_0,W_1) +   H \left(W_{1}, \left( X_{\boldsymbol{c}_j} \right)_{j \in \mathcal{B}_{K-1} } \right)   
	+ \sum_{i=1}^{K-2}  H\left( W_1, Z_{i},\left(  X_{\boldsymbol{c}_j} \right)_{j \in \mathcal{B}_i }\right)  \nonumber \\ 
	\ge  &   \dots  \overset {(c)} \ge       (K-1)H(W_0,W_1) +  H \left(W_{1}, X_{\boldsymbol{c}_{K-1}}  \right),  
\end{align}
where $(a)$ follows from the submodularity of the entropy function, 
$(b)$   follows from the correctness constraint \eqref{decoding},     
and $(c)$ follows from repeating steps $(a)$ and $(b)$.

Finally, for $N=2$ and  $K>2N-2=2$, we have 
\begin{align*}
	&  \frac{(K+1)(K+2)}{4} MF + \frac{K(K+1)}{2}   RF  \nonumber \\   
	= &  KRF + \frac{K+2}{2}MF  + \sum_{i=1}^{K-1} \left( iRF   + \frac{i+1}{2} MF \right)  \\ 
	\overset {(a)}  { \ge } 
	& \sum_{j \in [K]} H \left( Z_0, X_{\boldsymbol{c}_j }\right) - \frac{K-2}{2} H(Z_0) 
	+ \sum_{i=1}^{K-1} \left( \sum_{j \in \mathcal{B}_i } H \left( Z_{i},X_{\boldsymbol{c}_j}\right) -  \frac{i-1}{2} H(Z_{i})    \right) \nonumber \\   
	\overset {(b)} { = } 
	& \sum_{j \in [K]} H  \left( W_1, Z_0, X_{\boldsymbol{c}_j } \right) - \frac{K-2}{2} H(Z_0) 
	+ \sum_{i=1}^{K-1} \left( \sum_{j \in \mathcal{B}_i } H \left( W_1, Z_{i},X_{\boldsymbol{c}_j} \right) -  \frac{i-1}{2} H(Z_{i})    \right)\nonumber \\ 
	\overset {(c)} { \ge } 
	&  H \left(W_{1}, \left( X_{\boldsymbol{c}_j} \right)_{j \in [K]}\right)  + \frac{K-2}{2}  \left(2H(W_1,Z_0) -   H(Z_0)  \right)  
	+  H(W_1,Z_0)  \nonumber \\ 
	& + \sum_{i=1}^{K-1} \bigg(   H \left(W_1, Z_{i},\left(  X_{\boldsymbol{c}_j} \right)_{j \in \mathcal{B}_i } \right) 
	+   \frac{i-1}{2}  \left( 2H(W_1, Z_{i}) -   H(Z_{i})  \right)    \bigg)  \nonumber \\   
	\overset {(d)} { \ge } 
	&    (K-1)H \left( W_0,W_1\right) +  H \left(W_{1}, X_{\boldsymbol{c}_{K-1}}  \right)  +  H(W_1,Z_0) 
	+ \left( \frac{K-2}{2} + \sum_{i=1}^{K-1} \frac{i-1}{2}  \right)   H \left( W_0,W_1\right)   \nonumber \\ 
	\overset {(e)} { =  } 
	&  (K-1)H(W_0,W_1) +  H \left(W_{1}, X_{\boldsymbol{a}_{1}}  \right)  +  H(W_1,Z_0) 
	+ \frac{(K+1)(K-2)}{4}H(W_0,W_1)   \nonumber \\
	\overset {(f)} { \ge }  
	& (K-1) H(W_0,W_1)   +  H(W_1) + H( W_1,Z_0, X_{\boldsymbol{a}_{1}} ) 
	+ \frac{(K+1)(K-2)}{4} H(W_0,W_1)  \nonumber \\  
	\overset {(g)} { =   } &  \frac{K(K+3)}{2} F,    
\end{align*} 
where $(a)$ follows from the facts that $H(Z_j) \leq MF$ and $H(X_{\boldsymbol{D}}) \leq RF$,   for $\forall j \in [K], \forall \boldsymbol{D} \in \mathcal{D}$, 
$(b)$ follows from the correctness constraint \eqref{decoding},    
$(c)$ and $(f)$ follow from the submodularity of the entropy function,  
the first two terms of $(d)$ follow from \eqref{conLemma4},  the last two terms of $(d)$ follow from \eqref{symmetryfile}, more specifically,  
\begin{align}
	2H(W_1,Z_i) \overset {\text{{\color{red} }}}   =  & H(W_0,Z_i) +  H(W_1,Z_i)  
	\ge 
	H(W_0,W_1) + H(Z_i), \quad \forall i \in [K],  \nonumber  
\end{align}   
$(e)$  follows from the privacy constraint, more specifically, Lemma \ref{Lemconp},  and   
$(g)$ follows from the correctness constraint \eqref{decoding}, and  the assumption that the files are independent,  uniformly distributed and $F$ bits each.

In addition, for the case of $N=2, K=2$, the inequality $3M+3R \ge 5$ is proved in Theorem \ref{con0}.  
	
The proof of Lemma \ref{con1} is thus complete.   
 
\section{Proof of Theorem  \ref{Theop2} } \label{pfCoro}  
\subsection {Proof of  \eqref{cor1} } \label{pfCoro1}
From Theorem \ref{achD}, let $N =2$ and  $r=K+1,K,K-1$,  the $(M,R)$ given by $(0,2), \Big( \frac{1}{K+1},\frac{2K}{K+1}\Big)$ and $ \Big(\frac{2}{K},\frac{2(K-1)}{K+1} \Big)$ 
are achievable. 
From \cite[Theorem 2]{Chinmay2022}, let $N =2$ and  $r=K+1,K,K-1$,  the $(M,R)$ given by $(2,0), \Big(\frac{2K}{K+1},\frac{1}{K+1} \Big)$ and $ \Big(\frac{2(K-1)}{K+1},\frac{2}{K} \Big)$ are achievable.  
The lower convex envelope of the above memory-rate pairs can be achieved by memory-sharing.   

Next, from Theorem \ref{con0} and Lemma \ref{con1}, let $N=2$, we get  
\begin{align*}
(K+1)(K+2) M + 2K(K+1)R &\ge 2K(K+3), \text{ and }   \\
2K(K+1)M + (K+1)(K+2) R &\ge 2K(K+3)  .
\end{align*} 
From \cite[Theorem 2]{MaddahAli2014}, which is a converse result for the traditional coded caching problem not considering privacy, let $N=2, s=1,2$, we get $2M + R \ge 2$ and $M+2R \ge 2$. 
Thus, for the $(2,K)$ demand private coded caching problem, any $(M, R)$ pair must satisfy
\begin{align*} 
R^{*p}_{N,K}(M) \ge & \max \bigg\{2-2M, \frac{2K(K+3)}{(K+1)(K+2)}-\frac{2K}{K+2}M ,
 \frac{K+3}{K+1}-\frac{K+2}{2K}M ,1- \frac{1}{2}M
\bigg\}.
\end{align*}
The above converse and achievability results meet when $M\in \left[0,\frac{2}{K}\right] \cup \big[\frac{2(K-1)}{K+1},2\big]$. 

The proof of \eqref{cor1} is thus complete. 

\subsection {Proof of  \eqref{cor2}  } \label{pfCoro2} 
From Theorem \ref{con0}, let $N=2,K=2$,  we get $3M + 3R \ge 5 $.     
Following the definition of privacy, i.e., \eqref{privacy2}, we note that any two users among the   $K = 3 $ users in the system and $N = 2$ files can be viewed as an $ (N, K)= (2,2)$ demand private coded caching system.   Thus, $3M + 3R \ge 5 $ also serves as a converse bound for the $ (N, K) = (2,3)$ demand private coded caching problem.  
From Theorem \ref{con0} and Lemma \ref{con1}, let $N=2,K=3$, we get $6M + 5R \ge 9 $  and $ 5M + 6R \ge 9 $.  
From \cite[Theorem 2]{MaddahAli2014}, which is a converse result for the traditional coded caching problem not considering privacy, let $N=2, s=1,2$, we get $2M + R \ge 2$ and $M+2R \ge 2$. 
Combining the above converse results, for the $(2,3)$ demand private coded caching problem, any $(M, R)$ pair must satisfy 
\begin{align*} 
& R_{N,K}^{*p}(M)   \ge \max \bigg \{ 2-2M, \frac{9-6M}{5}, \frac{5-3M}{3}, 
\frac{9-5M}{6} ,  \frac{2-M}{2} \bigg \}.
\end{align*} 
The six corner points, i.e., $(0,2), \big(\frac{1}{4},\frac{3}{2}\big),\big(\frac{2}{3},1\big),\big(1,\frac{2}{3}\big) ,\big(\frac{3}{2},\frac{1}{4}\big)$ and $(2,0)$ can be obtained from the following schemes.  
From Theorem \ref{achD}, let $N =2, K=3$ and  $r=4,3,2$,  the $(M,R)$ given by $(0,2), \big(\frac{1}{4},\frac{3}{2}\big)$ and $\big(\frac{2}{3},1\big)$ are achievable.
From \cite[Theorem 2]{Chinmay2022}, let $N =2, K=3$ and  $r=4,3,2$,  the $(M,R)$ given by $(2,0)$, $\big(\frac{3}{2},\frac{1}{4}\big)$ and $\big(1,\frac{2}{3}\big)$ are achievable.

The proof of  \eqref{cor2} is thus complete.

\bibliographystyle{IEEEtran}
\bibliography{ref}

\end{document}